\documentclass[acmsmall,nonacm]{acmart}

\AtBeginDocument{%
  \providecommand\BibTeX{{%
    \normalfont B\kern-0.5em{\scshape i\kern-0.25em b}\kern-0.8em\TeX}}}

\renewcommand\footnotetextcopyrightpermission[1]{} %

\setcopyright{acmcopyright}
\copyrightyear{2021}
\acmYear{2021}
\acmDOI{10.1145/1122445.1122456}

\acmJournal{CSUR}
\acmVolume{37}
\acmNumber{4}
\acmArticle{0}
\acmMonth{8}

\usepackage{array}
\usepackage[acronyms,nonumberlist,nopostdot,nomain,nogroupskip]{glossaries}
\usepackage{makecell}
\usepackage{pifont}
\usepackage[normalem]{ulem}
\usepackage{xspace}
\usepackage{enumitem}

\newcommand{\myhref}[3][blue]{\href{#2}{\color{#1}{#3}}}%

\newacronym[plural=\gls{dnn}s,firstplural=deep neural networks (DNNs)]{dnn}{DNN}{deep neural network}
\newacronym[plural=\gls{cnn}s,firstplural=convolutional neural networks (CNNs)]{cnn}{CNN}{convolutional neural network}
\newacronym{ae}{AE}{Autoencoder}

\newacronym{kd}{KD}{knowledge distillation}
\newacronym{hnd}{HND}{head network distillation}
\newacronym{ghnd}{GHND}{generalized head network distillation}
\newacronym{cv}{CV}{computer vision}
\newacronym{nlp}{NLP}{natural language processing}
\newacronym{dl}{DL}{deep learning}
\newacronym{dsp}{DSP}{digital signal processing}
\newacronym{iot}{IoT}{Internet of Things}
\newacronym{sc}{SC}{split computing}
\newacronym{ee}{EE}{early exiting}
\newacronym{lc}{LC}{local computing}
\newacronym{ec}{EC}{edge computing}

\newcommand{\accmetric}{\textsf{\bf A}\xspace} %
\newcommand{\compmetric}{\textsf{\bf C}\xspace} %
\newcommand{\datametric}{\textsf{\bf D}\xspace} %
\newcommand{\energymetric}{\textsf{\bf E}\xspace} %
\newcommand{\latencymetric}{\textsf{\bf L}\xspace} %
\newcommand{\traincostmetric}{\textsf{\bf T}\xspace} %

\makeatletter

\newcommand{\dynscriptsize}{\check@mathfonts\fontsize{\sf@size}{\z@}\selectfont}
\makeatother
\newcommand\textunderset[2]{%
  \leavevmode
  \vtop{\offinterlineskip
    \halign{%
      \hfil##\hfil\cr %
      \strut#2\cr
      \noalign{\kern-.3ex}
      \dynscriptsize\strut#1\cr
    }%
  }%
}

\begin{document}

\title{Split Computing and Early Exiting for Deep Learning Applications: Survey and Research Challenges}

\author{Yoshitomo Matsubara}
\email{yoshitom@uci.edu}
\orcid{0000-0002-5620-0760}
\author{Marco Levorato}
\email{levorato@uci.edu}
\affiliation{%
  \institution{University of California, Irvine}
  \city{Irvine}
  \state{California}
  \country{USA}
  \postcode{92697}
}

\author{Francesco Restuccia}
\affiliation{%
  \institution{Northeastern University}
  \streetaddress{360 Huntington Ave}
  \city{Boston}
  \state{Massachusetts}
  \country{USA}}
  \postcode{02115}
\email{f.restuccia@northeastern.edu}

\renewcommand{\shortauthors}{Matsubara et al.}

\begin{abstract}
Mobile devices such as smartphones and autonomous vehicles increasingly rely on \glspl{dnn} to execute complex inference tasks such as image classification and speech recognition, among others. However, continuously executing the entire \gls{dnn} on mobile devices can quickly deplete their battery. Although task offloading to cloud/edge servers may decrease the mobile device's computational burden, erratic patterns in channel quality, network, and edge server load can lead to a significant delay in task execution. Recently, approaches based on \gls{sc} have been proposed, where the \gls{dnn} is split into a head and a tail model, executed respectively on the mobile device and on the edge server. Ultimately, this may reduce bandwidth usage as well as energy consumption. Another approach, called \gls{ee}, trains models to embed multiple ``exits'' earlier in the architecture, each providing increasingly higher target accuracy. Therefore, the trade-off between accuracy and delay can be tuned according to the current conditions or application demands. In this paper, we provide a comprehensive survey of the state of the art in \gls{sc} and \gls{ee} strategies by presenting a comparison of the most relevant approaches. We conclude the paper by providing a set of compelling research challenges.
\end{abstract}

\ccsdesc[500]{Human-centered computing~Ubiquitous and mobile computing}
\ccsdesc[300]{Computer systems organization~Embedded and cyber-physical systems}
\ccsdesc[500]{Computing methodologies~Neural networks}
\keywords{Split Computing, Edge Computing, Early Exit, Neural Networks, Deep Learning}

\maketitle

\glsresetall

\section{Introduction}
The field of \gls{dl} has evolved at an impressive pace over the last few years \cite{lecun2015deep}, with new breakthroughs continuously appearing in domains such as \gls{cv}, \gls{nlp}, \gls{dsp}, and wireless networking \cite{jagannath2019machine,restuccia2020deep} among others -- we refer to \cite{pouyanfar2018survey} for a comprehensive survey on \gls{dl}. For example, today's state of the art \glspl{dnn} can classify thousands of images with unprecedented accuracy \cite{huang2017densely}, while bleeding-edge advances in deep reinforcement learning (DRL) have shown to provide near-human capabilities in a multitude of complex optimization tasks, from playing dozens of Atari video games \cite{mnih2013playing} to winning games of Go against top-tier players \cite{silver2017mastering}. 

As \gls{dl}-based classifiers improve their predictive accuracy, mobile applications such as speech recognition in smartphones \cite{deng2013new,hinton2012deep}, real-time unmanned navigation \cite{padhy2018deep} and drone-based surveillance \cite{singh2018eye,zhang2020person} are increasingly using \glspl{dnn} to perform complex inference tasks.
However, \textit{state-of-the-art \gls{dnn} models present computational requirements that cannot be satisfied by the majority of the mobile devices available today}.
In fact, many state-of-the-art \gls{dnn} models for difficult tasks -- such as computer vision and natural language processing -- are extremely complex.
For instance, the EfficientDet~\cite{tan2020efficientdet} family offers the best performance for object detection tasks. While EfficientDet-D7 achieves a mean average precision (mAP) of 52.2\%, it involves 52M parameters and will take seconds to be executed on strong embedded devices equipped with GPUs such as the NVIDIA Jetson Nano and Raspberry Pi.
Notably, the execution of such complex models significantly increases energy consumption. While lightweight models specifically designed for mobile devices exist~\cite{tan2019mnasnet,sandler2018mobilenetv2}, the reduced computational burden usually comes to the detriment of the model accuracy. For example, compared to ResNet-152~\cite{he2016deep}, the networks MnasNet~\cite{tan2019mnasnet} and MobileNetV2~\cite{sandler2018mobilenetv2} present up to 6.4\% accuracy loss on the ImageNet dataset. 
YOLO-Lite~\cite{redmon2018yolov3} achieves a frame rate of 22 frames per second on some embedded devices but has a mean average precision (mAP) of 12.36\% on the COCO dataset~\cite{lin2014microsoft}.
To achieve 33.8\% mAP on the COCO dataset, even the simplest model in the EfficientDet family, EfficientDet-D0, requires 3 times more FLOPs (2.5B)~\footnote{In \citet{tan2020efficientdet}, FLOP denotes number of multiply-adds.} than SSD-MobileNetV2~\cite{sandler2018mobilenetv2} (0.8B FLOPs).
While SSD-MobileNetV2 is a lower-performance DNN specifically designed for mobile platforms and can process up to 6~fps, its mAP on COCO dataset is 20\% and keeping the model running on a mobile device significantly increases power consumption.
On the other hand, due to excessive end-to-end latency, cloud-based approaches are hardly applicable in most of the latency-constrained applications where mobile devices usually operate.
Most of the techniques we overview in the survey can be applied to both mobile device to edge server and edge server to cloud offloading. For the sake of clarity, we primarily refer to the former to explain the frameworks.

Recently, \gls{ec} approaches~\cite{mao2017survey,chen2019deep} have attempted to address the ``latency vs computation'' conundrum by completely offloading the \gls{dnn} execution to servers located very close to the mobile device, \textit{i.e.}, at the ``edge'' of the network.
However, canonical \gls{ec} does not consider that the quality of wireless links -- although providing high throughput on the average -- can suddenly fluctuate due to the presence of erratic noise and interference patterns, which may impair performance in latency-bound applications. For example, mobility and impaired propagation have been shown to decrease throughput even in high-bandwidth wireless links~\cite{zhang2019will,mateo2019analysis} while many \gls{iot} systems are based on communication technologies such as Long Range (LoRa)~\cite{samie2016iot}, which has a maximum data rate of 37.5 Kbps due to duty cycle limitations \cite{adelantado2017understanding}. 

The severe offloading limitations of some mobile devices, coupled with the instability of the wireless channel (\emph{e.g.}, UAV network~\cite{gupta2015survey}), imply that the amount of data offloaded to edge should be decreased, while at the same time keep the model accuracy as close as possible to the original. For this reason, \textit{\gls{sc}} \cite{kang2017neurosurgeon} and \textit{\gls{ee}} strategies \cite{teerapittayanon2016branchynet} have been proposed to provide an intermediate option between \gls{ec} and local computing. The key intuition behind \gls{sc} and \gls{ee} is similar to the one behind model pruning~\cite{han2016deep,li2016pruning,he2017channel,yang2017designing} and knowledge distillation \cite{hinton2014distilling,kim2016sequence,mirzadeh2020improved} -- since modern \glspl{dnn} are heavily over-parameterized \cite{yu2020understanding,yu2019understanding}, their accuracy can be preserved even with substantial reduction in the number of weights and filters, and thus representing the input with fewer parameters.
Specifically, \gls{sc} divide a larger \gls{dnn} into head and tail models, which are respectively executed by the mobile device and edge server.
\gls{ee}, on the other hand, proposes the introduction of ``subbranches'' into the early layers of \gls{dnn} models, so that the full computation of the model can be halted -- and a prediction result provided -- if the classifiers in the current subbranches have high confidence with the specific model input.\smallskip

\textbf{Motivation and Novel Contributions}.~The proliferation of \gls{dl}-based mobile applications in the \gls{iot} and 5G landscapes implies that techniques such as \gls{sc} and \gls{ee} are not simply ``nice-to-have'' features, but will become fundamental computational components in the years to come. Although a significant amount of research work has been done in \gls{sc} and \gls{ee}, to the best of our knowledge, a comprehensive survey of the state of the art has not been conducted yet. Moreover, there are still a series of research challenges that need to be addressed to take \gls{sc} and \gls{ee} to the next level. For this reason, this paper makes the following novel contributions:\smallskip

\begin{itemize}[leftmargin=1em]
    \item We summarize \gls{sc} and \gls{ee} studies with respect to approaches, tasks, and models. We first provide an overview of local, edge, split computing, and early-exit models in Section \ref{sec:overview}, by highlighting similarities and difference among them;
    \item We then discuss and compare the various approaches to \gls{sc} and \gls{ee} in Sections \ref{sec:split_comp} and \ref{sec:earlyexiting}, by highlighting the training strategies and applications.  Since code availability is fundamental for replicability/reproducibility~\cite{gundersen2018state}\footnote{To address this problem, major machine learning venues (\emph{e.g.}, ICML, NeurIPS, CVPR, ECCV, NAACL, ACL, and EMNLP) adopt a reproducibility checklist as part of official review process such as ML Code Completeness Checklist. See \label{fn:ml_code_checklist}\url{https://github.com/paperswithcode/releasing-research-code}.}, we provide for each work its corresponding code repository, if available, so that interested readers can reproduce and learn from existing studies;
    \item We conclude the paper by discussing in Section \ref{sec:research_challenges} a compelling agenda of research challenges in \gls{sc} and \gls{ee}, hoping to spur further contributions in these exciting and timely fields.
\end{itemize}

\section{Overview of Local, Edge, Split Computing and Early-Exit Models}
\label{sec:overview}

\begin{figure}[h]
    \centering
    \includegraphics[width=0.95\linewidth]{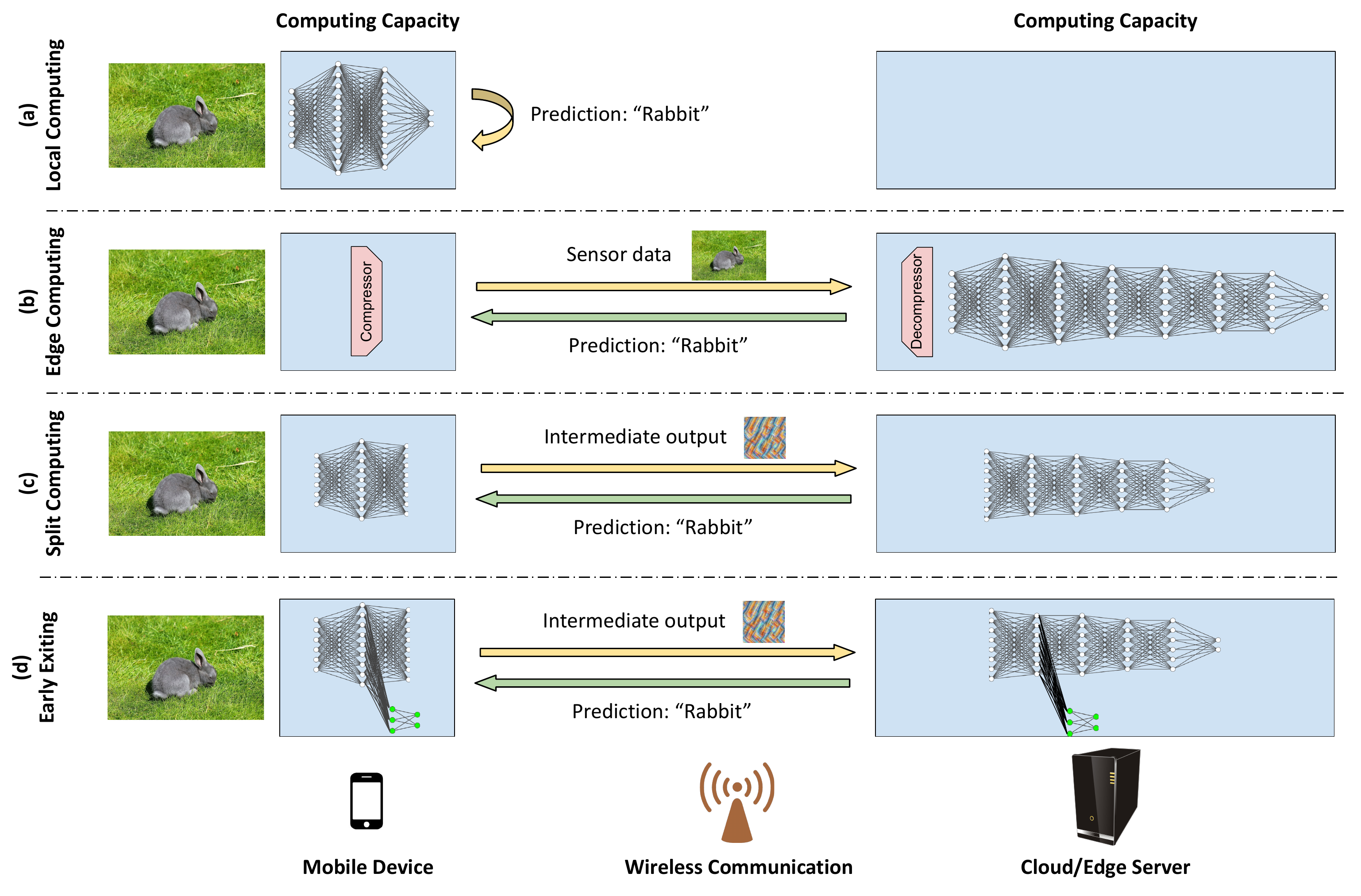}
    \vspace{-1em}
    \caption{Overview of (a) local, (b) edge, (c) split computing, and (d) early exiting: image classification as an example.}
    \label{fig:local_edge_split_computing}
\end{figure}

In this section, we provide an overview of local, edge, split computing, and early-exit models, which are the main computational paradigms that will be discussed in the paper. Figure~\ref{fig:local_edge_split_computing} provides a graphical overview of the approaches.

All these techniques operate on a \gls{dnn} model $\mathcal{M}(\cdot)$ whose task is to produce the inference output $\mathbf{y}$ from an input $\mathbf{x}$. Typically, $\mathbf{x}$ is a high-dimensional variable, whereas the output $\mathbf{y}$ has significantly lower dimensionality~\cite{tishby2015deep}. Split computing and early exit approaches are contextualized in a setting where the system is composed of a mobile device and an edge server interconnected via a wireless channel. The overall goal of the system is to produce the inference output $\mathbf{y}$ from the input $\mathbf{x}$ acquired by the mobile device, by means of the \gls{dnn} $\mathbf{y}{=}\mathcal{M}(\mathbf{x})$ under -- possibly time varying -- constraints on:

\vspace{1.5mm}
\noindent
{\bf Resources:} (\emph{i}) the computational capacity (roughly expressed as number operations per second) $C_{\rm md}$ and $C_{\rm es}$ of the mobile device and edge server, respectively, (\emph{ii}) the capacity $\phi$, in bits per second, of the wireless channel connecting the mobile device to the edge server;

\vspace{1.5mm}
\noindent
{\bf Performance:} (\emph{i}) the absolute of average value of the time from the generation of $\mathbf{x}$ to the availability of $\mathbf{y}$, (\emph{ii}) the degradation of the ``quality'' of the output $\mathbf{y}$.

Split, edge, local, and early-exiting strategies strive to find suitable operating points with respect to accuracy, end-to-end delay, and energy consumption, which are inevitably influenced by the characteristics of the underlying system. It is generally assumed that the computing and energy capacities of the mobile device are smaller than that of the edge server.
As a consequence, if part of the workload is allocated to the mobile device, then the execution time increases while battery lifetime decreases. However, as explained later, the workload executed by the mobile device may result in a reduced amount of data to be transferred over the wireless channel, possibly compensating for the larger execution time and leading to smaller end-to-end delays.

\subsection{Local and Edge Computing}
We start with an overview of local and edge computing. In \gls{lc}, the function $\mathcal{M}(\mathbf{x})$ is entirely executed by the mobile device. This approach eliminates the need to transfer data over the wireless channel. However, the complexity of the best performing \glspl{dnn} most likely exceeds the computing capacity and energy consumption available at the mobile device. Usually, simpler models $\hat{\mathcal{M}}(\mathbf{x})$ are used, such as MobileNet~\cite{sandler2018mobilenetv2} and MnasNet~\cite{tan2019mnasnet} which often have a degraded accuracy performance.
Besides designing lightweight neural models executable on mobile devices, the widely used techniques to reduce the complexity of models are knowledge distillation~\cite{hinton2014distilling} and model pruning/quantization~\cite{jacob2018quantization,li2018auto-tuning} -- described in Section~\ref{subsec:model_compression}.
Some of the techniques are also leveraged in \gls{sc} studies to introduce bottlenecks without sacrificing model accuracy as will be described in the following sections.

In \gls{ec}, the input $\mathbf{x}$ is transferred to the edge server, which then executes the original model $\mathcal{M}(\mathbf{x})$. In this approach, which preserves full accuracy, the mobile device is not allocated computing workload, but the full input $\mathbf{x}$ needs to be transferred to the edge server. This may lead to an excessive end-to-end delay in degraded channel conditions and erasure of the task in extreme conditions. A possible approach to reduce the load imposed to the wireless channel, and thus also transmission delay and erasure probability, is to compress the input $\mathbf{x}$. We define, then, the encoder and decoder models $\mathbf{z}{=}F(\mathbf{x})$ and $\hat{\mathbf{x}}{=}G(\mathbf{z})$, which are executed at the mobile device and edge server, respectively. The distance $d(\mathbf{x},\hat{\mathbf{x}})$ defines the performance of the encoding-decoding process $\hat{\mathbf{x}}{=}G(F(\mathbf{x}))$, a metric which is separate, but may influence, the accuracy loss of $\mathcal{M}(\hat{\mathbf{x}})$ with respect to $\mathcal{M}(\mathbf{x})$, that is, of the model executed with the reconstructed input with respect to the model executed with the original input. Clearly, the encoding/decoding functions increase the computing load both at the mobile device and edge server side. A broad range of different compression approaches exists ranging from low-complexity traditional compression (\emph{e.g.}, JPEG compression for images in \gls{ec}~\cite{nakahara2021retransmission}) to neural compression models~\cite{balle2016end,balle2018variational,yang2020variational}.
We remark that while the compressed input data \emph{e.g.}, JPEG objects, can reduce the data transfer time in \gls{ec}, those representations are designed to allow the accurate reconstruction of the input signal. Therefore, these approaches may (\emph{i}) decrease privacy as a ``reconstructable'' representation is transferred to the edge server~\cite{wang2020data}; (\emph{ii}) result in larger amount of data to be transmitted over the channel compared to representation specifically designed for the computing task as in bottleneck-based \gls{sc} as explained in the following sections.

\subsection{Split Computing and Early Exiting}
Split computing (SC) aims at achieving the following goals: (\emph{i}) the computing load is distributed across the mobile device and edge server; and (\emph{ii}) establishes a task-oriented compression to reduce data transfer delays.
We consider a neural model $\mathcal{M}(\cdot)$ with $L$ layers, and define $\mathbf{z}_{\ell}$ the output of the $\ell$-th layer. Early implementations of \gls{sc} select a layer $\ell$ and divide the model $\mathcal{M}(\cdot)$ to define the head and tail submodels $\mathbf{z}_{\ell}{=}\mathcal{M}_{H}(\mathbf{x})$ and $\mathbf{\hat{y}}{=}\mathcal{M}_{T}(\mathbf{z}_{\ell})$, executed at the mobile device and edge server, respectively. In early instances of \gls{sc}, the architecture and weights of the head and tail model are exactly the same as the first $\ell$ layers and last $L-\ell$ layers of $\mathcal{M}(\cdot)$. This simple approach preserves accuracy but allocates part of the execution of $\mathcal{M}(\cdot)$ to the mobile device, whose computing power is expected to be smaller than that of the edge server, so that the total execution time may be larger. The transmission time of $\mathbf{z}_{\ell}$ may be larger or smaller compared to that of transmitting the input $\mathbf{x}$, depending on the size of the tensor $\mathbf{z}_{\ell}$. However, we note that in most relevant applications the size of $\mathbf{z}_{\ell}$ becomes smaller than that of $\mathbf{x}$ only in later layers, which would allocate most of the computing load to the mobile device. 
More recent \gls{sc} frameworks introduce the notion of \emph{bottleneck} to achieve \emph{in-model} compression toward the global task~\cite{matsubara2019distilled}.
As formally described in the next section, a bottleneck is a compression point at one layer in the model, which can be realized by reducing the number of nodes of the target layer, and/or by quantizing its output.
We note that as \gls{sc} realizes a task-oriented compression, it guarantees a higher degree of privacy compared to \gls{ec}. In fact, the representation may lack information needed to fully reconstruct the original input data.

Another approach to enable mobile computing is referred to early exiting (EE). The core idea is to create models with multiple ``exits'' across the model, where each exit can produce the model output. Then, the first exit providing a target confidence on the output is selected. This approach tunes the computational complexity, determined by the exit point, to the sample or to system conditions. Formally, we can define a sequence of models $\mathcal{M}_i$ and $\mathcal{B}_i$, $i{=}1,\ldots,N$. Model $\mathcal{M}_i$ takes as input $\mathbf{z}_{i-1}$ (the output of model $\mathcal{M}_{i-1}$) and outputs $\mathbf{z}_i$, where we set $\mathbf{z}_{0}{=}\mathbf{x}$. The branch models $\mathcal{B}_i$ take as input $\mathbf{z}_i$ and produce the estimate of the desired output $\mathbf{y}_i$. 
Thus, the concatenation of $\mathcal{M}_1,\ldots,\mathcal{M}_N$ results into an output analogous to that of the original model.
Intuitively, the larger the number of models used to produce the output $\mathbf{y}_i$, the better the accuracy. Thus, while \gls{sc} optimizes intermediate representations to preserve information toward the final task (\emph{e.g.}, classification) for the whole dataset, early exit models take a ``per sample'' control perspective. Each sample will be sequentially analyzed by concatenations of $\mathcal{M}_i$ and $\mathcal{B}_i$ sections until a predefined confidence level is reached. The hope is that a portion of the samples will require a smaller number of sections compared to executing the whole sequence.

\section{Background of Deep Learning for Mobile Applications}
\label{sec:backgrond}

In this section, we provide an overview of recent approaches to reduce the computational complexity of \gls{dnn} models for resource-constrained mobile devices.
These approaches can be categorized into two main classes: (\emph{i}) approaches that attempt to directly design lightweight models and (\emph{ii}) model compression.

\subsection{Lightweight Models}
\label{subsec:lightweight}

From a conceptual perspective, The design of small deep learning models is one of the simplest ways to reduce inference cost.
However, there is a trade-off between model complexity and model accuracy, which makes this approach practically challenging when aiming at high model performance.
The MobileNet series~\cite{howard2017mobilenets,sandler2018mobilenetv2,howard2019searching} is one among the most popular lightweight models for computer vision tasks, where
\citet{howard2017mobilenets} describes the first version MobileNetV1.
By using a pair of depth-wise and point-wise convolution layers in place of standard convolution layers, the design drastically reduces model size, and thus computing load.
Following this study, \citet{sandler2018mobilenetv2} proposed MobileNetV2, which achieves an improved accuracy.
The design is based on MobileNetV1~\cite{howard2017mobilenets}, and uses the bottleneck residual block, a resource-efficient block with inverted residuals and linear bottlenecks.
\citet{howard2019searching} presents MobileNetV3, which further improves the model accuracy and is designed by a hardware-aware neural architecture search~\cite{tan2019mnasnet} with NetAdapt~\cite{yang2018netadapt}. The largest variant of MobileNetV3, MobileNetV3-Large 1.0, achieves a comparable accuracy of ResNet-34~\cite{he2016deep} for the ImageNet dataset, while reducing by about 75\% the model parameters.

While many of the lightweight neural networks are often manually designed, there are also studies on automating the neural architecture search (NAS)~\cite{zoph2017neural}.
For instance, \citet{zoph2018learning} designs a novel search space through experiments with the CIFAR-10 dataset~\cite{krizhevsky2009learning}, that is then scaled to larger, higher resolution image datasets such as the ImageNet dataset~\cite{russakovsky2015imagenet}, to design their proposed model: NASNet.
Leveraging the concept of NAS, some studies design lightweight models in a platform-aware fashion.
\citet{dong2018dpp} proposes the Device-aware Progressive Search for Pareto-optimal Neural Architectures (DDP-Net) framework, that optimizes the network design with respect to two objectives: device-related (\emph{e.g.}, inference latency and memory usage) and device-agnostic (\emph{e.g.}, accuracy and model size) objectives.
Similarly, \citet{tan2019mnasnet} propose an automated mobile neural architecture search (MNAS) method and design the MnasNet models by optimizing both model accuracy and inference time.

\subsection{Model Compression}
\label{subsec:model_compression}

A different approach to produce small \gls{dnn} models is to ``compress'' a large model.
Model pruning and quantization~\cite{han2015learning,han2016deep,jacob2018quantization,li2020train} are the dominant model compression approaches. The former removes parameters from the model, while the latter uses fewer bits to represent them. In both these approaches, a large model is trained first and then compressed, rather than directly designing a lightweight model followed by training.
In \citet{jacob2018quantization}, the authors empirically show that their quantization technique leads to an improved tradeoff between inference time and accuracy on MobileNet~\cite{howard2017mobilenets} for image classification tasks on Qualcomm Snapdragon 835 and 821 compared to the original, float-only MobileNet.
For what concerns model pruning, \citet{li2017pruning,liu2021ebert} demonstrates that it is difficult for model pruning itself to accelerate inference while achieving strong performance guarantees on general-purpose hardware due to the unstructured sparsity of the pruned model and/or kernels in layers.

Knowledge distillation~\cite{bucilua2006model,hinton2014distilling} is another popular model compression method.
While model pruning and quantization make trained models smaller, the concept of knowledge distillation is to provide outputs extracted from the trained model (called ``teacher'') as informative signals to train smaller models (called ``student'') in order to improve the accuracy of predesigned small models.
Thus, the goal of the process is that of \emph{distilling knowledge of a trained teacher model into a smaller student model} for boosting accuracy of the smaller model without increasing model complexity.
For instance, \citet{ba2014do} proposes a method to train small neural networks by mimicking the detailed behavior of larger models. The experimental results show that models trained by this mimic learning method achieve performance close to that of deeper neural networks on some phoneme recognition and image recognition tasks.
The formulation of some knowledge distillation methods will be described in Section~\ref{sec:split_bottle_training}.

\section{Split Computing: A Survey} \label{sec:split_comp}

\begin{figure}[h]
    \centering
    \includegraphics[width=0.95\linewidth]{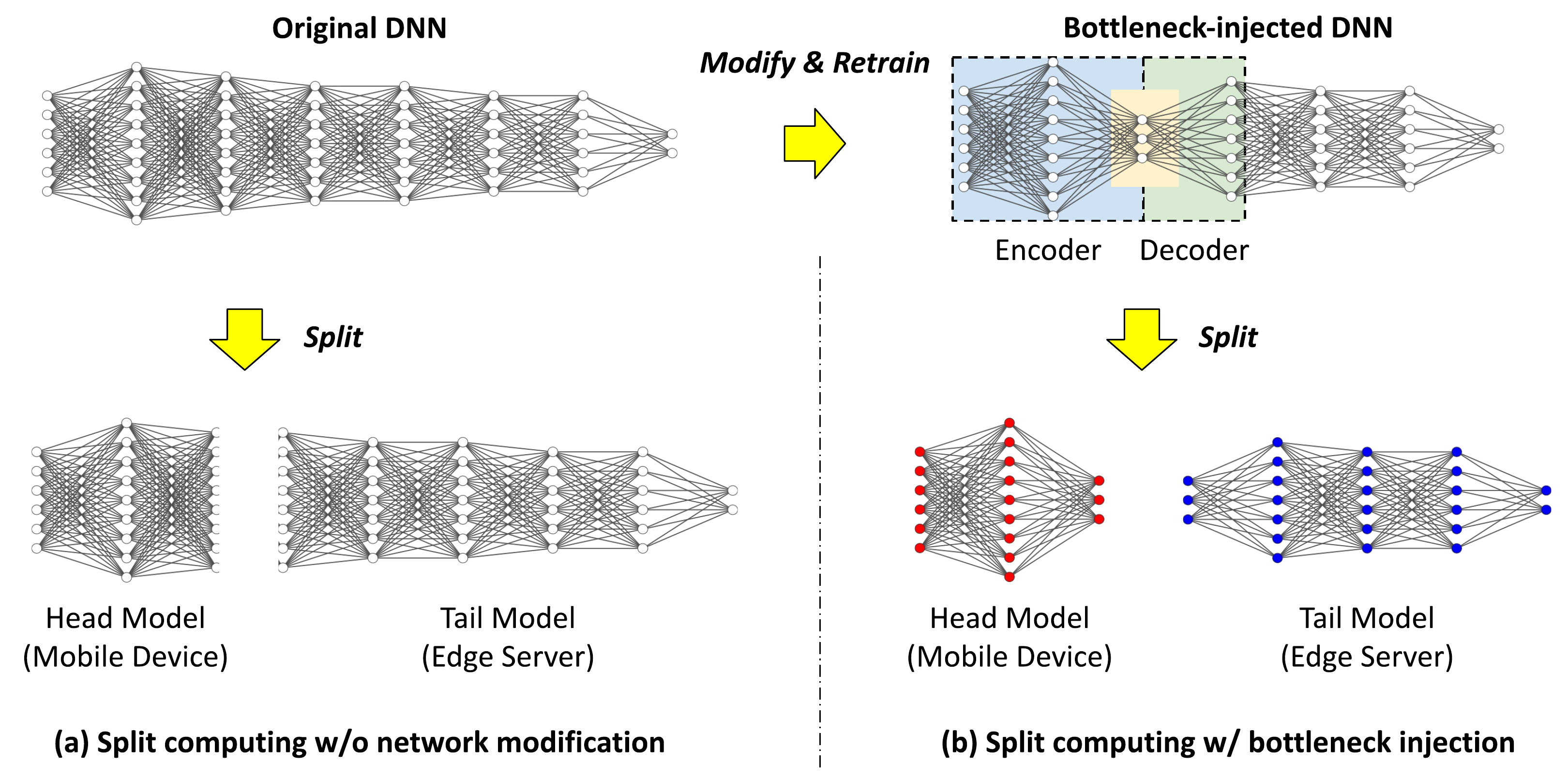}
    \vspace{-1em}
    \caption{Two different \gls{sc} approaches.}
    \label{fig:split_computing}
\end{figure}

This section discusses existing state of of the art in \gls{sc}. Figure~\ref{fig:split_computing} illustrates the existing \gls{sc} approaches. They can be categorized into either (i) \emph{without network modification} or (ii) \emph{with bottleneck injection}. We first present \gls{sc} approaches without \gls{dnn} modification in Section \ref{sec:split_nomod}. We then discuss the motivations behind the introduction of \gls{sc} with bottlenecks in Section \ref{sec:need_for_bottle}, which are then discussed in details in Section \ref{subsec:sc_with_bottleneck}. Since the latter require specific training procedures, we devote Section \ref{sec:split_bottle_training} to their discussion.

\subsection{Split Computing without \gls{dnn} Modification}\label{sec:split_nomod}

In this class of approaches, the architecture and weights of the head $\mathcal{M}_{H}(\cdot)$ and tail $\mathcal{M}_T(\cdot)$ models are exactly the same as the first $\ell$ layers and last $L-\ell$ layers of $\mathcal{M}(\cdot)$.
To the best of our knowledge,~\citet{kang2017neurosurgeon} proposed the first \gls{sc} approach (called ``Neurosurgeon''), which searches for the best partitioning layer in a \gls{dnn} model for minimizing total (end-to-end) latency or energy consumption. Formally, inference time in \gls{sc} is the sum of processing time on mobile device, delay of communication between mobile device and edge server, and the processing time on edge server.

Interestingly, their experimental results show that the best partitioning (splitting) layers in terms of energy consumption and total latency for most of the considered models result in either their input or output layers.
In other words, deploying the whole model on either a mobile device or an edge server (\emph{i.e.,} local computing or \gls{ec}) would be the best option for such \gls{dnn} models.
Following the work by~\citet{kang2017neurosurgeon}, the research communities explored various \gls{sc} approaches mainly focused on \gls{cv} tasks such as image classification.
Table~\ref{table:split_without_bottleneck} summarizes the studies on \gls{sc} without architectural modifications.

\begin{table}
    \caption{Studies on \gls{sc} without architectural modifications.}
    \vspace{-1em}
    \label{table:split_without_bottleneck}
    \def\arraystretch{1.5}
    \begin{center}
        \scriptsize
        \begin{tabular}{|c|c|c|c|c|c|}
            \hline
            \multicolumn{1}{|c|}{\bf Work} & \multicolumn{1}{c|}{\bf Task(s)} & \multicolumn{1}{c|}{\bf Dataset(s)} & \multicolumn{1}{c|}{\bf Model(s)} & \multicolumn{1}{c|}{\bf Metrics} & \multicolumn{1}{c|}{\bf Code}\\ \hline
            \hline
            \textunderset{(2017)}{\citet{kang2017neurosurgeon}} & \makecell{Image classification\\Speech recognition\\Part-of-speech tagging\\Named entity recognition\\Word chunking} & \makecell{N/A\\(No task-specific metrics)} & \makecell{AlexNet~\cite{krizhevsky2012imagenet}\\VGG-19~\cite{simonyan2014very}\\DeepFace~\cite{taigman2014deepface}\\LeNet-5~\cite{lecun1998gradient-based}\\Kaldi~\cite{povey2011kaldi}\\SENNA~\cite{collobert2011natural}} & \datametric, \energymetric, \latencymetric & \\ \hline
            \textunderset{(2018)}{\citet{li2018learning}} & Image classification & \makecell{N/A\\(No task-specific metrics)} & AlexNet~\cite{krizhevsky2012imagenet} & \compmetric, \datametric & \\ \hline
            \textunderset{(2018)}{\citet{jeong2018computation}} & Image classification & \makecell{N/A\\(No task-specific metrics)} & \makecell{GoogLeNet~\cite{szegedy2015going}\\AgeNet~\cite{levi2015age}\\GenderNet~\cite{levi2015age}} & \datametric, \latencymetric & \\ \hline
            \textunderset{(2018)}{\citet{li2018auto-tuning}} & Image classification & ImageNet~\cite{russakovsky2015imagenet} & \makecell{AlexNet~\cite{krizhevsky2012imagenet}\\VGG-16~\cite{simonyan2014very}\\ResNet-18~\cite{he2016deep}\\GoogLeNet~\cite{szegedy2015going}} & \accmetric, \datametric, \latencymetric & \\ \hline
            \textunderset{(2018)}{\citet{choi2018deep}} & Object detection & VOC 2007~\cite{everingham2007pascal} & YOLO9000~\cite{redmon2017yolo9000} & \accmetric, \compmetric, \datametric, \latencymetric & \\ \hline
            \textunderset{(2019)}{\citet{eshratifar2019jointdnn}} & \makecell{Image classification\\Speech recognition} & \makecell{N/A\\(No task-specific metrics)} & \makecell{AlexNet~\cite{krizhevsky2012imagenet}\\OverFeat~\cite{sermanet2014overfeat}\\NiN~\cite{lin2014network}\\VGG-16~\cite{simonyan2014very}\\ResNet-50~\cite{he2016deep}} & \datametric, \energymetric, \latencymetric & \\ \hline
            \textunderset{(2019)}{\citet{zeng2019boomerang}} & Image classification & CIFAR-10~\cite{krizhevsky2009learning} & AlexNet~\cite{krizhevsky2012imagenet} & \accmetric, \datametric, \latencymetric & \\ \hline
            \textunderset{(2020)}{\citet{cohen2020lightweight}} & \makecell{Image classification\\Object detection} & \makecell{ImageNet (2012)~\cite{russakovsky2015imagenet}\\COCO 2017~\cite{lin2014microsoft}} & \makecell{VGG-16~\cite{simonyan2014very}\\ResNet-50~\cite{he2016deep}\\YOLOv3~\cite{redmon2018yolov3}} & \accmetric, \datametric & \\ \hline
            \textunderset{(2020)}{\citet{pagliari2020crime}} & \makecell{Natural language inference\\Reading comprehension\\Sentiment analysis} & \makecell{N/A\\(No task-specific metrics)} & RNNs & \energymetric, \latencymetric & \\ 
            \hline
            \textunderset{(2021)}{\citet{itahara2021packet}} & Image classification & CIFAR-10~\cite{krizhevsky2009learning} & VGG-16~\cite{simonyan2014very} & \accmetric, \datametric & \\ 
            \hline
        \end{tabular}
        \\\footnotesize \accmetric: Model accuracy, \compmetric: Model complexity, \datametric: Transferred data size, \energymetric: Energy consumption, \latencymetric: Latency, \traincostmetric: Training cost
    \end{center}
\end{table}

\citet{jeong2018computation} used this partial offloading approach as a privacy-preserving way for computation offloading to blind the edge server to the original data captured by client. Leveraging neural network quantization techniques, \citet{li2018auto-tuning} discussed best splitting point in \gls{dnn} models to minimize inference latency, and showed quantized \gls{dnn} models did not degrade accuracy comparing to the (pre-quantized) original models.
\citet{choi2018deep} proposed a feature compression strategy for object detection models that introduces a quantization/video-coding based compressor to the intermediate features in YOLO9000~\cite{redmon2017yolo9000}.

\citet{eshratifar2019jointdnn} propose JointDNN for collaborative computation between mobile device and cloud, and demonstrate that using either local computing only or cloud computing only is not an optimal solution in terms of inference time and energy consumption.
Different from~\cite{kang2017neurosurgeon}, they consider not only discriminative deep learning models (\emph{e.g.}, classifiers), but also generative deep learning models and autoencoders as benchmark models in their experimental evaluation. 
\citet{cohen2020lightweight} introduce a technique to code the output of the head portion in a split \gls{dnn} to a wide range of bit-rates, and demonstrate the performance for image classification and object detection tasks.
\citet{pagliari2020crime} first discuss the collaborative inference for simple recurrent neural networks, and their proposed scheme is designed to automatically select the best inference device for each input data in terms of total latency or end-device energy.
\citet{itahara2021packet} use dropout layers~\cite{srivastava2014dropout} to emulate a packet loss scenario rather than for the sake of compression and discuss the robustness of VGG-based models~\cite{simonyan2014very} for split computing.

While only a few studies in Table~\ref{table:split_without_bottleneck} heuristically choose splitting points~\cite{choi2018deep,cohen2020lightweight}, most of the other studies~\cite{kang2017neurosurgeon,li2018learning,jeong2018computation,li2018auto-tuning,eshratifar2019jointdnn,zeng2019boomerang,pagliari2020crime} in Table~\ref{table:split_without_bottleneck} analyze various types of cost (\emph{e.g.}, computational load and energy consumption on mobile device, communication cost, and/or privacy risk) to partition \gls{dnn} models at each of their splitting points.
Based on the analysis, performance profiles of the split \gls{dnn} models are derived to inform selection.
Concerning metrics, many of the studies in Table~\ref{table:split_without_bottleneck} do not discuss task-specific performance metrics such as accuracy.
This is in part because the proposed approaches do not modify the input or intermediate representations in the models (\emph{i.e.}, the final prediction will not change).
On the other hand, \citet{li2018auto-tuning,choi2018deep,cohen2020lightweight} introduce lossy compression techniques to intermediate stages in DNN models, which may affect the final prediction results.
Thus, discussing trade-off between compression rate and task-specific performance metrics would be essential for such studies.
As shown in the table, such trade-off is discussed only for \gls{cv} tasks, and many of the models considered in such studies have weak performance compared with state-of-the-art models and complexity within reach of modern mobile devices.
Specific to image classification tasks, most of the models considered in the studies listed in Table~\ref{table:split_without_bottleneck} are more complex and/or the accuracy is comparable to or lower than that of lightweight baseline models such as MobileNetV2~\cite{sandler2018mobilenetv2} and MnasNet~\cite{tan2019mnasnet}.
Thus, in future work, more accurate models should be considered to discuss the performance trade-off and further motivate \gls{sc} approaches. 

\subsection{The Need for Bottleneck Injection}\label{sec:need_for_bottle}

\begin{table}
    \caption{Statistics of image classification datasets in \gls{sc} studies}
    \vspace{-1em}
    \label{table:input_sizes}
    \def\arraystretch{1.5}
    \begin{center}
        \begin{tabular}{|c|c|c|c|c|}
            \hline
            \multicolumn{1}{|c|}{\bf } & \multicolumn{1}{c|}{\bf MNIST} & \multicolumn{1}{c|}{\bf CIFAR-10} & \multicolumn{1}{c|}{\bf CIFAR-100} & \multicolumn{1}{c|}{\bf ImageNet (2012)}\\ \hline
            \hline
            \# labeled train/dev(test) samples: & 60k/10k & 50k/10k & 50k/10k & 1,281k/50k \\\hline
            \# object categories & 10 & 10 & 100 & 1,000 \\\hline
            Input tensor size & $1 \times 32 \times 32$ & $3 \times 32 \times 32$ & $3 \times 32 \times 32$ & $3 \times 224 \times 224$* \\\hline
            JPEG data size [KB/sample] & 0.9657 & 1.790 & 1.793 & 44.77 \\
            \hline
        \end{tabular}
        \\\footnotesize \raggedright * A standard (resized) input tensor size for \gls{dnn} models.
    \end{center}
\end{table}

While~\citet{kang2017neurosurgeon} empirically show that executing the whole model on either mobile device or edge server would be best in terms of total inference and energy consumption for most of their considered \gls{dnn} models, their proposed approach find the best partitioning layers inside some of their considered \gls{cv} models (\glspl{cnn}) to minimize the total inference time.
There are a few trends observed from their experimental results: (i) communication delay to transfer data from mobile device to edge server is a key component in \gls{sc} to reduce total inference time; (ii) all the neural models they considered for NLP tasks are relatively small (consisting of only a few layers), that potentially resulted in finding the output layer is the best partition point (\emph{i.e.}, local computing) according to their proposed approach; (iii) similarly, not only \gls{dnn} models they considered (except VGG~\cite{simonyan2014very}) but also the size of the input data to the models (See Table~\ref{table:input_sizes}) are relatively small, which gives more advantage to \gls{ec} (fully offloading computation).
In other words, it highlights that complex \gls{cv} tasks requiring large (high-resolution) images for models to achieve high accuracy such as ImageNet and COCO datasets would be essential to discuss the trade-off between accuracy and execution metrics to be minimized (\emph{e.g.}, total latency, energy consumption) for \gls{sc} studies. The key issue is that straightforward \gls{sc} approaches like~\citet{kang2017neurosurgeon} rely on the existence of \emph{natural bottlenecks} -- that is, intermediate layers whose output $\mathbf{z}_{\ell}$ tensor size is smaller than the input -- inside the model.
Without such natural bottlenecks in the model, straightforward splitting approaches would fail to improve performance in most settings~\cite{arbera2013to,guo2018cloud}.

Some models, such as AlexNet~\cite{krizhevsky2012imagenet}, VGG~\cite{simonyan2014very} and DenseNet~\cite{huang2017densely}, possess such layers~\cite{matsubara2019distilled}. However, recent \gls{dnn} models such as ResNet~\cite{he2016deep}, Inception-v3~\cite{szegedy2016rethinking}, Faster R-CNN~\cite{ren2015faster} and Mask R-CNN~\cite{he2017mask} do not have natural bottlenecks in the early layers, that is, splitting the model would result in compression only when assigning a large portion of the workload to the mobile device. As discussed earlier, reducing the communication delay is a key to minimize total inference time in \gls{sc}.
For these reasons, introducing \emph{artificial bottlenecks} to \gls{dnn} models by modifying their architecture is a recent trend and has been attracting attention from the research community.
Since the main role of such encoders in \gls{sc} is to compress intermediate features rather than to complete inference, the encoders usually consist of only a few layers.
Also, the resulting encoders in \gls{sc} to be executed on constrained mobile devices are often much smaller (\emph{e.g.,} 10K parameters in the encoder of ResNet-based \gls{sc} model~\cite{matsubara2020neural}), than lightweight models such as MobileNetV2~\cite{sandler2018mobilenetv2} (3.5M parameters) and MnasNet~\cite{tan2019mnasnet} (4.4M parameters).
Thus, even if the model accuracy is either degraded or comparable to such small models, \gls{sc} models are still beneficial in terms of computational burden and energy consumption at the mobile devices.

\subsection{Split Computing with Bottleneck Injection}
\label{subsec:sc_with_bottleneck}

This class of models can be described as composed of $3$ sections: $\mathcal{M}_{E}$, $\mathcal{M}_{D}$ and $\mathcal{M}_{T}$. We define $\mathbf{z}_{\ell}|\mathbf{x}$ as the output of the $\ell$-th layer of the original model given the input $\mathbf{x}$. The concatenation of the $\mathcal{M}_{E}$ and $\mathcal{M}_{D}$ models is designed to produce a possibly noisy version $\hat{\mathbf{z}}_{\ell}|\mathbf{x}$ of $\mathbf{z}_{\ell}|\mathbf{x}$, which is taken as input by $\mathcal{M}_{T}$ to produce the output $\hat{\mathbf{y}}$, on which the accuracy degradation with respect to $\mathbf{y}$ is measured.
The models $\mathcal{M}_{E}$, $\mathcal{M}_{D}$ function as specialized encoders and decoders in the form $\hat{\mathbf{z}}_{\ell}{=}\mathcal{M}_{D}(\mathcal{M}_E(\mathbf{x}))$, where $\mathcal{M}_{E}(\mathbf{x})$ produces the latent variable $\mathbf{z}$. In worlds, the two first sections of the modified model transform the input $\mathbf{x}$ into a version of the output of the $\ell$-th layer via the intermediate representation $\mathbf{z}$, thus functioning as encoder/decoder functions. The model is split after the first section, that is, $\mathcal{M}_{E}$ is the head model, and the concatenation of $\mathcal{M}_{D}$ and $\mathcal{M}_{T}$ is the tail model. Then, the tensor $\mathbf{z}$ is transmitted over the channel. The objective of the architecture is to minimize the size of $\mathbf{z}$ to reduce the communication time while also minimizing the complexity of $\mathcal{M}_E$ (that is, the part of the model executed at the -- weaker -- mobile device) and the discrepancy between $\mathbf{y}$ and $\hat{\mathbf{y}}$. The layer between $\mathcal{M}_E$ and $\mathcal{M}_D$ is the injected bottleneck.

Table~\ref{table:split_with_bottleneck} summarizes \gls{sc} studies with bottleneck injected strategies.
To the best of our knowledge, the papers in \cite{eshratifar2019bottlenet} and \cite{matsubara2019distilled} were the first to propose altering existing \gls{dnn} architectures to design relatively small bottlenecks at early layers in \gls{dnn} models, instead of introducing compression techniques (\emph{e.g.}, quantization, autoencoder) to the models, so that communication delay (cost) and total inference time can be further reduced.
Following these studies, \citet{hu2020fast} introduce bottlenecks to MobileNetV2~\cite{sandler2018mobilenetv2} (modified for CIFAR datasets) in a similar way for \gls{sc}, and discuss end-to-end performance evaluation.
\citet{choi2020back} combine multiple compression techniques such as quantization and tiling besides convolution/deconvolution layers, and design a feature compression approach for object detectors.
Similar to the concept of bottleneck injection,
\citet{shao2020bottlenet++} find that over-compression of intermediate features and inaccurate communication between computing devices can be tolerated unless the prediction performance of the models are significantly degraded by them.
Also, \citet{jankowski2020joint} propose introducing a reconstruction-based bottleneck to \gls{dnn} models, which is similar to the concept of BottleNet~\cite{eshratifar2019bottlenet}.
A comprehensive discussion on the delay/complexity/accuracy tradeoff can be found in~\cite{yao2020deep,matsubara2020head}.

\begin{table}
    \caption{Studies on \gls{sc} \uline{with bottleneck injection strategies.}}
    \vspace{-1em}
    \label{table:split_with_bottleneck}
    \def\arraystretch{1.5}
    \begin{center}
        \bgroup
        \setlength{\tabcolsep}{0.1em}
        \scriptsize
        \begin{tabular}{|c|c|c|c|c|c|c|}
            \hline
            \multicolumn{1}{|c|}{\bf Work} & \multicolumn{1}{c|}{\bf Task(s)} & \multicolumn{1}{c|}{\bf Dataset(s)} & \multicolumn{1}{c|}{\bf Base Model(s)} & \multicolumn{1}{c|}{\bf Training} & \multicolumn{1}{c|}{\bf Metrics} & \multicolumn{1}{c|}{\bf Code}\\ \hline
            \hline
            \textunderset{(2019)}{\citet{eshratifar2019bottlenet}} & Image classification & miniImageNet~\cite{snell2017prototypical} & \makecell{ResNet-50~\cite{he2016deep}\\VGG-16~\cite{simonyan2014very}} & CE-based & \accmetric, \datametric, \latencymetric & \\ \hline
            \textunderset{(2019, 2020)}{Matsubara et al.~\cite{matsubara2019distilled,matsubara2020head}} & Image classification & \makecell{Caltech 101~\cite{fei2006one}\\ImageNet (2012)~\cite{russakovsky2015imagenet}} & \makecell{DenseNet-169~\cite{huang2017densely}\\DenseNet-201~\cite{huang2017densely}\\ResNet-152~\cite{he2016deep}\\Inception-v3~\cite{szegedy2016rethinking}} & \makecell{HND\\KD\\CE-based} & \accmetric, \compmetric, \datametric, \latencymetric, \traincostmetric & \myhref{https://github.com/yoshitomo-matsubara/head-network-distillation}{Link} \\ \hline
            \textunderset{(2020)}{\citet{hu2020fast}} & Image classification & CIFAR-10/100~\cite{krizhevsky2009learning} & MobileNetV2~\cite{sandler2018mobilenetv2} & CE-based & \accmetric, \datametric, \latencymetric & \\ \hline
            \textunderset{(2020)}{\citet{choi2020back}} & Object detection & COCO 2014~\cite{lin2014microsoft} & YOLOv3~\cite{redmon2018yolov3} & Reconstruct. & \accmetric, \datametric & \\ \hline
            \textunderset{(2020)}{\citet{shao2020bottlenet++}} & Image classification & CIFAR-100~\cite{krizhevsky2009learning} & \makecell{ResNet-50~\cite{he2016deep}\\VGG-16~\cite{simonyan2014very}} & \makecell{CE-based\\(Multi-stage)} & \accmetric, \compmetric, \datametric & \\ \hline
            \textunderset{(2020)}{\citet{jankowski2020joint}} & Image classification & CIFAR-100~\cite{krizhevsky2009learning} & VGG-16~\cite{simonyan2014very} & \makecell{CE + $\mathcal{L}_{2}$\\(Multi-stage)} & \accmetric, \compmetric, \datametric & \\ \hline
            \textunderset{(2020)}{Matsubara et al.~\cite{matsubara2020split,matsubara2020neural}} & \makecell{Object detection\\Keypoint detection} & COCO 2017~\cite{lin2014microsoft} & \makecell{Faster R-CNN~\cite{ren2015faster}\\Mask R-CNN~\cite{he2017mask}\\Keypoint R-CNN~\cite{he2017mask}} & \makecell{HND\\GHND} & \accmetric, \compmetric, \datametric, \latencymetric & \myhref{https://github.com/yoshitomo-matsubara/hnd-ghnd-object-detectors}{Link} \\ \hline
            \textunderset{(2020)}{\citet{yao2020deep}} & \makecell{Image classification\\Speech recognition} & \makecell{ImageNet (2012)~\cite{russakovsky2015imagenet}\\LibriSpeech~\cite{panayotov2015librispeech}} & \makecell{ResNet-50~\cite{he2016deep}\\Deep Speech~\cite{hannun2014deep}} & \makecell{Reconstruct.\\+ KD} & \accmetric, \datametric, \energymetric, \latencymetric, \traincostmetric & \myhref{https://github.com/CPS-AI/Deep-Compressive-Offloading}{Link}* \\ \hline
            \textunderset{(2021)}{\citet{assine2021single}} & Object detection & COCO 2017~\cite{lin2014microsoft} & EfficientDet~\cite{tan2020efficientdet} & GHND-based & \accmetric, \compmetric, \datametric & \myhref{https://github.com/jsiloto/adaptive-cod}{Link} \\ \hline
            \textunderset{(2021)}{\citet{sbai2021cut}} & Image classification & \makecell{Subset of ImageNet~\cite{russakovsky2015imagenet}\\(700 out of 1,000 classes)} & \makecell{MobileNetV1~\cite{howard2017mobilenets}\\VGG-16~\cite{simonyan2014very}} & \makecell{Reconstruct.\\+ KD} & \accmetric, \compmetric, \datametric & \\ \hline
            \textunderset{(2021)}{\citet{lee2021splittable}} & Object detection & COCO 2017~\cite{lin2014microsoft} & YOLOv5~\cite{ultralytics2021yolov5} & CE-based & \accmetric, \compmetric, \datametric, \latencymetric & \\ \hline
            \textunderset{(2022)}{\citet{matsubara2022bottlefit}} & Image Classification & ImageNet (2012)~\cite{russakovsky2015imagenet} & \makecell{DenseNet-169~\cite{huang2017densely}\\DenseNet-201~\cite{huang2017densely}\\ResNet-152~\cite{he2016deep}} & \makecell{Reconst.\\HND\\GHND\\CE/KD\\(Multi-stage)} & \accmetric, \compmetric, \datametric, \energymetric \latencymetric & \myhref{https://github.com/yoshitomo-matsubara/bottlefit-split_computing}{Link} \\ \hline
            \textunderset{(2021, 2022)}{Matsubara et al.~\cite{matsubara2022supervised,matsubara2022sc2}} & \makecell{Image Classification\\Object detection\\Semantic Segmentation} & \makecell{ImageNet (2012)~\cite{russakovsky2015imagenet}\\COCO 2017~\cite{lin2014microsoft}\\PASCAL VOC (2012)~\cite{everingham2012pascal}} & \makecell{ResNet-50~\cite{he2016deep}\\ResNet-101~\cite{he2016deep}\\RegNetY-6.4GF~\cite{radosavovic2020designing}\\Hybrid ViT~\cite{steiner2021train}\\RetinaNet~\cite{lin2017focal}\\Faster R-CNN~\cite{ren2015faster}\\DeepLabv3~\cite{chen2017rethinking}} & \makecell{GHND\\CE/KD+Rate\\(Multi-stage)} & \accmetric, \compmetric, \datametric, \latencymetric & \makecell{\myhref{https://github.com/yoshitomo-matsubara/supervised-compression}{\makecell{Link\\(2021)}}\\\\\myhref{https://github.com/yoshitomo-matsubara/sc2-benchmark}{\makecell{Link\\(2022)}}} \\ \hline
        \end{tabular}
        \egroup
        \\\raggedright \footnotesize \accmetric: Model accuracy, \compmetric: Model complexity, \datametric: Transferred data size, \energymetric: Energy consumption, \latencymetric: Latency, \traincostmetric: Training cost\\
        \raggedright * The repository is incomplete and lacks of instructions to reproduce the reported results for vision and speech datasets.
    \end{center}
\end{table}

These studies are all focused on image classification. Other \gls{cv} tasks present further challenges. For instance, state of the art object detectors such as R-CNN models have more narrow range of layers that we can introduce bottlenecks due to the network architecture, which has multiple forward paths to forward outputs from intermediate layers to feature pyramid network (FPN)~\cite{lin2017feature}. The head network distillation training approach -- discussed later in this section -- was used in~\citet{matsubara2020neural} to address some of these challenges and reduce the amount of data transmitted over the channel by $94$\% while degrading mAP (mean average precision) loss by $1$ point.
\citet{assine2021single} introduce bottlenecks to the EfficientDet-D2~\cite{tan2020efficientdet} object detector, and apply the  training method based on the generalized head network distillation~\cite{matsubara2020neural} and mutual learning~\cite{yang2020mutualnet} to the modified model.
Following the studies on \gls{sc} for resource-constrained edge computing systems~\cite{matsubara2019distilled,matsubara2020head,yao2020deep}, \citet{sbai2021cut} introduce autoencoder to small classifiers and train them on a subset of the ImageNet dataset in a similar manner.
These studies discuss the trade-off between accuracy and memory size on mobile devices, considering communication constraints based 3G and LoRa technologies~\cite{samie2016iot}.
Similar to~\cite{matsubara2020split,matsubara2020neural,assine2021single},~\citet{lee2021splittable} design a lightweight encoder for object detector on mobile device followed by both a module to amplify the compressed feature and the object detector to be executed on edge server.
\citet{matsubara2022bottlefit} empirically show that bottleneck-injected models can be further improved by elaborating the methods to train the models.
The resulting models outperform models with autoencoder-based feature compression (\emph{e.g.}, Fig.~\ref{fig:autoencoder}) in terms of the tradeoff between model accuracy and transferred data size.

\citet{matsubara2022supervised} propose a supervised compression method for resource-constrained edge computing systems, which adapts ideas from knowledge distillation and neural image compression~\cite{balle2016end,balle2018variational}.
Their student model (namely, \emph{Entropic Student}) contains a lightweight encoder with a learnable prior, which quantizes and entropy-codes latent representations under a prior probability model for efficiently saving the size of data to be offloaded to edge server.
By adjusting a balancing weight in their loss function during training, we can control the tradeoff between data size (rate) and model accuracy (distortion).
The performance of the entropic student model was demonstrated for three large-scale downstream supervised tasks: image classification (ImageNet), object detection (COCO), and semantic segmentation (COCO, PASCAL VOC).
Notably, the representation produced by a single trained encoder of the entropic student model can serve multiple downstream tasks.
Following the study,~\citet{matsubara2022sc2} further investigate this approach and empirically show that it generalizes to other reference models (\emph{e.g.}, ResNet-101~\cite{he2016deep}, RegNetY-6.4GF~\cite{radosavovic2020designing}, Hybrid ViT~\cite{steiner2021train}).
Through experiments, the study also points out that simply introducing such bottleneck layers at later layers in a model can improve the conventional rate-distortion (R-D) tradeoff, which will result in most of the computational load will be assigned to a weak mobile device.

In contrast to \gls{sc} studies without bottlenecks in Table~\ref{table:split_without_bottleneck}, many of the studies on bottleneck injection strategies in Table~\ref{table:split_with_bottleneck} are published with code that would help the research communities replicate/reproduce the experimental results and build on existing studies.

\subsection{\gls{sc} with Bottlenecks: Training Methodologies}
\label{sec:split_bottle_training}

Given that recent \gls{sc} studies with bottleneck injection strategies result in more or less accuracy loss comparing to the original models (\emph{i.e.}, without injected bottlenecks), various training methodologies are used and/or proposed in such studies.
Some of the training methods are designed specifically for architectures with injected bottlenecks. We now summarize the differences between the various training methodologies used in recent \gls{sc} studies.

We recall that $\mathbf{x}$ and $\mathbf{y}$ are an input (\emph{e.g.}, an RGB image) and the corresponding label (\emph{e.g.}, one-hot vector) respectively.
Given an input $\mathbf{x}$, a \gls{dnn} model $\mathcal{M}$ returns its output $\mathbf{\hat{y}} = \mathcal{M}(\mathbf{x})$ such as class probabilities in classification task.
Each of the $L$ layers of model $\mathcal{M}$ can be either low-level (\emph{e.g.}, convolution~\cite{lecun1998gradient-based}, batch normalization~\cite{ioffe2015batch}), ReLU~\cite{nair2010rectified}) or high-level layers (\emph{e.g.}, residual block in ResNet~\cite{he2016deep} and dense block in DenseNet~\cite{huang2017densely}) which are composed by multiple low-level layers. 
$\mathcal{M}(\mathbf{x})$ is a sequence of the $L$ layer functions $\mathrm{f}_{j}$'s, and the $j^\text{th}$ layer transforms $\mathbf{z}_{j-1}$, the output from the previous ${(j-1)}^\text{th}$ layer:

\begin{equation}
    \mathbf{z}_{j} = \left\{
    \begin{array}{ll}
        \mathbf{x}  & j = 0 \\
        \mathrm{f}_j(\mathbf{z}_{j-1}, \mathbf{\theta}_j) \hspace{1cm} & 1 \le j < L~, \\
        \mathrm{f}_L(\mathbf{z}_{L-1}, \mathbf{\theta}_L) = \mathcal{M}(\mathbf{x}) = \mathbf{\hat{y}} & j = L
    \end{array}
    \right.
    \label{eq:dnn_layers}
\end{equation}
\noindent where $\mathbf{\theta}_{j}$ denotes the $j^\text{th}$ layer's hyperparameters and parameters to be optimized during training.

\subsection*{Cross entropy-based training}

To optimize parameters in a \gls{dnn} model, we first need to define a loss function and update the parameters by minimizing the loss value with an optimizer such as stochastic gradient descent and Adam~\cite{kingma2015adam} during training.
In image classification, a standard method is to train a \gls{dnn} model $\mathcal{M}$ in an end-to-end manner using the cross entropy like many of the studies~\cite{eshratifar2019bottlenet,hu2020fast,matsubara2020head} in Table~\ref{table:split_with_bottleneck}.
For simplicity, here we focus on the categorical cross entropy and assume $c \equiv \mathbf{y}$ is the correct class index given a model input $\mathbf{x}$.  
Given a pair of $\mathbf{x}$ and $c$, we obtain the model output $\mathbf{\hat{y}} = \mathcal{M}(\mathbf{x})$, and then the (categorical) cross entropy loss is defined as

\begin{equation}
    \mathcal{L}_\text{CE}(\mathbf{\hat{y}}, c) = -\log \left( \frac{\exp \left(\hat{\mathbf{y}}_{c} \right)}{\sum_{j \in \mathcal{C}} \exp\left(\hat{\mathbf{y}}_j \right)} \right),
    \label{eq:ce_loss}
\end{equation}

\noindent where $\hat{\mathbf{y}}_{j}$ is the class probability for the class index $j$, and $\mathcal{C}$ is a set of considered classes ($c \in \mathcal{C}$).

\begin{figure}[t]
    \centering
    \includegraphics[width=0.8\linewidth]{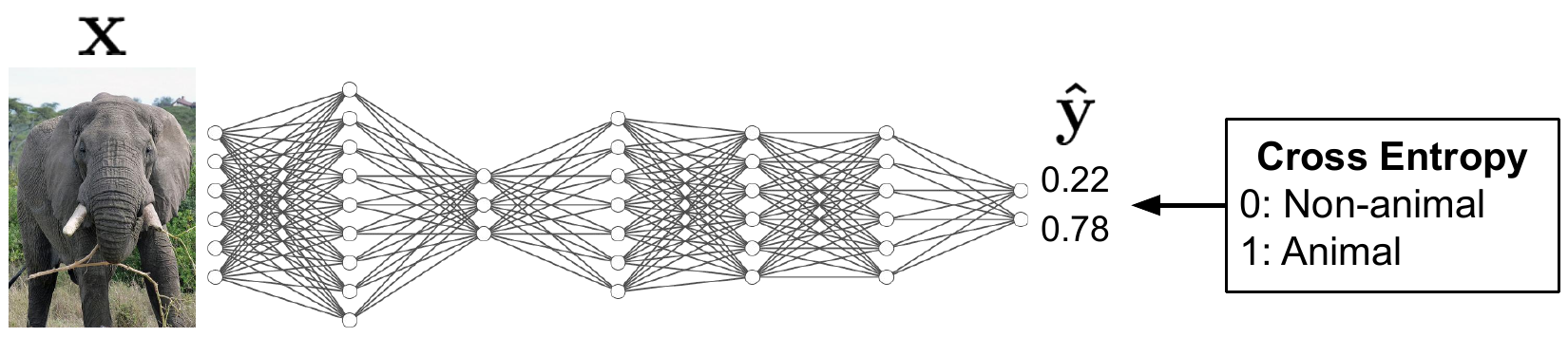}
    \vspace{-1em}
    \caption{Cross entropy-based training for bottleneck-injected \gls{dnn}.}
    \label{fig:bottleneck_ce}
\end{figure}

As shown in Eq.~(\ref{eq:ce_loss}), the loss function used in cross entropy-based training methods are used as a function of the final output $\mathbf{\hat{y}}$, and thus are not designed for \gls{sc} frameworks.
While~\citet{eshratifar2019bottlenet,hu2020fast,shao2020bottlenet++,lee2021splittable} use cross entropy to train bottleneck-injected \gls{dnn} models in end-to-end manners (Fig.~\ref{fig:bottleneck_ce}),~\citet{matsubara2020head} empirically show that these methods cause a larger accuracy loss in complex tasks such as ImageNet dataset~\cite{russakovsky2015imagenet} compared to other more advanced techniques, including knowledge distillation.

\subsection*{Knowledge distillation}
Complex \gls{dnn} models are usually trained to learn parameters for discriminating between a large number of classes (\emph{e.g.}, $1,000$ in ImageNet dataset), and are often overparameterized.
\Gls{kd}~\cite{li2014learning,ba2014do,hinton2014distilling} is a training scheme to address this problem, and trains a \gls{dnn} model (called ``student'') using additional signals from a pretrained \gls{dnn} model (called ``teacher'' and often larger than the student). 
In standard cross entropy-based training -- that is, using ``hard targets'' (\emph{e.g.}, one-hot vectors) -- we face a side-effect that the trained models assign probabilities to all of the incorrect classes.
From the relative probabilities of incorrect classes, we can see how large models tend to generalize.

\begin{figure}[t]
    \centering
    \includegraphics[width=0.8\linewidth]{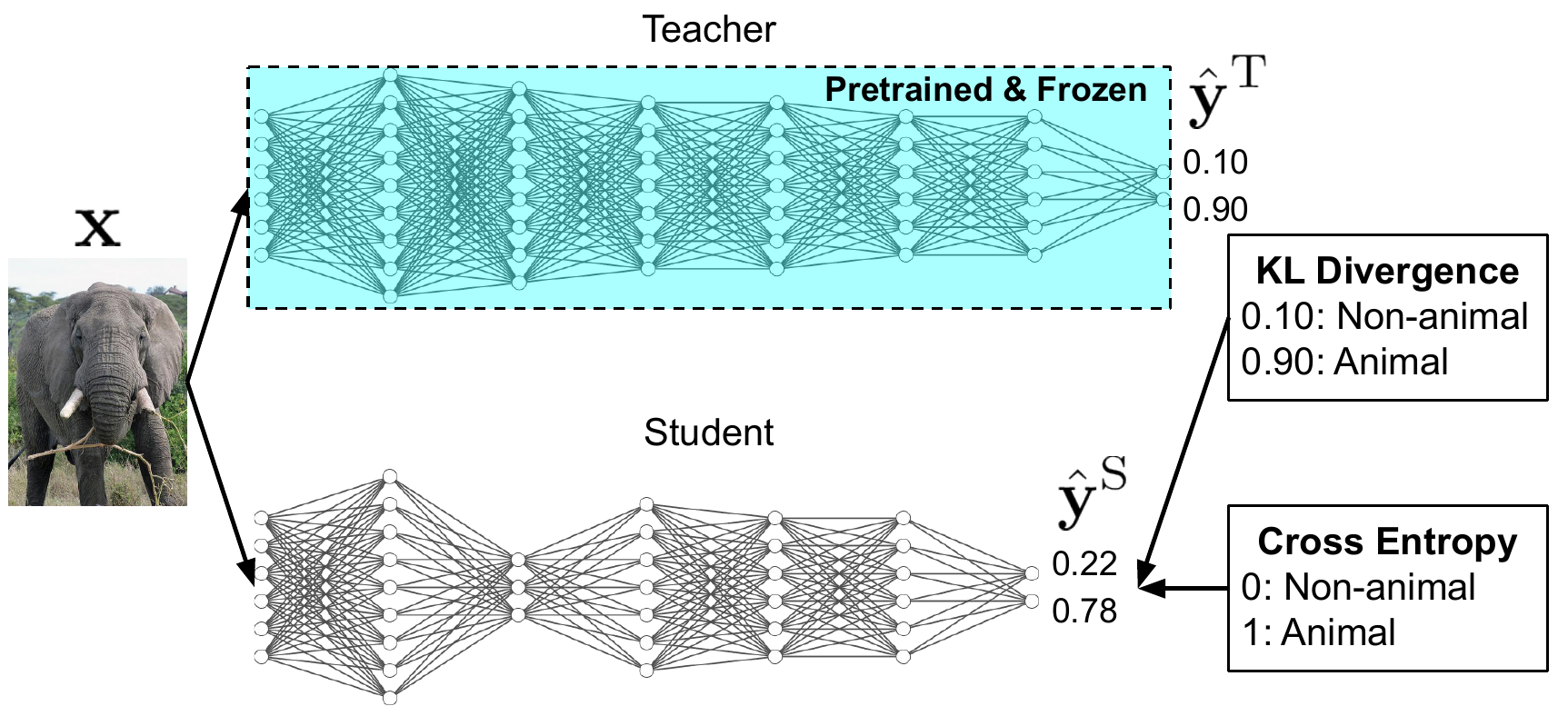}
    \vspace{-1em}
    \caption{Knowledge distillation for bottleneck-injected \gls{dnn} (student), using a pretrained model as teacher.}
    \label{fig:bottleneck_kd}
\end{figure}

As illustrated in Fig.~\ref{fig:bottleneck_kd}, by distilling the knowledge from a pretrained complex model (teacher), a student model can be more generalized and avoid overfitting to the training dataset, using the outputs of the teacher model as ``soft targets'' in addition to the hard targets~\cite{hinton2014distilling}.

\begin{equation}
    \mathcal{L}_\text{KD}(\hat{\mathbf{y}}^\text{S}, \hat{\mathbf{y}}^\text{T}, \mathbf{y}) = \alpha \mathcal{L}_\text{task}(\hat{\mathbf{y}}^\text{S}, \mathbf{y}) + (1 - \alpha) \tau^2  \mathrm{KL} \left( \mathrm{q}(\hat{\mathbf{y}}^\text{S}), \mathrm{p}(\hat{\mathbf{y}}^\text{T}) \right),
    \label{eq:kd_loss}
\end{equation}

\noindent
where $\alpha$ is a balancing factor (hyperparameter) between \emph{hard target} (left term) and \emph{soft target} (right term) losses, and $\tau$ is another hyperparameter called \emph{temperature} to soften the outputs of teacher and student models in Eq.~(\ref{eq:pq}).
$\mathcal{L}_\text{task}$ is a task-specific loss function, and it is a cross entropy loss in image classification tasks \emph{i.e.}, $\mathcal{L}_\text{task} = \mathcal{L}_\text{CE}$.
$\mathrm{KL}$ is the Kullback-Leibler divergence function, where $\mathrm{q}(\hat{\mathbf{y}}^\text{S})$ and $\mathrm{p}(\hat{\mathbf{y}}^\text{T})$ are probability distributions of student and teacher models for an input $\mathbf{x}$, that is, $\mathrm{q}(\hat{\mathbf{y}}^\text{S}) = [\mathrm{q}_{1}(\hat{\mathbf{y}}^\text{S}), \cdots, \mathrm{q}_{|\mathcal{C}|}(\hat{\mathbf{y}}^\text{S})]$ and $\mathrm{p}(\hat{\mathbf{y}}^\text{T}) = [\mathrm{p}_{1}(\hat{\mathbf{y}}^\text{S}), \cdots, \mathrm{p}_{|C|}(\hat{\mathbf{y}}^\text{T})]$:

\begin{equation}
    \mathrm{q}_{k}(\hat{\mathbf{y}}^\text{S}) = \frac{\exp \left( \frac{\hat{\mathbf{y}}^\text{S}_{k}}{\tau} \right)}{\sum_{j \in \mathcal{C}} \exp\left( \frac{\hat{\mathbf{y}}^\text{S}_{j}}{\tau} \right)}, ~~\mathrm{p}_{k}(\hat{\mathbf{y}}^\text{T}) = \frac{\exp \left( \frac{\hat{\mathbf{y}}^\text{T}_{k}}{\tau} \right)}{\sum_{j \in \mathcal{C}} \exp\left( \frac{\hat{\mathbf{y}}^\text{T}_{j}}{\tau} \right)},
    \label{eq:pq}
\end{equation}

Using the ImageNet dataset, it is empirically shown in~\citet{matsubara2020head} that all the considered bottleneck-injected student models trained with their teacher models (original models without injected bottlenecks) consistently outperform those trained without the teacher models. This result matches a widely known trend in knowledge distillation reported in~\citet{ba2014do}.
However, similar to cross entropy, the knowledge distillation is still not aware of bottlenecks we introduce to \gls{dnn} models and may result in significant accuracy loss as suggested by \citet{matsubara2020head}. 

\subsection*{Reconstruction-based training}

As illustrated in Fig.~\ref{fig:autoencoder}, \citet{choi2020back,jankowski2020joint,yao2020deep,sbai2021cut} inject \gls{ae} models into existing \gls{dnn} models, and train the injected components by minimizing the reconstruction error.
First manually an intermediate layer in a \gls{dnn} model (say its $j^\text{th}$ layer) is chosen, and the output of the $j^\text{th}$ layer $\mathbf{z}_{j}$ is fed to the encoder $\mathrm{f}_\text{enc}$ whose role is to compress $\mathbf{z}_{j}$.
The encoder's output $\mathbf{z}_\text{enc}$ is a compressed representation, \emph{i.e.}, bottleneck to be transferred to edge server and the following decoder $\mathrm{f}_\text{dec}$ decompresses the compressed representation and returns $\mathbf{z}_\text{dec}$.
As the decoder is designed to reconstruct $\mathbf{z}_{j}$, its output $\mathbf{z}_\text{dec}$ should share the same dimensionality with $\mathbf{z}_{j}$.
Then, the injected \gls{ae} are trained by minimizing the following reconstruction loss:

\begin{figure}[t]
    \centering
    \includegraphics[width=0.8\linewidth]{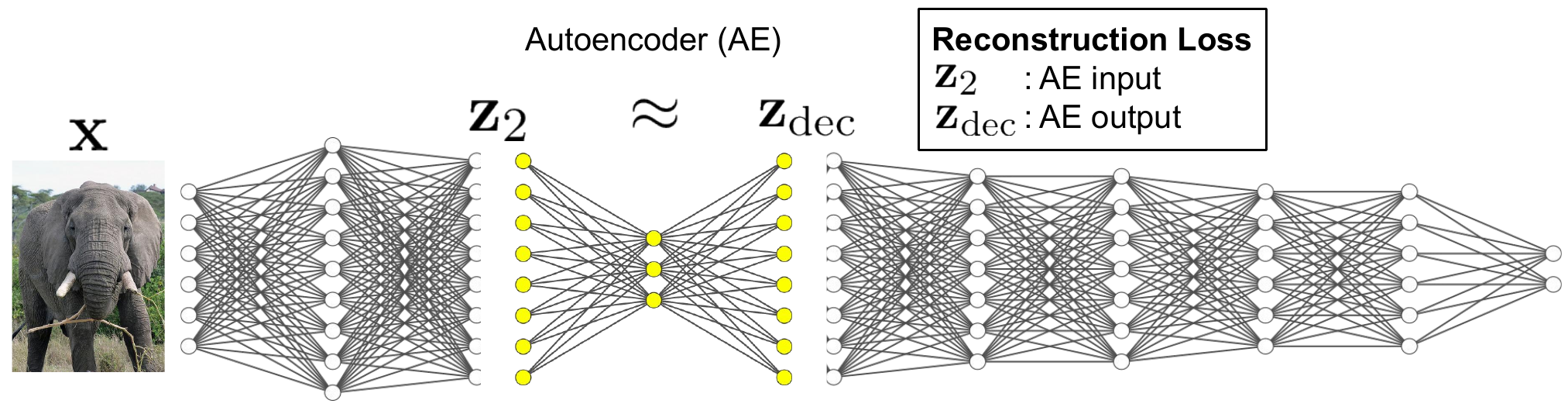}
    \vspace{-1em}
    \caption{Reconstruction-based training to compress intermediate output (here $\mathbf{z}_{2}$) in \gls{dnn} by \gls{ae} (yellow).}
    \label{fig:autoencoder}
\end{figure}

\begin{eqnarray}
    \mathcal{L}_\text{Recon.}\left( \mathbf{z}_{j} \right) &=& \| \mathbf{z}_{j} - \mathrm{f}_\text{dec}\left( \mathrm{f}_\text{enc}\left( \mathbf{z}_{j}; \theta_\text{enc} \right); \mathbf{\theta_\text{dec}} \right) + \epsilon \|_{n}^{m}, \\
    &=& \| \mathbf{z}_{j} - \mathbf{z}_\text{dec} + \epsilon \|_{n}^{m} \nonumber,
    \label{eq:l2_loss}
\end{eqnarray}

\noindent where $\|\mathbf{z}\|_n^m$ denotes $m^\text{th}$ power of $n$-norm of $\mathbf{z}$, and $\epsilon$ is an optional regularization constant.
For example, \citet{choi2020back} set $m = 1$, $n = 2$ and $\epsilon = 10^{-6}$, and \citet{jankowski2020joint} use $m = n = 1$ and $\epsilon = 0$.
Inspired by the idea of knowledge distillation~\cite{hinton2014distilling}, \citet{yao2020deep} also consider additional squared errors between intermediate feature maps from models with and without bottlenecks as additional loss terms like generalized head network distillation~\cite{matsubara2020neural} described later.
While \citet{yao2020deep} shows high compression rate with small accuracy loss by injecting encoder-decoder architectures to existing \gls{dnn} models, such strategies~\cite{choi2020back,jankowski2020joint,yao2020deep,sbai2021cut} increase computational complexity as a result.
Suppose the encoder and decoder consist of $L_\text{enc}$ and $L_\text{dec}$ layers respectively, then the total number of layers in the altered \gls{dnn} model is $L + L_\text{enc} + L_\text{dec}$.

\subsection*{Head network distillation}
\begin{figure}[t]
    \centering
    \includegraphics[width=0.75\linewidth]{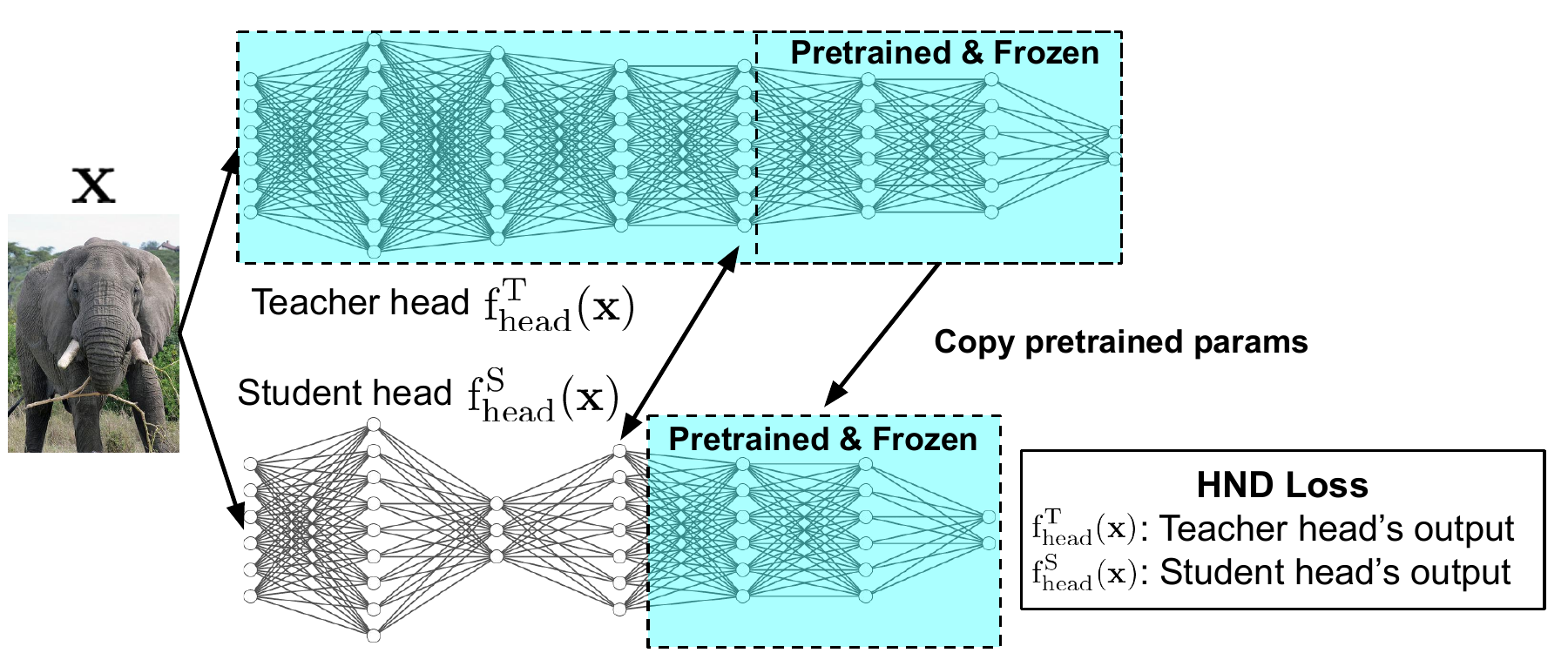}
    \vspace{-1em}
    \caption{Head network distillation for bottleneck-injected \gls{dnn} (student), using a pretrained model as teacher. The student model's tail portion is copied from that of its teacher model with respect to the architecture and pretrained parameters.}
    \label{fig:bottleneck_hnd}
\end{figure}

The training methods described above are focused on either end-to-end or encoder-decoder training.
The first approach often requires hard targets such as one-hot vectors and more training cost while the latter can focus on the injected components (encoder and decoder) during training, but the additional components (layers) will increase the complexity of the \gls{dnn} model.
To reduce both training cost and model complexity while preserving accuracy, it is proposed in~\citet{matsubara2019distilled} to use \gls{hnd} to distill the head portion of the \gls{dnn} -- which contains a bottleneck -- leveraging pretrained \gls{dnn} models.
Figure~\ref{fig:bottleneck_hnd} illustrates this approach.

The original pretrained \gls{dnn} (consisting of $L$ layers) is used as a starting point, whose architecture (in the head part) is simplified.
As only the teacher's head portion is altered, the tail portion of the student model is identical to that of the teacher model with respect to architecture and the same pretrained parameters can be maintained.
Thus, head network distillation requires only the first layers of the teacher and student models in training session as the student head model $\mathrm{f}_\text{head}^\text{S}$ will be trained to mimic behavior of teacher's head model $\mathrm{f}_\text{head}^\text{T}$ given an input $\mathbf{x}$:

\begin{equation}
    \mathcal{L}_\text{HND}(\mathbf{x}) = \| \mathrm{f}_\text{head}^\text{S}(\mathbf{x}; \mathbf{\theta}_\text{head}^\text{S}) - \mathrm{f}_\text{head}^\text{T}(\mathbf{x}; \mathbf{\theta}_\text{head}^\text{T}) \|^2 ,
    \label{eq:hnd_loss}
\end{equation}

\noindent where $\mathrm{f}_\text{head}^\text{S}$ and $\mathrm{f}_\text{head}^\text{T}$ are sequences of the first $L_\text{head}^\text{S}$ and $L_\text{head}^\text{T}$ layers in student and teacher models ($L_\text{head}^\text{S} \ll L^\text{S}$, and $L_\text{head}^\text{T} \ll L$ ), respectively.

Experimental results with the ImageNet (ILSVRC 2012) dataset show that given a bottleneck-introduced model, the head network distillation method consistently outperforms cross entropy-based training~\cite{eshratifar2019bottlenet,hu2020fast,shao2020bottlenet++} and knowledge distillation methods in terms of not only training cost but also accuracy of the trained model.
This method is extended in~\citet{matsubara2020neural}, where the generalized head network distillation technique (GHND) is proposed for complex object detection tasks and models. We note that these tasks require finer feature maps mimicking those at intermediate layers in the original pretrained object detectors. The loss function in this approach is

\begin{equation}
    \mathcal{L}_\text{GHND}(\mathbf{x}) = \sum_{j \in \mathcal{J}} \lambda_{j} \cdot \mathcal{L}_{j}(\mathbf{x}, \mathrm{f}_{1-L_j^\text{S}}^\text{S}, \mathrm{f}_{1-L_j}^\text{T}),
    \label{eq:ghnd_loss}
\end{equation}

\noindent where $j$ is loss index, $\lambda_{j}$ is a scale factor (hyperparameter) associated with loss $\mathcal{L}_{j}$, and $\mathrm{f}_{1-L_j^\text{S}}^\text{S}$ and $\mathrm{f}_{1-L_j^\text{T}}^\text{T}$ indicate the corresponding sequences of the first $L_j^\text{S}$ and $L_j^\text{T}$ layers in the student and teacher models (functions of input data $\mathbf{x}$), respectively.
The total loss, then, is a linear combination of $|\mathcal{J}|$ weighted losses.
Following Eq.~(\ref{eq:ghnd_loss}), the previously proposed head network distillation technique~\cite{matsubara2019distilled} can be seen as a special case of \gls{ghnd}.
\gls{ghnd} significantly improved the object detection performance in bottleneck-injected R-CNN models on COCO 2017 dataset while achieving a high compression rate.

\section{Early Exiting: A Survey}
\label{sec:earlyexiting}

This section presents a survey of the state of the art in \gls{ee} strategies. We first provide a compendium of work focused on \gls{cv} and \gls{nlp} applications in Sections~\ref{subsec:vision_app} and~\ref{subsec:nlp_app}, respectively.
Section~\ref{subsec:train_early_exits} summarizes training methodologies used in the \gls{ee} studies.

\subsection{Rationale behind \gls{ee}}

The core idea of \gls{ee}, first proposed in~\citet{teerapittayanon2016branchynet}, is to circumvent the need to make \gls{dnn} models smaller by introducing early exits in the \gls{dnn}, where execution is terminated at the first exit achieving the desired confidence on the input sample. For instance, some samples in test datasets (and in real-world problems) will be easy for a \gls{dnn} model, but others may not be, depending on ML models we use. Thus, \gls{ee} ends the inference process with fewer transforms (layers) for such easy samples so that the overall inference time and computation cost are reduced.

\begin{figure}[h]
    \centering
    \includegraphics[width=0.95\linewidth]{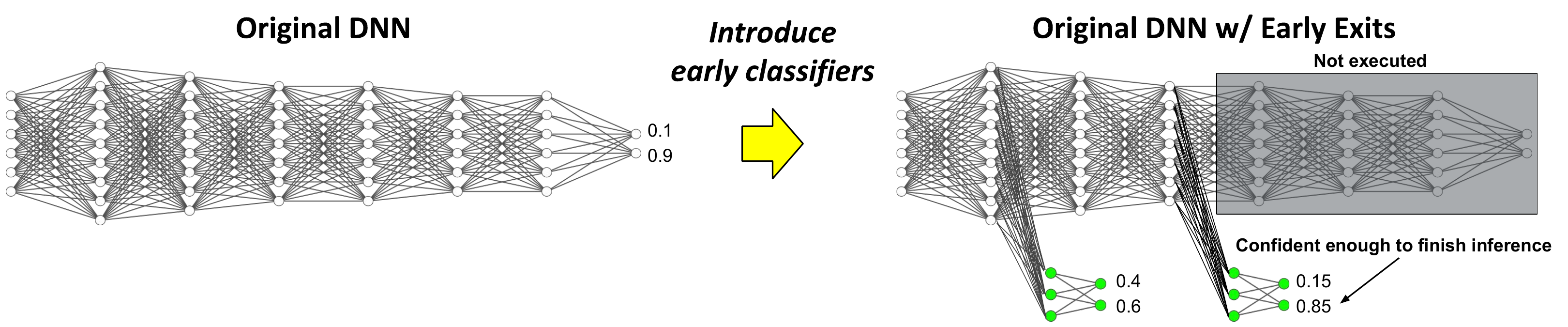}
    \caption{Illustration of two early exits (green) introduced to \gls{dnn}.}
    \label{fig:early_exiting}
\end{figure}

Figure~\ref{fig:early_exiting} illustrates an example of early classifiers (subbranches) introduced in a \gls{dnn} model. In this example, the second early classifier has sufficient confidence in its output (class probability is 0.85 out of 1.0) to terminate the inference for the input sample so that the following layers are not executed. Note that all the exits are executed until the desired confidence is reached, that is, the computational complexity up to that point increases. Thus, the classifiers added to the \gls{dnn} model need to be simple, that is, they need to have fewer layers than the layers after the branches. Otherwise, the overall inference cost will increase on average rather than decrease.  \citet{teerapittayanon2017distributed} also applies this idea to mobile-edge-cloud computing systems; the smallest neural model is allocated to the mobile device, and if that model's confidence for the input is not large enough, the intermediate output is forwarded to the edge server, where inference will continue using a mid-sized neural model with another exit. If the output still does not reach the target confidence, the intermediate layer's output is forwarded to the cloud, which executes the largest neural model. \gls{ee} strategies have been widely investigated in the literature, as summarized in Table~\ref{table:early_exits}.

As shown in Tables~\ref{table:split_without_bottleneck} and~\ref{table:split_with_bottleneck}, most of the studies on \gls{sc} were focused on computer vision.
For \gls{ee}, we can confirm a good balance between the studies with computer vision and \gls{nlp} applications as summarized in Table~\ref{table:early_exits}, with structural/conceptual differences between the two domains.
Moreover, \gls{cnn} (\emph{e.g.}, AlexNet~\cite{krizhevsky2012imagenet} and ResNet~\cite{he2016deep}) and Transformer-based models (\emph{e.g.}, BERT~\cite{devlin2019bert}) are mostly discussed in the \gls{ee} studies for computer vision and \gls{nlp}, respectively.
For these reasons, we categorize the \gls{ee} papers by task domain in Sections~\ref{subsec:vision_app} and~\ref{subsec:nlp_app}.

\begin{table}
    \caption{Studies on \uline{early exiting strategies.}}
    \vspace{-1em}
    \label{table:early_exits}
    \def\arraystretch{1.5}
    \begin{center}
        \bgroup
        \setlength{\tabcolsep}{0.3em}
        \scriptsize
        \begin{tabular}{|c|c|c|c|c|c|}
            \hline
            \multicolumn{1}{|c|}{\bf Work} & \multicolumn{1}{c|}{\bf Task(s)} & \multicolumn{1}{c|}{\bf Dataset(s)} & \multicolumn{1}{c|}{\bf Base Model(s)} & \multicolumn{1}{c|}{\bf Metrics} & \multicolumn{1}{c|}{\bf Code}\\ \hline
            \hline
            \textunderset{(2016)}{\citet{teerapittayanon2016branchynet}} & Image classification & \makecell{MNIST~\cite{lecun1998gradient-based}\\CIFAR-10~\cite{krizhevsky2009learning}} & \makecell{LeNet-5~\cite{lecun1998gradient-based}\\AlexNet~\cite{krizhevsky2012imagenet}\\ResNet~\cite{he2016deep}} & \accmetric, \latencymetric & \myhref{https://gitlab.com/kunglab/branchynet}{Link} \\ \hline
            \textunderset{(2017)}{\citet{teerapittayanon2017distributed}} & Image classification* & \makecell{Multi-camera \\multi-object detection~\cite{roig2011conditional}} & Distributed DNNs & \accmetric, \datametric & \myhref{https://github.com/kunglab/ddnn}{Link} \\\hline
            \textunderset{(2017)}{\citet{lo2017adynamic}} & Image classification & CIFAR-10/100~\cite{krizhevsky2009learning} & \makecell{NiN~\cite{lin2014network}\\ResNet~\cite{he2016deep}\\WRN~\cite{zagoruyko2016wide}} & \accmetric, \compmetric &  \\ \hline
            \textunderset{(2019)}{\citet{neshatpour2019exploiting}} & Image classification & ImageNet~\cite{russakovsky2015imagenet} & AlexNet~\cite{krizhevsky2012imagenet} & \accmetric, \compmetric, \latencymetric & \\ \hline
            \textunderset{(2019)}{\citet{zeng2019boomerang}} & Image classification & CIFAR-10~\cite{krizhevsky2009learning} & AlexNet~\cite{krizhevsky2012imagenet} & \accmetric, \datametric, \latencymetric & \\ \hline
            \textunderset{(2019)}{\citet{wang2019dynexit}} & Image classification & CIFAR-10/100~\cite{krizhevsky2009learning} & ResNet~\cite{he2016deep} & \accmetric, \compmetric &  \\ \hline
            \textunderset{(2019)}{\citet{li2019improved}} & Image classification & \makecell{CIFAR-10/100~\cite{krizhevsky2009learning}\\ImageNet (2012)~\cite{russakovsky2015imagenet}} & MSDNet~\cite{huang2018multiscale} & \accmetric, \compmetric & \myhref{https://github.com/kalviny/IMTA}{Link} \\ \hline
            \textunderset{(2019)}{\citet{phuong2019distillationbased}} & Image classification & \makecell{CIFAR-100~\cite{krizhevsky2009learning}\\ImageNet (2012)~\cite{russakovsky2015imagenet}} & MSDNet~\cite{huang2018multiscale} & \accmetric &  \myhref{https://github.com/mary-phuong/multiexit-distillation}{Link} \\ \hline
            \textunderset{(2020)}{\citet{elbayad2020depthadaptive}} & Machine translation & \makecell{IWSLT'14 De-En\\WMT'14 En-Fr} & Transformer~\cite{vaswani2017attention} & \accmetric, \compmetric & \\ \hline
            \textunderset{(2020)}{\citet{wang2020dual}} & Image classification & \makecell{CIFAR-100~\cite{krizhevsky2009learning}\\ImageNet (2012)~\cite{russakovsky2015imagenet}} & \makecell{ResNet~\cite{he2016deep}\\DenseNet~\cite{huang2017densely}} & \accmetric, \compmetric, \energymetric & \\ \hline
            \textunderset{(2020)}{\citet{yang2020resolution}} & Image classification & \makecell{CIFAR-10/100~\cite{krizhevsky2009learning}\\ImageNet~\cite{russakovsky2015imagenet}} & RANet & \accmetric, \compmetric & \myhref{https://github.com/yangle15/RANet-pytorch}{Link} \\ \hline
            \textunderset{(2020)}{\citet{soldaini2020cascade}} & Text ranking & \makecell{WikiQA~\cite{yang2015wikiqa}, TREC QA~\cite{wang2007jeopardy},\\ASNQ~\cite{garg2020tanda}, GPD} & RoBERTa~\cite{liu2019roberta} & \accmetric, \compmetric & \myhref{https://github.com/alexa/wqa-cascade-transformers}{Link} \\ \hline
            \textunderset{(2020)}{\citet{liu2020fastbert}} & Text classification & \makecell{ChnSentiCorp,\\Book review~\cite{qiu2018reevisiting},\\Shopping review,\\Weibo, THUCNews,\\Ag.News, Amz.F, DBpedia,\\Yahoo, Yelp.F, Yelp.P~\cite{zhang2015characterlevel}} & BERT~\cite{devlin2019bert} & \accmetric, \compmetric, \traincostmetric & \myhref{https://github.com/autoliuweijie/FastBERT}{Link} \\ \hline
            \textunderset{(2020)}{\citet{xin2020deebert}} & GLUE~\cite{wang2019glue} & \makecell{SST-2~\cite{socher2013recursive}, MRPC~\cite{dolan2005automatically},\\ QQP~\cite{iyer2017first}, MNLI~\cite{williams2018broad},\\ QNLI~\cite{rajpurkar2016squad}, \\RTE~\cite{dagan2005pascal,haim2006second,giampiccolo2007third,bentivogli2009fifth}} & \makecell{BERT~\cite{devlin2019bert}\\RoBERTa~\cite{liu2019roberta}} & \accmetric, \compmetric & \myhref{https://github.com/castorini/DeeBERT}{Link} \\ \hline
            \textunderset{(2020)}{\citet{xing2020early}} & Quality enhancement & RAISE~\cite{dang2015raise} & Dynamic DNN & \accmetric, \compmetric & \myhref{https://github.com/RyanXingQL/RBQE}{Link} \\ \hline
            \textunderset{(2020)}{\citet{laskaridis2020spinn}} & Image classification & \makecell{CIFAR-100~\cite{krizhevsky2009learning}\\ImageNet (2012)~\cite{russakovsky2015imagenet}} & \makecell{ResNet-56~\cite{he2016deep}\\ResNet-50~\cite{he2016deep}\\Inception-v3~\cite{szegedy2016rethinking}} & \accmetric, \energymetric, \latencymetric & \\ \hline
            \textunderset{(2020)}{\citet{xin2020early}} & Text ranking & \makecell{MS MARCO~\cite{nguyen2016ms}\\ASNQ~\cite{garg2020tanda}} & BERT~\cite{devlin2019bert} & \accmetric, \latencymetric & \myhref{https://github.com/castorini/earlyexiting-monobert}{Link} \\ \hline
            \textunderset{(2020)}{\citet{zhou2020bert}} & GLUE~\cite{wang2019glue} & \makecell{CoLA~\cite{warstadt2019neural}, SST-2~\cite{socher2013recursive},\\ MRPC~\cite{dolan2005automatically}, STS-B~\cite{cer2017semeval},\\ QQP~\cite{iyer2017first}, MNLI~\cite{williams2018broad},\\ QNLI~\cite{rajpurkar2016squad}, WNLI~\cite{levesque2012winograd},\\RTE~\cite{dagan2005pascal,haim2006second,giampiccolo2007third,bentivogli2009fifth}} & \makecell{BERT~\cite{devlin2019bert}\\ALBERT~\cite{lan2019albert}} & \accmetric, \compmetric, \latencymetric, \traincostmetric & \myhref{https://github.com/JetRunner/PABEE}{Link} \\ \hline
            \textunderset{(2020)}{\citet{matsubara2020neural}} & \makecell{Keypoint detection} & COCO 2017~\cite{lin2014microsoft} & \makecell{Keypoint R-CNN~\cite{he2017mask}} & \accmetric, \datametric, \latencymetric & \myhref{https://github.com/yoshitomo-matsubara/hnd-ghnd-object-detectors}{Link} \\
            \hline
            \textunderset{(2021)}{\citet{garg2021will}} & \makecell{Text ranking\\Question answering} & \makecell{WikiQA~\cite{yang2015wikiqa}, ASNQ~\cite{garg2020tanda}\\SQuAD 1.1~\cite{rajpurkar2016squad}} & \makecell{BERT~\cite{devlin2019bert}\\RoBERTa~\cite{liu2019roberta}\\ELECTRA~\cite{clark2019electra}} & \accmetric, \latencymetric & \myhref{https://github.com/alexa/wqa-question-filtering}{Link} \\ \hline
            \textunderset{(2021)}{\citet{wolczyk2021zero}} & Image classification & \makecell{CIFAR-10/100~\cite{krizhevsky2009learning}, Tiny ImageNet} & \makecell{ResNet-56~\cite{he2016deep}\\MobileNet~\cite{howard2017mobilenets}\\WideResNet~\cite{zagoruyko2016wide}\\VGG-16BN~\cite{simonyan2014very}} & \accmetric, \latencymetric & \myhref{https://github.com/gmum/Zero-Time-Waste}{Link} \\ \hline
            \textunderset{(2021)}{\citet{chiang2021optimal}} & Image classification & CIFAR-100~\cite{krizhevsky2009learning} & \makecell{VGG-11~\cite{simonyan2014very}\\VGG-13~\cite{simonyan2014very}\\VGG-16~\cite{simonyan2014very}\\VGG-19~\cite{simonyan2014very}} & \accmetric, \latencymetric & \\ \hline
            \textunderset{(2021)}{\citet{pomponi2021probabilistic}} & Image classification & \makecell{SVHN~\cite{netzer2011reading}, CIFAR-10/100~\cite{krizhevsky2009learning}} & \makecell{AlexNet~\cite{krizhevsky2012imagenet}\\VGG-11~\cite{simonyan2014very}\\ResNet-20~\cite{he2016deep}} & \accmetric & \myhref{https://github.com/jaryP/ConfidenceBranchNetwok}{Link} \\ \hline
        \end{tabular}
        \\\footnotesize \accmetric: Model accuracy, \compmetric: Model complexity, \datametric: Transferred data size, \energymetric: Energy consumption, \latencymetric: Latency, \traincostmetric: Training cost
        \\\raggedright *~ The authors extract annotated objects from the original dataset for multi-camera object detection, and use the extracted images for an image classification task.\\
        \egroup
    \end{center}
\end{table}

\subsection{\gls{ee} for \gls{cv} Applications}
\label{subsec:vision_app}

Similar to the \gls{sc} studies we discussed in Session~\ref{sec:split_comp}, the research community mainly focused on \gls{ee} approaches applied to \gls{cv} tasks.

\subsection*{Design approaches}
\citet{wang2020dual} propose a unified Dual Dynamic Inference that introduces the following features to a \gls{dnn} model: Input-Adaptive Dynamic Inference (IADI) and Resource-Adaptive Dynamic Inference (RADI).
The IADI dynamically determines which sub-networks to be executed for cost-efficient inference, and the RADI leverages the concept of \gls{ee} to offer ``anytime classification''.
Using the concept of \gls{ee}, \citet{lo2017adynamic} proposes two different methods: (i) authentic operation, and (ii) dynamic network sizing.
The first approach is used to determine whether the model input is transferred to the edge server, and the latter dynamically adjusts the number of layers to be used as an auxiliary neural model deployed on mobile device for efficient usage of communication channels in \gls{ec} systems.
\citet{neshatpour2019exploiting} decomposes a \gls{dnn}'s inference pipeline into multiple stages, and introduce \gls{ee} (termination) points for energy-efficient inference.

\subsection*{Training approaches}
\citet{wang2019dynexit} focus on training methods for \glspl{dnn} with an early exit and observes that prior \gls{ee} approaches suffered from the burden of manually tuning balancing weights of early exit losses to find a good trade-off between computational complexity and overall accuracy.
To address this problem, the authors propose a strategy to dynamically adjust the loss weights for the ResNet models they consider.
\citet{li2019improved} and \citet{phuong2019distillationbased} introduce multiple early exits to \gls{dnn} models and apply knowledge distillation to each of the early exits as students, using their final classifiers as teacher models.
Similar to other studies, the \glspl{dnn} with early exits are designed to finish inference for ``easy'' samples by early sub-classifiers based on confidence thresholds defined beforehand.

\subsection*{Inference approaches}
\citet{yang2020resolution} leverage \gls{ee} strategies for multi-scale inputs, and propose an approach to classify ``easy'' samples with smaller neural models.
Different from prior studies, their proposed approach scales up the input image (use higher-resolution image as input), depending on the classification difficulty of the sample.
\citet{laskaridis2020spinn} design a distributed inference system that employs synergistic device-cloud computation for collaborative inference, including an \gls{ee} strategy (referred to as progressive inference in their work).
\citet{xing2020early} apply \gls{ee} strategies to quality enhancement tasks and propose a resource-efficient blind quality enhancement approach for compressed images.
By identifying ``easy'' samples in the tasks, they dynamically process input samples with/without early exits. \citet{zeng2019boomerang} combine \gls{ee} and \gls{sc} approaches, and propose a framework named Boomerang, which is designed to automate end-to-end \gls{dnn} inference planning for IoT scenarios; they introduce multiple early exits in AlexNet~\cite{krizhevsky2012imagenet}. Their proposed framework profiles the model to decide its partition (splitting) point.

In addition to introducing and training bottleneck points for object detector, \citet{matsubara2020neural} introduce a \emph{neural filter} in an early stage of the head-distilled Keypoint R-CNN model. Similarly to \gls{ee} frameworks, the filter identifies pictures without objects of interest and trigger termination of the execution before the output of the bottleneck is forwarded.
\citet{wolczyk2021zero} propose Zero Time Waste, a method in which each early exit reuses predictions returned by its predecessors.
The method adds direct connections between early exits and combines outputs of the previous early exits like an ensemble model.
Through experiments with multiple image classification datasets and model architectures, they demonstrate that their proposed method improves a tradeoff between accuracy and inference time comparing to other early exit methods.
Extending the idea of BranchyNet~\cite{teerapittayanon2016branchynet}, \citet{chiang2021optimal} formulate the early-exit (branch) placement problem.
They propose a dynamic programming algorithm to address the problem and discuss the tradeoff between model accuracy and inference time.
\citet{pomponi2021probabilistic} introduce multiple early exits to a classifier and train the entire multi-exit model jointly.
Using multiple base models, they discuss various early-exit stopping criteria.
Many studies on \gls{ee} for \gls{cv} tasks publish their source code to ensure replicability of their work.

\subsection{\gls{ee} for \gls{nlp} Applications}
\label{subsec:nlp_app}
Interestingly, \gls{ee} approaches have been widely studied not only in \gls{cv} tasks -- the main application of \gls{sc} -- but also \gls{nlp} tasks. Recent studies introduce subbranches (early exits) to transformer-based models such as BERT~\cite{devlin2019bert}. While these transformer-based models achieve state of the art performance in \gls{nlp} tasks, they have an extremely large number of parameters, \emph{e.g.}, BERT~\cite{devlin2019bert} has up to 355 million parameters where the largest image classification model used in \gls{sc} studies (Tables ~\ref{table:split_without_bottleneck} and ~\ref{table:split_with_bottleneck}), ResNet-152, has 60 million parameters.

In~\citet{elbayad2020depthadaptive} an \gls{ee} technique for \gls{nlp} tasks is developed for transformer sequence-to-sequence models~\cite{vaswani2017attention} in machine translation tasks. The decoder networks in the considered transformer models can be trained by either aligned training or mixed training methods. The former method optimizes all classifiers in the decoder network simultaneously. However, when a different classifier (exit) is chosen for each token (\emph{e.g.}, word) at test time, some of the hidden states from previous time steps may be missed and then the input states to the following decoder network will be misaligned (mismatched). The latter method addresses this issue. In mixed sample training,  several paths of random exits are sampled at which the model is assumed to have exited for reducing the mismatch by feeding hidden states from different decoder depths of previous time steps.

For different tasks, \citet{soldaini2020cascade}, \citet{xin2020deebert} and \citet{liu2020fastbert} propose \gls{ee} frameworks based on BERT~\cite{devlin2019bert} and RoBERTa~\cite{liu2019roberta} that share almost the same network architecture.
Focused on text ranking, specifically answer sentence selection tasks with question answering datasets, \citet{soldaini2020cascade} add classification layers to intermediate stages of RoBERTa to build sequential (neural) rerankers~\cite{matsubara2020reranking} inside as early exits, and propose the Cascade Transformer models.
Focusing on powerful transformer models for industrial scenarios, \citet{liu2020fastbert} discuss the effectiveness on twelve (six English and six Chinese) NLP datasets of BERT models when early classifiers are introduced.
Similar to the studies by \citet{li2019improved} and \citet{phuong2019distillationbased}, \citet{liu2020fastbert} leverage knowledge distillation~\cite{hinton2014distilling} to train early classifiers, treating the final classifier of the BERT model and their introduced early classifiers as a teacher and student classifiers, respectively.
\citet{xin2020deebert} target general language understanding evaluation (GLUE) tasks~\cite{wang2019glue}, and introduce early exits after each of 12 transformer blocks in BERT and RoBERTa models.

While the Cascade Transformer~\cite{soldaini2020cascade} disregards a fixed portion of candidates (samples) given a query in answer sentence selection tasks, \citet{xin2020early} use a score-based \gls{ee} strategy for a BERT architecture for text ranking tasks.
\citet{zhou2020bert} introduce early classifiers to BERT and ALBERT~\cite{lan2019albert} models and discusses adversarial robustness using the ALBERT models with and without the early exits.
Using an adversarial attack method~\cite{jin2020bert}, the authors feed perturbed input data (called adversarial examples~\cite{kurakin2016adversarial}) to their trained models and show how robust their models are against the adversarial attack, compared to those without early classifiers.
\citet{garg2021will} propose an approach to filter out questions in answer sentence selection and question answering tasks.
Leveraging the concept of knowledge distillation, they train a question filter model (student), whose input is a query, by mimicking the top-1 candidate score of the answer model (teacher), whose input is a pair of query and the list of the candidate answers.
When the trained question filter model finds a query answerable for the answer model, the subsequent inference pipeline will be executed.
Otherwise, the question filter model terminates the inference process for the query (\emph{i.e.}, early exit) to save the overall inference cost.

Most of the studies on \gls{ee} for \gls{nlp} tasks in Table~\ref{table:early_exits} are published with source code to assure replicable results.
Notably, this application domain enjoys a well-generalized open source framework -- Huggingface's Transformers~\cite{wolf2020transformers} -- which provides state-of-the-art (pretrained) Transformer models, including the BERT, RoBERTa, and ALBERT models used in the above studies.

\subsection{Training Methodologies for \gls{ee} Strategies}
\label{subsec:train_early_exits}

\begin{figure}[t]
    \centering
    \includegraphics[width=1.0\linewidth]{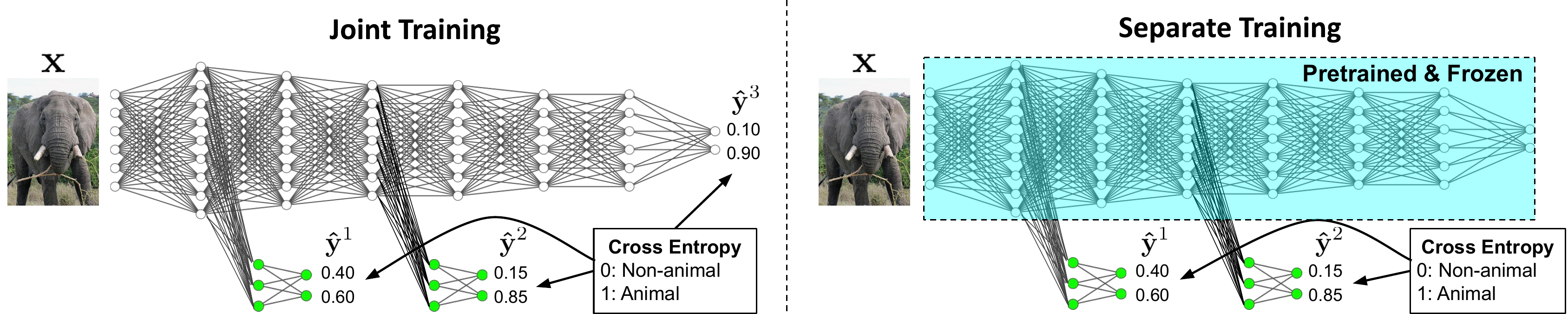}
    \vspace{-1em}
    \caption{Examples of joint and separate training methods for \gls{dnn} with early exits.}
    \label{fig:train_early_exits}
\end{figure}

To introduce \gls{ee} strategies, the early classifiers need to be trained in addition to the base models.
We can categorize the training methodologies used in \gls{ee} studies into two main classes: \emph{joint training} and \emph{separate training}, illustrated in Fig.~\ref{fig:train_early_exits} and described in the next sections.

\subsection*{Joint training}
Most of the training methods used in existing works belong to this category. Joint training trains all the (early) classifiers in a model simultaneously (left part of Fig.~\ref{fig:train_early_exits}).
Specifically, these studies~\cite{teerapittayanon2016branchynet,teerapittayanon2017distributed,lo2017adynamic,zeng2019boomerang,wang2019dynexit,elbayad2020depthadaptive,wang2020dual,yang2020resolution,soldaini2020cascade,xing2020early,laskaridis2020spinn,xin2020early,zhou2020bert,pomponi2021probabilistic} define a loss function for each of the classifier, and minimize the weighted sum of cross entropy losses per sample as follows:

\begin{equation}
    \mathcal{L}_\text{Joint}([\mathbf{\hat{y}}^{1}, \cdots, \mathbf{\hat{y}}^{N}], c) = \sum_{j=1}^{N} \lambda_{j} \mathcal{L}_\text{CE}(\mathbf{\hat{y}}^{j},  c),
    \label{eq:weighted_sum}
\end{equation}
\noindent where $[\mathbf{\hat{y}}^{1}, \cdots, \mathbf{\hat{y}}^{N}]$ indicates outputs from $N$ (early) classifiers, and the correct label $c$ is shared across all the classifiers in a model.
Note that the base model (final classifier) is also counted as one of the $N$ classifiers, and $N-1$ early classifiers are introduced to the base model.

Instead, \citet{li2019improved,phuong2019distillationbased} use a knowledge distillation-based loss such as Eq.~(\ref{eq:kd_loss}) by treating the final classifier (last exit) as teacher model and all the early classifiers as student models.
This approach is based on the assumption that the last classifier will achieve the highest accuracy among all the (early) classifiers in the model, and early classifiers (students) could learn from the last classifier as a teacher model.

\subsection*{Separate training}
A few studies~\cite{liu2020fastbert,xin2020deebert,matsubara2020neural,garg2021will} suggest training the early classifiers separately.
This approach can be interpreted as a two-stage training paradigm that trains a model in the first stage and then trains the early classifiers introduced to the pretrained model whose parameters are fixed in the second stage (See Fig.~\ref{fig:train_early_exits} (right)).
For instance, \citet{xin2020deebert} fine-tune a BERT model in the first stage following~\citet{devlin2019bert}. Then, the early classifiers are introduced in the model and trained while all the parameters of the BERT model learnt in the first stage are kept frozen.
\citet{liu2020fastbert} adopt a similar approach, but in the second training stage, knowledge distillation is used to train the early classifiers.
Different from \gls{sc} studies using knowledge distillation, the teacher model is fixed, and only the additional parameters corresponding to the early classifiers are trained.
\citet{wolczyk2021zero} introduce early exits to a pretrained model.
Using the cross entropy loss, they train the introduced early exits.

\section{Split Computing and Early Exiting: Research Challenges}\label{sec:research_challenges}

In this section, we describe some of the research challenges in the \gls{sc} and \gls{ee} domains.

\subsection*{Evaluation of \gls{sc} and \gls{ee} in more practical settings}

Due to the cross-disciplinary nature of this research area, it is essential to design practical and convincing evaluation settings to demonstrate the effectiveness of proposed approaches. As shown in Tables~\ref{table:split_with_bottleneck} and \ref{table:early_exits}, the techniques proposed in many of the recent related studies are validated only on small-scale datasets such as MNIST and CIFAR datasets, which leads to some concerns on the input data size in relation with compression. Indeed, Table~\ref{table:input_sizes} suggests that the input data size in many of such datasets is relatively small (\emph{e.g.}, smaller than 2 kilobytes per image with a resolution of $32 \times 32$ pixels).
The low-resolution of the input size may enable conventional \gls{ec}, where the mobile device fully offloads the computing task by transferring the input data to an edge server. In fact, the transmission of such a small amount of data would require a short time even in settings with limited communication capacity. As a consequence, executing even small head models on a resource-limited mobile device could lead to an overall delay increase.

Based on the above discussion, it becomes apparent that the models and datasets, in addition to the wireless and computing environments, are of paramount importance when assessing the performance of \gls{sc} and \gls{ee} schemes. Of particular relevance is the evaluation of accuracy, which is not provided in some of the early studies (\emph{e.g.},~\cite{sandler2018mobilenetv2,he2016deep,simonyan2014very}) and the consideration of state-of-the-art models and datasets which are largely used in the machine learning community. For instance, the use of small models, such as
MobileNetV2, ResNet-50, VGG-16, which are likely overparameterized for simple classification tasks, could lead to wrong conclusions when injecting bottlenecks. Conversely, it was shown in~\cite{matsubara2019distilled} how challenging it is to inject bottlenecks when considering complex vision tasks such as classification on the ImageNet dataset~\cite{russakovsky2015imagenet}.

\subsection*{Optimization of bottleneck design and placement in \gls{sc}}

The study of the architecture and placement of the bottleneck in a \gls{dnn} model is also of considerable importance. As suggested in~\cite{matsubara2022sc2}, important metrics include: (i) bottleneck data size (or compression rate), (ii) complexity of head model executed on mobile device, and (iii) resulting model accuracy. As a principle, the smaller the bottleneck representation is, the lower the communication cost between mobile device and edge server will be.
In general, the objective of \gls{sc} is to generate a bottleneck whose data size is smaller than that of input data such as JPEG file size of input data, which is in turn much smaller than data size of input tensor (32-bit floating point), as the communication delay is a key component to reduce overall inference time~\cite{matsubara2019distilled,yang2020resolution,matsubara2020head,matsubara2020neural}. Secondly, since mobile devices often have limited computing resources and may have other constraints such as energy consumption due to their battery capacities, \gls{sc} should aim at minimizing their computational load by making head models as lightweight as possible.
For instance, designing a small bottleneck at a very early stage of the \gls{dnn} model enables a reduction in the computational complexity of the head model~\cite{matsubara2020split,matsubara2020neural}.

On top of these two criteria, the resulting model accuracy by the bottleneck injection should not be compromised as the introduced bottleneck removes more or less information at the placement compared to the original model. A reasonable lower bound of the model accuracy in \gls{sc} would be that of widely recognized lightweight models \emph{e.g.,} MobileNetV2~\cite{sandler2018mobilenetv2} for ImageNet dataset, considering a local computing system where such lightweight models can be efficiently executed. In general, it is challenging to optimize bottleneck design and placement with respect to all the three different metrics, and existing studies empirically design the bottlenecks and determine the placements.
Thus, theoretical discussion on bottleneck design and placement should be an interesting research topic for future work.

\subsection*{Dynamic control of exits in \gls{ee}}

In most of the recent studies, early exits are used when one of the introduced early classifiers (exits) is confident enough in its prediction. However, users are required to determine a threshold for each of the classifiers beforehand at least for one early classifier in the original model where we introduce the early classifier to. For example, if the first classifier's prediction score is greater than 0.9 in range of 0.0 and 1.0, then the inference for the input is terminated.

To achieve more efficient inference without significantly sacrificing the accuracy of the original model, the system needs to find a balance between (early) classifiers. As recent studies introduce multiple early exits to a model at different stages, such optimizations are challenging. In addition to manually defining such a threshold for each of the classifiers based on empirical results, a possibly interesting direction is the optimization of the decision-making process, that is, at which (early) classifier we should terminate the inference for a given input, without a set of thresholds defined beforehand based on system characteristics.

\subsection*{Expanding the Application Domain of \gls{sc} and \gls{ee}}

The application domains of \gls{sc} and (in minor part) \gls{ee} remain primarily focused on image classification. This focus may be explained by the size of the input, which makes compression a relevant problem in many settings and the complexity of the models and tasks. However, there are many other unexplored domains which \gls{sc} would benefit. Real-time health conditions monitoring via wearable sensors is a notable example of application where a significant amount of data is transferred from sensors to edge servers such as cellular phones and home hubs. For instance, the detection and monitoring of heart anomalies (\emph{e.g.}, arrhythmia) from  (ECG)~\cite{gadaleta2018deep} require the processing of high-rate samples (\emph{e.g.}, $100$-$1000$ per heart cycle) using high complexity DNN models\cite{hannun2019cardiologist}.  Health monitoring applications pose different challenges compared to \gls{cv}-based applications. Indeed, in the former, both the computing capacity and the bandwidth available to the system are often smaller compared to the latter scenario, and conceptual advancements are required.

\subsection*{Toward an Information-Theoretic Perspective}
\label{sec:inf_theory}

The key intuition behind the success of \gls{sc} and \gls{ee} is similar to what has led to the success of techniques such as model pruning~\cite{han2016deep,li2016pruning,he2017channel,yang2017designing} and knowledge distillation \cite{hinton2014distilling,kim2016sequence,mirzadeh2020improved}: most state-of-the-art \glspl{dnn} are significantly over-parameterized \cite{yu2020understanding,yu2019understanding}. A possible approach to justify \gls{sc} and \gls{ee} can be found in the study of \emph{information bottlenecks} (IB), which were introduced in~\cite{tishby2000information} as a compression technique in which a random variable $\mathbf{X}$ is compressed while preserving relevant information about another random variable $\mathbf{Y}$. The IB method has been applied in~\cite{tishby2015deep} to quantify mutual information between the network layers and derive an information theory limit on \gls{dnn} efficiency. This has led to attempts at explaining the behavior of deep neural networks with the information bottleneck formalism~\cite{saxe2019information}. 

Despite these early attempts, a strong connection between this relatively new perspective and the techniques described in this paper is still elusive. Some of the approaches and architectures discussed in this paper are meaningful attempts to efficiently extract a compressed representation of the input and provide sufficient information toward a certain task early in the network layers. The emerging IB formalism is a promising approach to enable the first moves in the information theoretic analysis of neural networks-based transformations. We believe that this interpretation could serve as a foundation for an in-depth study of structural properties for both \gls{sc} and \gls{ee}.

\glsresetall
\section{Conclusions}

Mobile devices such as smartphones and drones have now become an integral part of our daily lives. These devices increasingly utilize \glspl{dnn} to execute complex inference tasks such as image classification and speech recognition, among others. For this reason, in this paper, we provided a comprehensive survey of the state of the art in \gls{sc} and \gls{ee} by presenting a thorough comparison of the most relevant approaches. We also provided a set of compelling research challenges that need to be addressed to improve existing work in the field. We hope this survey will elicit further research in these emerging fields.

\begin{acks}
This work was partially supported by the NSF under grant IIS-1724331, MLWiNS-2003237, CCF-2140154, and CNS-2134567.
\end{acks}

\bibliographystyle{ACM-Reference-Format}
\bibliography{references}


\begin{thebibliography}{174}


\ifx \showCODEN    \undefined \def \showCODEN     #1{\unskip}     \fi
\ifx \showDOI      \undefined \def \showDOI       #1{#1}\fi
\ifx \showISBNx    \undefined \def \showISBNx     #1{\unskip}     \fi
\ifx \showISBNxiii \undefined \def \showISBNxiii  #1{\unskip}     \fi
\ifx \showISSN     \undefined \def \showISSN      #1{\unskip}     \fi
\ifx \showLCCN     \undefined \def \showLCCN      #1{\unskip}     \fi
\ifx \shownote     \undefined \def \shownote      #1{#1}          \fi
\ifx \showarticletitle \undefined \def \showarticletitle #1{#1}   \fi
\ifx \showURL      \undefined \def \showURL       {\relax}        \fi
\providecommand\bibfield[2]{#2}
\providecommand\bibinfo[2]{#2}
\providecommand\natexlab[1]{#1}
\providecommand\showeprint[2][]{arXiv:#2}

\bibitem[\protect\citeauthoryear{Adelantado, Vilajosana, Tuset-Peiro, Martinez,
  Melia-Segui, and Watteyne}{Adelantado et~al\mbox{.}}{2017}]%
        {adelantado2017understanding}
\bibfield{author}{\bibinfo{person}{Ferran Adelantado}, \bibinfo{person}{Xavier
  Vilajosana}, \bibinfo{person}{Pere Tuset-Peiro}, \bibinfo{person}{Borja
  Martinez}, \bibinfo{person}{Joan Melia-Segui}, {and} \bibinfo{person}{Thomas
  Watteyne}.} \bibinfo{year}{2017}\natexlab{}.
\newblock \showarticletitle{{Understanding the Limits of LoRaWAN}}.
\newblock \bibinfo{journal}{\emph{IEEE Communications Magazine}}
  \bibinfo{volume}{55}, \bibinfo{number}{9} (\bibinfo{year}{2017}),
  \bibinfo{pages}{34--40}.
\newblock


\bibitem[\protect\citeauthoryear{Assine, Valle, et~al\mbox{.}}{Assine
  et~al\mbox{.}}{2021}]%
        {assine2021single}
\bibfield{author}{\bibinfo{person}{Juliano~S Assine}, \bibinfo{person}{Eduardo
  Valle}, {et~al\mbox{.}}} \bibinfo{year}{2021}\natexlab{}.
\newblock \showarticletitle{Single-Training Collaborative Object Detectors
  Adaptive to Bandwidth and Computation}.
\newblock \bibinfo{journal}{\emph{arXiv preprint arXiv:2105.00591}}
  (\bibinfo{year}{2021}).
\newblock


\bibitem[\protect\citeauthoryear{Ba and Caruana}{Ba and Caruana}{2014}]%
        {ba2014do}
\bibfield{author}{\bibinfo{person}{Jimmy Ba} {and} \bibinfo{person}{Rich
  Caruana}.} \bibinfo{year}{2014}\natexlab{}.
\newblock \showarticletitle{{Do deep nets really need to be deep?}}. In
  \bibinfo{booktitle}{\emph{NIPS 2014}}. \bibinfo{pages}{2654--2662}.
\newblock


\bibitem[\protect\citeauthoryear{Ball{\'e}, Laparra, and Simoncelli}{Ball{\'e}
  et~al\mbox{.}}{2017}]%
        {balle2016end}
\bibfield{author}{\bibinfo{person}{Johannes Ball{\'e}}, \bibinfo{person}{Valero
  Laparra}, {and} \bibinfo{person}{Eero~P Simoncelli}.}
  \bibinfo{year}{2017}\natexlab{}.
\newblock \showarticletitle{End-to-end Optimized Image Compression}. In
  \bibinfo{booktitle}{\emph{International Conference on Learning
  Representations}}.
\newblock


\bibitem[\protect\citeauthoryear{Ball{\'e}, Minnen, Singh, Hwang, and
  Johnston}{Ball{\'e} et~al\mbox{.}}{2018}]%
        {balle2018variational}
\bibfield{author}{\bibinfo{person}{Johannes Ball{\'e}}, \bibinfo{person}{David
  Minnen}, \bibinfo{person}{Saurabh Singh}, \bibinfo{person}{Sung~Jin Hwang},
  {and} \bibinfo{person}{Nick Johnston}.} \bibinfo{year}{2018}\natexlab{}.
\newblock \showarticletitle{Variational image compression with a scale
  hyperprior}. In \bibinfo{booktitle}{\emph{International Conference on
  Learning Representations}}.
\newblock


\bibitem[\protect\citeauthoryear{Barbera, Kosta, Mei, and Stefa}{Barbera
  et~al\mbox{.}}{2013}]%
        {arbera2013to}
\bibfield{author}{\bibinfo{person}{Marco~V Barbera}, \bibinfo{person}{Sokol
  Kosta}, \bibinfo{person}{Alessandro Mei}, {and} \bibinfo{person}{Julinda
  Stefa}.} \bibinfo{year}{2013}\natexlab{}.
\newblock \showarticletitle{{To offload or not to offload? the bandwidth and
  energy costs of mobile cloud computing}}. In
  \bibinfo{booktitle}{\emph{Proceedings of IEEE INFOCOM 2013}}.
  \bibinfo{pages}{1285--1293}.
\newblock


\bibitem[\protect\citeauthoryear{Bentivogli, Clark, Dagan, and
  Giampiccolo}{Bentivogli et~al\mbox{.}}{2009}]%
        {bentivogli2009fifth}
\bibfield{author}{\bibinfo{person}{Luisa Bentivogli}, \bibinfo{person}{Peter
  Clark}, \bibinfo{person}{Ido Dagan}, {and} \bibinfo{person}{Danilo
  Giampiccolo}.} \bibinfo{year}{2009}\natexlab{}.
\newblock \showarticletitle{The Fifth PASCAL Recognizing Textual Entailment
  Challenge.}. In \bibinfo{booktitle}{\emph{Proceedings of Text Analysis
  Conference (TAC’09}}.
\newblock


\bibitem[\protect\citeauthoryear{Buciluǎ, Caruana, and
  Niculescu-Mizil}{Buciluǎ et~al\mbox{.}}{2006}]%
        {bucilua2006model}
\bibfield{author}{\bibinfo{person}{Cristian Buciluǎ}, \bibinfo{person}{Rich
  Caruana}, {and} \bibinfo{person}{Alexandru Niculescu-Mizil}.}
  \bibinfo{year}{2006}\natexlab{}.
\newblock \showarticletitle{Model compression}. In
  \bibinfo{booktitle}{\emph{Proceedings of the 12th ACM SIGKDD international
  conference on Knowledge discovery and data mining}}.
  \bibinfo{pages}{535--541}.
\newblock


\bibitem[\protect\citeauthoryear{Cer, Diab, Agirre, Lopez-Gazpio, and
  Specia}{Cer et~al\mbox{.}}{2017}]%
        {cer2017semeval}
\bibfield{author}{\bibinfo{person}{Daniel Cer}, \bibinfo{person}{Mona Diab},
  \bibinfo{person}{Eneko Agirre}, \bibinfo{person}{I{\~n}igo Lopez-Gazpio},
  {and} \bibinfo{person}{Lucia Specia}.} \bibinfo{year}{2017}\natexlab{}.
\newblock \showarticletitle{{SemEval-2017 Task 1: Semantic Textual Similarity
  Multilingual and Cross-lingual Focused Evaluation}}. In
  \bibinfo{booktitle}{\emph{Proceedings of the 11th International Workshop on
  Semantic Evaluation (SemEval-2017)}}. \bibinfo{pages}{1--14}.
\newblock


\bibitem[\protect\citeauthoryear{Chen and Ran}{Chen and Ran}{2019}]%
        {chen2019deep}
\bibfield{author}{\bibinfo{person}{Jiasi Chen} {and} \bibinfo{person}{Xukan
  Ran}.} \bibinfo{year}{2019}\natexlab{}.
\newblock \showarticletitle{{Deep Learning With Edge Computing: A Review}}.
\newblock \bibinfo{journal}{\emph{Proc. IEEE}} \bibinfo{volume}{107},
  \bibinfo{number}{8} (\bibinfo{year}{2019}), \bibinfo{pages}{1655--1674}.
\newblock


\bibitem[\protect\citeauthoryear{Chen, Papandreou, Schroff, and Adam}{Chen
  et~al\mbox{.}}{2017}]%
        {chen2017rethinking}
\bibfield{author}{\bibinfo{person}{Liang-Chieh Chen}, \bibinfo{person}{George
  Papandreou}, \bibinfo{person}{Florian Schroff}, {and}
  \bibinfo{person}{Hartwig Adam}.} \bibinfo{year}{2017}\natexlab{}.
\newblock \showarticletitle{{Rethinking Atrous Convolution for Semantic Image
  Segmentation}}.
\newblock \bibinfo{journal}{\emph{arXiv preprint arXiv:1706.05587}}
  (\bibinfo{year}{2017}).
\newblock


\bibitem[\protect\citeauthoryear{Chiang, Liu, Wang, Hong, and Wu}{Chiang
  et~al\mbox{.}}{2021}]%
        {chiang2021optimal}
\bibfield{author}{\bibinfo{person}{Chang-Han Chiang}, \bibinfo{person}{Pangfeng
  Liu}, \bibinfo{person}{Da-Wei Wang}, \bibinfo{person}{Ding-Yong Hong}, {and}
  \bibinfo{person}{Jan-Jan Wu}.} \bibinfo{year}{2021}\natexlab{}.
\newblock \showarticletitle{{Optimal Branch Location for Cost-effective
  Inference on Branchynet}}. In \bibinfo{booktitle}{\emph{2021 IEEE
  International Conference on Big Data (Big Data)}}. IEEE,
  \bibinfo{pages}{5071--5080}.
\newblock


\bibitem[\protect\citeauthoryear{Choi and Baji{\'c}}{Choi and
  Baji{\'c}}{2018}]%
        {choi2018deep}
\bibfield{author}{\bibinfo{person}{Hyomin Choi} {and} \bibinfo{person}{Ivan~V
  Baji{\'c}}.} \bibinfo{year}{2018}\natexlab{}.
\newblock \showarticletitle{{Deep Feature Compression for Collaborative Object
  Detection}}. In \bibinfo{booktitle}{\emph{2018 25th IEEE International
  Conference on Image Processing (ICIP)}}. IEEE, \bibinfo{pages}{3743--3747}.
\newblock


\bibitem[\protect\citeauthoryear{Choi, Cohen, and Baji{\'c}}{Choi
  et~al\mbox{.}}{2020}]%
        {choi2020back}
\bibfield{author}{\bibinfo{person}{Hyomin Choi}, \bibinfo{person}{Robert~A
  Cohen}, {and} \bibinfo{person}{Ivan~V Baji{\'c}}.}
  \bibinfo{year}{2020}\natexlab{}.
\newblock \showarticletitle{{Back-And-Forth Prediction for Deep Tensor
  Compression}}. In \bibinfo{booktitle}{\emph{ICASSP 2020-2020 IEEE
  International Conference on Acoustics, Speech and Signal Processing
  (ICASSP)}}. IEEE, \bibinfo{pages}{4467--4471}.
\newblock


\bibitem[\protect\citeauthoryear{Clark, Luong, Le, and Manning}{Clark
  et~al\mbox{.}}{2019}]%
        {clark2019electra}
\bibfield{author}{\bibinfo{person}{Kevin Clark}, \bibinfo{person}{Minh-Thang
  Luong}, \bibinfo{person}{Quoc~V Le}, {and} \bibinfo{person}{Christopher~D
  Manning}.} \bibinfo{year}{2019}\natexlab{}.
\newblock \showarticletitle{{ELECTRA: Pre-training Text Encoders as
  Discriminators Rather Than Generators}}. In
  \bibinfo{booktitle}{\emph{International Conference on Learning
  Representations}}.
\newblock


\bibitem[\protect\citeauthoryear{Cohen, Choi, and Baji{\'c}}{Cohen
  et~al\mbox{.}}{2020}]%
        {cohen2020lightweight}
\bibfield{author}{\bibinfo{person}{Robert~A Cohen}, \bibinfo{person}{Hyomin
  Choi}, {and} \bibinfo{person}{Ivan~V Baji{\'c}}.}
  \bibinfo{year}{2020}\natexlab{}.
\newblock \showarticletitle{{Lightweight Compression Of Neural Network Feature
  Tensors For Collaborative Intelligence}}. In \bibinfo{booktitle}{\emph{2020
  IEEE International Conference on Multimedia and Expo (ICME)}}. IEEE,
  \bibinfo{pages}{1--6}.
\newblock


\bibitem[\protect\citeauthoryear{Collobert, Weston, Bottou, Karlen,
  Kavukcuoglu, and Kuksa}{Collobert et~al\mbox{.}}{2011}]%
        {collobert2011natural}
\bibfield{author}{\bibinfo{person}{Ronan Collobert}, \bibinfo{person}{Jason
  Weston}, \bibinfo{person}{L{\'e}on Bottou}, \bibinfo{person}{Michael Karlen},
  \bibinfo{person}{Koray Kavukcuoglu}, {and} \bibinfo{person}{Pavel Kuksa}.}
  \bibinfo{year}{2011}\natexlab{}.
\newblock \showarticletitle{Natural language processing (almost) from scratch}.
\newblock \bibinfo{journal}{\emph{Journal of machine learning research}}
  \bibinfo{volume}{12} (\bibinfo{year}{2011}), \bibinfo{pages}{2493--2537}.
\newblock


\bibitem[\protect\citeauthoryear{Dagan, Glickman, and Magnini}{Dagan
  et~al\mbox{.}}{2005}]%
        {dagan2005pascal}
\bibfield{author}{\bibinfo{person}{Ido Dagan}, \bibinfo{person}{Oren Glickman},
  {and} \bibinfo{person}{Bernardo Magnini}.} \bibinfo{year}{2005}\natexlab{}.
\newblock \showarticletitle{The PASCAL recognising textual entailment
  challenge}. In \bibinfo{booktitle}{\emph{Proceedings of the First
  international conference on Machine Learning Challenges: evaluating
  Predictive Uncertainty Visual Object Classification, and Recognizing Textual
  Entailment}}. \bibinfo{pages}{177--190}.
\newblock


\bibitem[\protect\citeauthoryear{Dang-Nguyen, Pasquini, Conotter, and
  Boato}{Dang-Nguyen et~al\mbox{.}}{2015}]%
        {dang2015raise}
\bibfield{author}{\bibinfo{person}{Duc-Tien Dang-Nguyen},
  \bibinfo{person}{Cecilia Pasquini}, \bibinfo{person}{Valentina Conotter},
  {and} \bibinfo{person}{Giulia Boato}.} \bibinfo{year}{2015}\natexlab{}.
\newblock \showarticletitle{{RAISE}: A raw images dataset for digital image
  forensics}. In \bibinfo{booktitle}{\emph{Proceedings of the 6th ACM
  multimedia systems conference}}. \bibinfo{pages}{219--224}.
\newblock


\bibitem[\protect\citeauthoryear{Deng, Hinton, and Kingsbury}{Deng
  et~al\mbox{.}}{2013}]%
        {deng2013new}
\bibfield{author}{\bibinfo{person}{Li Deng}, \bibinfo{person}{Geoffrey Hinton},
  {and} \bibinfo{person}{Brian Kingsbury}.} \bibinfo{year}{2013}\natexlab{}.
\newblock \showarticletitle{{New types of deep neural network learning for
  speech recognition and related applications: An overview}}. In
  \bibinfo{booktitle}{\emph{2013 IEEE international conference on acoustics,
  speech and signal processing}}. IEEE, \bibinfo{pages}{8599--8603}.
\newblock


\bibitem[\protect\citeauthoryear{Devlin, Chang, Lee, and Toutanova}{Devlin
  et~al\mbox{.}}{2019}]%
        {devlin2019bert}
\bibfield{author}{\bibinfo{person}{Jacob Devlin}, \bibinfo{person}{Ming-Wei
  Chang}, \bibinfo{person}{Kenton Lee}, {and} \bibinfo{person}{Kristina
  Toutanova}.} \bibinfo{year}{2019}\natexlab{}.
\newblock \showarticletitle{{BERT: Pre-training of Deep Bidirectional
  Transformers for Language Understanding}}. In
  \bibinfo{booktitle}{\emph{Proceedings of the 2019 Conference of the North
  American Chapter of the Association for Computational Linguistics: Human
  Language Technologies, Volume 1 (Long and Short Papers)}}.
  \bibinfo{pages}{4171--4186}.
\newblock


\bibitem[\protect\citeauthoryear{Dolan and Brockett}{Dolan and
  Brockett}{2005}]%
        {dolan2005automatically}
\bibfield{author}{\bibinfo{person}{William~B Dolan} {and}
  \bibinfo{person}{Chris Brockett}.} \bibinfo{year}{2005}\natexlab{}.
\newblock \showarticletitle{Automatically constructing a corpus of sentential
  paraphrases}. In \bibinfo{booktitle}{\emph{Proceedings of the Third
  International Workshop on Paraphrasing (IWP2005)}}.
\newblock


\bibitem[\protect\citeauthoryear{Dong, Cheng, Juan, Wei, and Sun}{Dong
  et~al\mbox{.}}{2018}]%
        {dong2018dpp}
\bibfield{author}{\bibinfo{person}{Jin-Dong Dong}, \bibinfo{person}{An-Chieh
  Cheng}, \bibinfo{person}{Da-Cheng Juan}, \bibinfo{person}{Wei Wei}, {and}
  \bibinfo{person}{Min Sun}.} \bibinfo{year}{2018}\natexlab{}.
\newblock \showarticletitle{{DPP-Net: Device-aware Progressive Search for
  Pareto-optimal Neural Architectures}}. In
  \bibinfo{booktitle}{\emph{Proceedings of the European Conference on Computer
  Vision (ECCV)}}. \bibinfo{pages}{517--531}.
\newblock


\bibitem[\protect\citeauthoryear{Elbayad, Gu, Grave, and Auli}{Elbayad
  et~al\mbox{.}}{2020}]%
        {elbayad2020depthadaptive}
\bibfield{author}{\bibinfo{person}{Maha Elbayad}, \bibinfo{person}{Jiatao Gu},
  \bibinfo{person}{E. Grave}, {and} \bibinfo{person}{M. Auli}.}
  \bibinfo{year}{2020}\natexlab{}.
\newblock \showarticletitle{Depth-Adaptive Transformer}. In
  \bibinfo{booktitle}{\emph{International Conference on Learning
  Representations}}.
\newblock


\bibitem[\protect\citeauthoryear{Eshratifar, Abrishami, and Pedram}{Eshratifar
  et~al\mbox{.}}{2019a}]%
        {eshratifar2019jointdnn}
\bibfield{author}{\bibinfo{person}{Amir~Erfan Eshratifar},
  \bibinfo{person}{Mohammad~Saeed Abrishami}, {and} \bibinfo{person}{Massoud
  Pedram}.} \bibinfo{year}{2019}\natexlab{a}.
\newblock \showarticletitle{{JointDNN: An Efficient Training and Inference
  Engine for Intelligent Mobile Cloud Computing Services}}.
\newblock \bibinfo{journal}{\emph{IEEE Transactions on Mobile Computing}}
  (\bibinfo{year}{2019}).
\newblock


\bibitem[\protect\citeauthoryear{Eshratifar, Esmaili, and Pedram}{Eshratifar
  et~al\mbox{.}}{2019b}]%
        {eshratifar2019bottlenet}
\bibfield{author}{\bibinfo{person}{Amir~Erfan Eshratifar},
  \bibinfo{person}{Amirhossein Esmaili}, {and} \bibinfo{person}{Massoud
  Pedram}.} \bibinfo{year}{2019}\natexlab{b}.
\newblock \showarticletitle{{BottleNet: A Deep Learning Architecture for
  Intelligent Mobile Cloud Computing Services}}. In
  \bibinfo{booktitle}{\emph{2019 IEEE/ACM Int. Symposium on Low Power
  Electronics and Design (ISLPED)}}. \bibinfo{pages}{1--6}.
\newblock


\bibitem[\protect\citeauthoryear{Everingham, Van~Gool, Williams, Winn, and
  Zisserman}{Everingham et~al\mbox{.}}{2012}]%
        {everingham2012pascal}
\bibfield{author}{\bibinfo{person}{Mark Everingham}, \bibinfo{person}{Luc
  Van~Gool}, \bibinfo{person}{CKI Williams}, \bibinfo{person}{John Winn}, {and}
  \bibinfo{person}{Andrew Zisserman}.} \bibinfo{year}{2012}\natexlab{}.
\newblock \showarticletitle{{The PASCAL Visual Object Classes Challenge 2012
  (VOC2012)}}.
\newblock


\bibitem[\protect\citeauthoryear{Everingham, Van~Gool, Williams, Winn, and
  Zisserman}{Everingham et~al\mbox{.}}{[n.d.]}]%
        {everingham2007pascal}
\bibfield{author}{\bibinfo{person}{M. Everingham}, \bibinfo{person}{L.
  Van~Gool}, \bibinfo{person}{C.~K.~I. Williams}, \bibinfo{person}{J. Winn},
  {and} \bibinfo{person}{A. Zisserman}.} \bibinfo{year}{[n.d.]}\natexlab{}.
\newblock \bibinfo{title}{{The PASCAL Visual Object Classes Challenge 2007
  (VOC2007) Results}}.
\newblock
  \bibinfo{howpublished}{http://www.pascal-network.org/challenges/VOC/voc2007/workshop/index.html}.
\newblock


\bibitem[\protect\citeauthoryear{Fei-Fei, Fergus, and Perona}{Fei-Fei
  et~al\mbox{.}}{2006}]%
        {fei2006one}
\bibfield{author}{\bibinfo{person}{Li Fei-Fei}, \bibinfo{person}{Rob Fergus},
  {and} \bibinfo{person}{Pietro Perona}.} \bibinfo{year}{2006}\natexlab{}.
\newblock \showarticletitle{One-shot learning of object categories}.
\newblock \bibinfo{journal}{\emph{IEEE transactions on pattern analysis and
  machine intelligence}} \bibinfo{volume}{28}, \bibinfo{number}{4}
  (\bibinfo{year}{2006}), \bibinfo{pages}{594--611}.
\newblock


\bibitem[\protect\citeauthoryear{Gadaleta, Rossi, Steinhubl, and Quer}{Gadaleta
  et~al\mbox{.}}{2018}]%
        {gadaleta2018deep}
\bibfield{author}{\bibinfo{person}{Matteo Gadaleta}, \bibinfo{person}{Michele
  Rossi}, \bibinfo{person}{Steven~R Steinhubl}, {and} \bibinfo{person}{Giorgio
  Quer}.} \bibinfo{year}{2018}\natexlab{}.
\newblock \showarticletitle{Deep learning to detect atrial fibrillation from
  short noisy ECG segments measured with wireless sensors}.
\newblock \bibinfo{journal}{\emph{Circulation}} \bibinfo{volume}{138},
  \bibinfo{number}{Suppl\_1} (\bibinfo{year}{2018}),
  \bibinfo{pages}{A16177--A16177}.
\newblock


\bibitem[\protect\citeauthoryear{Garg and Moschitti}{Garg and
  Moschitti}{2021}]%
        {garg2021will}
\bibfield{author}{\bibinfo{person}{Siddhant Garg} {and}
  \bibinfo{person}{Alessandro Moschitti}.} \bibinfo{year}{2021}\natexlab{}.
\newblock \showarticletitle{{Will this Question be Answered? Question Filtering
  via Answer Model Distillation for Efficient Question Answering}}. In
  \bibinfo{booktitle}{\emph{Proceedings of the 2021 Conference on Empirical
  Methods in Natural Language Processing}}. \bibinfo{pages}{7329--7346}.
\newblock


\bibitem[\protect\citeauthoryear{Garg, Vu, and Moschitti}{Garg
  et~al\mbox{.}}{2020}]%
        {garg2020tanda}
\bibfield{author}{\bibinfo{person}{Siddhant Garg}, \bibinfo{person}{Thuy Vu},
  {and} \bibinfo{person}{Alessandro Moschitti}.}
  \bibinfo{year}{2020}\natexlab{}.
\newblock \showarticletitle{{TANDA}: Transfer and adapt pre-trained transformer
  models for answer sentence selection}. In
  \bibinfo{booktitle}{\emph{Proceedings of the AAAI Conference on Artificial
  Intelligence}}, Vol.~\bibinfo{volume}{34}. \bibinfo{pages}{7780--7788}.
\newblock


\bibitem[\protect\citeauthoryear{Giampiccolo, Magnini, Dagan, and
  Dolan}{Giampiccolo et~al\mbox{.}}{2007}]%
        {giampiccolo2007third}
\bibfield{author}{\bibinfo{person}{Danilo Giampiccolo},
  \bibinfo{person}{Bernardo Magnini}, \bibinfo{person}{Ido Dagan}, {and}
  \bibinfo{person}{William~B Dolan}.} \bibinfo{year}{2007}\natexlab{}.
\newblock \showarticletitle{The third pascal recognizing textual entailment
  challenge}. In \bibinfo{booktitle}{\emph{Proceedings of the ACL-PASCAL
  workshop on textual entailment and paraphrasing}}. \bibinfo{pages}{1--9}.
\newblock


\bibitem[\protect\citeauthoryear{Gundersen and Kjensmo}{Gundersen and
  Kjensmo}{2018}]%
        {gundersen2018state}
\bibfield{author}{\bibinfo{person}{Odd~Erik Gundersen} {and}
  \bibinfo{person}{Sigbj{\o}rn Kjensmo}.} \bibinfo{year}{2018}\natexlab{}.
\newblock \showarticletitle{State of the Art: Reproducibility in Artificial
  Intelligence}. In \bibinfo{booktitle}{\emph{Proceedings of the AAAI
  Conference on Artificial Intelligence}}, Vol.~\bibinfo{volume}{32}.
\newblock


\bibitem[\protect\citeauthoryear{Guo}{Guo}{2018}]%
        {guo2018cloud}
\bibfield{author}{\bibinfo{person}{Tian Guo}.} \bibinfo{year}{2018}\natexlab{}.
\newblock \showarticletitle{{Cloud-Based or On-Device: An Empirical Study of
  Mobile Deep Inference}}. In \bibinfo{booktitle}{\emph{2018 IEEE Int.
  Conference on Cloud Engineering (IC2E)}}. IEEE, \bibinfo{pages}{184--190}.
\newblock


\bibitem[\protect\citeauthoryear{Gupta, Jain, and Vaszkun}{Gupta
  et~al\mbox{.}}{2015}]%
        {gupta2015survey}
\bibfield{author}{\bibinfo{person}{Lav Gupta}, \bibinfo{person}{Raj Jain},
  {and} \bibinfo{person}{Gabor Vaszkun}.} \bibinfo{year}{2015}\natexlab{}.
\newblock \showarticletitle{{Survey of Important Issues in UAV Communication
  Networks}}.
\newblock \bibinfo{journal}{\emph{IEEE Communications Surveys \& Tutorials}}
  \bibinfo{volume}{18}, \bibinfo{number}{2} (\bibinfo{year}{2015}),
  \bibinfo{pages}{1123--1152}.
\newblock


\bibitem[\protect\citeauthoryear{Haim, Dagan, Dolan, Ferro, Giampiccolo,
  Magnini, and Szpektor}{Haim et~al\mbox{.}}{2006}]%
        {haim2006second}
\bibfield{author}{\bibinfo{person}{R~Bar Haim}, \bibinfo{person}{Ido Dagan},
  \bibinfo{person}{Bill Dolan}, \bibinfo{person}{Lisa Ferro},
  \bibinfo{person}{Danilo Giampiccolo}, \bibinfo{person}{Bernardo Magnini},
  {and} \bibinfo{person}{Idan Szpektor}.} \bibinfo{year}{2006}\natexlab{}.
\newblock \showarticletitle{The second pascal recognising textual entailment
  challenge}. In \bibinfo{booktitle}{\emph{Proceedings of the Second PASCAL
  Challenges Workshop on Recognising Textual Entailment}}.
\newblock


\bibitem[\protect\citeauthoryear{Han, Mao, and Dally}{Han
  et~al\mbox{.}}{2016}]%
        {han2016deep}
\bibfield{author}{\bibinfo{person}{Song Han}, \bibinfo{person}{Huizi Mao},
  {and} \bibinfo{person}{William~J Dally}.} \bibinfo{year}{2016}\natexlab{}.
\newblock \showarticletitle{{Deep Compression: Compressing Deep Neural Networks
  with Pruning, Trained Quantization and Huffman Coding}}. In
  \bibinfo{booktitle}{\emph{Fourth International Conference on Learning
  Representations}}.
\newblock


\bibitem[\protect\citeauthoryear{Han, Pool, Tran, and Dally}{Han
  et~al\mbox{.}}{2015}]%
        {han2015learning}
\bibfield{author}{\bibinfo{person}{Song Han}, \bibinfo{person}{Jeff Pool},
  \bibinfo{person}{John Tran}, {and} \bibinfo{person}{William Dally}.}
  \bibinfo{year}{2015}\natexlab{}.
\newblock \showarticletitle{Learning both Weights and Connections for Efficient
  Neural Network}.
\newblock \bibinfo{journal}{\emph{Advances in Neural Information Processing
  Systems}}  \bibinfo{volume}{28} (\bibinfo{year}{2015}).
\newblock


\bibitem[\protect\citeauthoryear{Hannun, Case, Casper, Catanzaro, Diamos,
  Elsen, Prenger, Satheesh, Sengupta, Coates, et~al\mbox{.}}{Hannun
  et~al\mbox{.}}{2014}]%
        {hannun2014deep}
\bibfield{author}{\bibinfo{person}{Awni Hannun}, \bibinfo{person}{Carl Case},
  \bibinfo{person}{Jared Casper}, \bibinfo{person}{Bryan Catanzaro},
  \bibinfo{person}{Greg Diamos}, \bibinfo{person}{Erich Elsen},
  \bibinfo{person}{Ryan Prenger}, \bibinfo{person}{Sanjeev Satheesh},
  \bibinfo{person}{Shubho Sengupta}, \bibinfo{person}{Adam Coates},
  {et~al\mbox{.}}} \bibinfo{year}{2014}\natexlab{}.
\newblock \showarticletitle{Deep speech: Scaling up end-to-end speech
  recognition}.
\newblock \bibinfo{journal}{\emph{arXiv preprint arXiv:1412.5567}}
  (\bibinfo{year}{2014}).
\newblock


\bibitem[\protect\citeauthoryear{Hannun, Rajpurkar, Haghpanahi, Tison, Bourn,
  Turakhia, and Ng}{Hannun et~al\mbox{.}}{2019}]%
        {hannun2019cardiologist}
\bibfield{author}{\bibinfo{person}{Awni~Y Hannun}, \bibinfo{person}{Pranav
  Rajpurkar}, \bibinfo{person}{Masoumeh Haghpanahi},
  \bibinfo{person}{Geoffrey~H Tison}, \bibinfo{person}{Codie Bourn},
  \bibinfo{person}{Mintu~P Turakhia}, {and} \bibinfo{person}{Andrew~Y Ng}.}
  \bibinfo{year}{2019}\natexlab{}.
\newblock \showarticletitle{Cardiologist-level arrhythmia detection and
  classification in ambulatory electrocardiograms using a deep neural network}.
\newblock \bibinfo{journal}{\emph{Nature medicine}} \bibinfo{volume}{25},
  \bibinfo{number}{1} (\bibinfo{year}{2019}), \bibinfo{pages}{65--69}.
\newblock


\bibitem[\protect\citeauthoryear{He, Gkioxari, Doll{\'a}r, and Girshick}{He
  et~al\mbox{.}}{2017a}]%
        {he2017mask}
\bibfield{author}{\bibinfo{person}{Kaiming He}, \bibinfo{person}{Georgia
  Gkioxari}, \bibinfo{person}{Piotr Doll{\'a}r}, {and} \bibinfo{person}{Ross
  Girshick}.} \bibinfo{year}{2017}\natexlab{a}.
\newblock \showarticletitle{{Mask R-CNN}}. In
  \bibinfo{booktitle}{\emph{Proceedings of the IEEE International Conference on
  Computer Vision}}. \bibinfo{pages}{2961--2969}.
\newblock


\bibitem[\protect\citeauthoryear{He, Zhang, Ren, and Sun}{He
  et~al\mbox{.}}{2016}]%
        {he2016deep}
\bibfield{author}{\bibinfo{person}{Kaiming He}, \bibinfo{person}{Xiangyu
  Zhang}, \bibinfo{person}{Shaoqing Ren}, {and} \bibinfo{person}{Jian Sun}.}
  \bibinfo{year}{2016}\natexlab{}.
\newblock \showarticletitle{{Deep residual learning for image recognition}}. In
  \bibinfo{booktitle}{\emph{Proceedings of the IEEE conference on computer
  vision and pattern recognition}}. \bibinfo{pages}{770--778}.
\newblock


\bibitem[\protect\citeauthoryear{He, Zhang, and Sun}{He et~al\mbox{.}}{2017b}]%
        {he2017channel}
\bibfield{author}{\bibinfo{person}{Yihui He}, \bibinfo{person}{Xiangyu Zhang},
  {and} \bibinfo{person}{Jian Sun}.} \bibinfo{year}{2017}\natexlab{b}.
\newblock \showarticletitle{{Channel Pruning for Accelerating Very Deep Neural
  Networks}}. In \bibinfo{booktitle}{\emph{Proceedings of the IEEE
  International Conference on Computer Vision}}. \bibinfo{pages}{1389--1397}.
\newblock


\bibitem[\protect\citeauthoryear{Hinton, Deng, Yu, Dahl, Mohamed, Jaitly,
  Senior, Vanhoucke, Nguyen, Sainath, et~al\mbox{.}}{Hinton
  et~al\mbox{.}}{2012}]%
        {hinton2012deep}
\bibfield{author}{\bibinfo{person}{Geoffrey Hinton}, \bibinfo{person}{Li Deng},
  \bibinfo{person}{Dong Yu}, \bibinfo{person}{George~E Dahl},
  \bibinfo{person}{Abdel-rahman Mohamed}, \bibinfo{person}{Navdeep Jaitly},
  \bibinfo{person}{Andrew Senior}, \bibinfo{person}{Vincent Vanhoucke},
  \bibinfo{person}{Patrick Nguyen}, \bibinfo{person}{Tara~N Sainath},
  {et~al\mbox{.}}} \bibinfo{year}{2012}\natexlab{}.
\newblock \showarticletitle{{Deep neural networks for acoustic modeling in
  speech recognition: The shared views of four research groups}}.
\newblock \bibinfo{journal}{\emph{IEEE Signal processing magazine}}
  \bibinfo{volume}{29}, \bibinfo{number}{6} (\bibinfo{year}{2012}),
  \bibinfo{pages}{82--97}.
\newblock


\bibitem[\protect\citeauthoryear{Hinton, Vinyals, and Dean}{Hinton
  et~al\mbox{.}}{2014}]%
        {hinton2014distilling}
\bibfield{author}{\bibinfo{person}{Geoffrey Hinton}, \bibinfo{person}{Oriol
  Vinyals}, {and} \bibinfo{person}{Jeff Dean}.}
  \bibinfo{year}{2014}\natexlab{}.
\newblock \showarticletitle{{Distilling the Knowledge in a Neural Network}}. In
  \bibinfo{booktitle}{\emph{Deep Learning and Representation Learning Workshop:
  NIPS 2014}}.
\newblock


\bibitem[\protect\citeauthoryear{Howard, Sandler, Chu, Chen, Chen, Tan, Wang,
  Zhu, Pang, Vasudevan, et~al\mbox{.}}{Howard et~al\mbox{.}}{2019}]%
        {howard2019searching}
\bibfield{author}{\bibinfo{person}{Andrew Howard}, \bibinfo{person}{Mark
  Sandler}, \bibinfo{person}{Grace Chu}, \bibinfo{person}{Liang-Chieh Chen},
  \bibinfo{person}{Bo Chen}, \bibinfo{person}{Mingxing Tan},
  \bibinfo{person}{Weijun Wang}, \bibinfo{person}{Yukun Zhu},
  \bibinfo{person}{Ruoming Pang}, \bibinfo{person}{Vijay Vasudevan},
  {et~al\mbox{.}}} \bibinfo{year}{2019}\natexlab{}.
\newblock \showarticletitle{{Searching for MobileNetV3}}. In
  \bibinfo{booktitle}{\emph{Proceedings of the IEEE/CVF International
  Conference on Computer Vision}}. \bibinfo{pages}{1314--1324}.
\newblock


\bibitem[\protect\citeauthoryear{Howard, Zhu, Chen, Kalenichenko, Wang, Weyand,
  Andreetto, and Adam}{Howard et~al\mbox{.}}{2017}]%
        {howard2017mobilenets}
\bibfield{author}{\bibinfo{person}{Andrew~G Howard}, \bibinfo{person}{Menglong
  Zhu}, \bibinfo{person}{Bo Chen}, \bibinfo{person}{Dmitry Kalenichenko},
  \bibinfo{person}{Weijun Wang}, \bibinfo{person}{Tobias Weyand},
  \bibinfo{person}{Marco Andreetto}, {and} \bibinfo{person}{Hartwig Adam}.}
  \bibinfo{year}{2017}\natexlab{}.
\newblock \showarticletitle{{MobileNets: Efficient Convolutional Neural
  Networks for Mobile Vision Applications}}.
\newblock \bibinfo{journal}{\emph{arXiv preprint arXiv:1704.04861}}
  (\bibinfo{year}{2017}).
\newblock


\bibitem[\protect\citeauthoryear{Hu and Krishnamachari}{Hu and
  Krishnamachari}{2020}]%
        {hu2020fast}
\bibfield{author}{\bibinfo{person}{Diyi Hu} {and} \bibinfo{person}{Bhaskar
  Krishnamachari}.} \bibinfo{year}{2020}\natexlab{}.
\newblock \showarticletitle{{Fast and Accurate Streaming CNN Inference via
  Communication Compression on the Edge}}. In \bibinfo{booktitle}{\emph{2020
  IEEE/ACM Fifth Int. Conference on Internet-of-Things Design and
  Implementation (IoTDI)}}. IEEE, \bibinfo{pages}{157--163}.
\newblock


\bibitem[\protect\citeauthoryear{Huang, Chen, Li, Wu, Maaten, and
  Weinberger}{Huang et~al\mbox{.}}{2018}]%
        {huang2018multiscale}
\bibfield{author}{\bibinfo{person}{Gao Huang}, \bibinfo{person}{Danlu Chen},
  \bibinfo{person}{T. Li}, \bibinfo{person}{Felix Wu},
  \bibinfo{person}{L.~V.~D. Maaten}, {and} \bibinfo{person}{Kilian~Q.
  Weinberger}.} \bibinfo{year}{2018}\natexlab{}.
\newblock \showarticletitle{{Multi-Scale Dense Networks for Resource Efficient
  Image Classification}}. In \bibinfo{booktitle}{\emph{International Conference
  on Learning Representations}}.
\newblock


\bibitem[\protect\citeauthoryear{Huang, Liu, Van Der~Maaten, and
  Weinberger}{Huang et~al\mbox{.}}{2017}]%
        {huang2017densely}
\bibfield{author}{\bibinfo{person}{Gao Huang}, \bibinfo{person}{Zhuang Liu},
  \bibinfo{person}{Laurens Van Der~Maaten}, {and} \bibinfo{person}{Kilian~Q
  Weinberger}.} \bibinfo{year}{2017}\natexlab{}.
\newblock \showarticletitle{{Densely connected convolutional networks}}. In
  \bibinfo{booktitle}{\emph{Proceedings of the IEEE conference on computer
  vision and pattern recognition}}. \bibinfo{pages}{4700--4708}.
\newblock


\bibitem[\protect\citeauthoryear{Ioffe and Szegedy}{Ioffe and Szegedy}{2015}]%
        {ioffe2015batch}
\bibfield{author}{\bibinfo{person}{Sergey Ioffe} {and}
  \bibinfo{person}{Christian Szegedy}.} \bibinfo{year}{2015}\natexlab{}.
\newblock \showarticletitle{Batch Normalization: Accelerating Deep Network
  Training by Reducing Internal Covariate Shift}. In
  \bibinfo{booktitle}{\emph{International Conference on Machine Learning}}.
  PMLR, \bibinfo{pages}{448--456}.
\newblock


\bibitem[\protect\citeauthoryear{Itahara, Nishio, and Yamamoto}{Itahara
  et~al\mbox{.}}{2021}]%
        {itahara2021packet}
\bibfield{author}{\bibinfo{person}{Sohei Itahara}, \bibinfo{person}{Takayuki
  Nishio}, {and} \bibinfo{person}{Koji Yamamoto}.}
  \bibinfo{year}{2021}\natexlab{}.
\newblock \showarticletitle{Packet-Loss-Tolerant Split Inference for
  Delay-Sensitive Deep Learning in Lossy Wireless Networks}.
\newblock \bibinfo{journal}{\emph{arXiv preprint arXiv:2104.13629}}
  (\bibinfo{year}{2021}).
\newblock


\bibitem[\protect\citeauthoryear{Iyer, Dandekar, and Csernai}{Iyer
  et~al\mbox{.}}{[n.d.]}]%
        {iyer2017first}
\bibfield{author}{\bibinfo{person}{Shankar Iyer}, \bibinfo{person}{Nikhil
  Dandekar}, {and} \bibinfo{person}{Korn{\'e}l Csernai}.}
  \bibinfo{year}{[n.d.]}\natexlab{}.
\newblock \bibinfo{title}{{First Quora Dataset Release: Question Pairs}}.
\newblock
\newblock
\newblock
\shownote{\url{https://www.quora.com/q/quoradata/First-Quora-Dataset-Release-Question-Pairs}
  [Online; Accessed on Januray 25, 2021].}


\bibitem[\protect\citeauthoryear{Jacob, Kligys, Chen, Zhu, Tang, Howard, Adam,
  and Kalenichenko}{Jacob et~al\mbox{.}}{2018}]%
        {jacob2018quantization}
\bibfield{author}{\bibinfo{person}{Benoit Jacob}, \bibinfo{person}{Skirmantas
  Kligys}, \bibinfo{person}{Bo Chen}, \bibinfo{person}{Menglong Zhu},
  \bibinfo{person}{Matthew Tang}, \bibinfo{person}{Andrew Howard},
  \bibinfo{person}{Hartwig Adam}, {and} \bibinfo{person}{Dmitry Kalenichenko}.}
  \bibinfo{year}{2018}\natexlab{}.
\newblock \showarticletitle{{Quantization and Training of Neural Networks for
  Efficient Integer-Arithmetic-Only Inference}}. In
  \bibinfo{booktitle}{\emph{Proceedings of the IEEE Conference on Computer
  Vision and Pattern Recognition}}. \bibinfo{pages}{2704--2713}.
\newblock


\bibitem[\protect\citeauthoryear{Jagannath, Polosky, Jagannath, Restuccia, and
  Melodia}{Jagannath et~al\mbox{.}}{2019}]%
        {jagannath2019machine}
\bibfield{author}{\bibinfo{person}{Jithin Jagannath}, \bibinfo{person}{Nicholas
  Polosky}, \bibinfo{person}{Anu Jagannath}, \bibinfo{person}{Francesco
  Restuccia}, {and} \bibinfo{person}{Tommaso Melodia}.}
  \bibinfo{year}{2019}\natexlab{}.
\newblock \showarticletitle{Machine Learning for Wireless Communications in the
  Internet of Things: A Comprehensive Survey}.
\newblock \bibinfo{journal}{\emph{Ad Hoc Networks}}  \bibinfo{volume}{93}
  (\bibinfo{year}{2019}), \bibinfo{pages}{101913}.
\newblock


\bibitem[\protect\citeauthoryear{Jankowski, G{\"u}nd{\"u}z, and
  Mikolajczyk}{Jankowski et~al\mbox{.}}{2020}]%
        {jankowski2020joint}
\bibfield{author}{\bibinfo{person}{Mikolaj Jankowski}, \bibinfo{person}{Deniz
  G{\"u}nd{\"u}z}, {and} \bibinfo{person}{Krystian Mikolajczyk}.}
  \bibinfo{year}{2020}\natexlab{}.
\newblock \showarticletitle{{Joint Device-Edge Inference over Wireless Links
  with Pruning}}. In \bibinfo{booktitle}{\emph{2020 IEEE 21st International
  Workshop on Signal Processing Advances in Wireless Communications (SPAWC)}}.
  IEEE, \bibinfo{pages}{1--5}.
\newblock


\bibitem[\protect\citeauthoryear{Jeong, Jeong, Lee, and Moon}{Jeong
  et~al\mbox{.}}{2018}]%
        {jeong2018computation}
\bibfield{author}{\bibinfo{person}{Hyuk-Jin Jeong}, \bibinfo{person}{InChang
  Jeong}, \bibinfo{person}{Hyeon-Jae Lee}, {and} \bibinfo{person}{Soo-Mook
  Moon}.} \bibinfo{year}{2018}\natexlab{}.
\newblock \showarticletitle{{Computation Offloading for Machine Learning Web
  Apps in the Edge Server Environment}}. In \bibinfo{booktitle}{\emph{2018 IEEE
  38th International Conference on Distributed Computing Systems (ICDCS)}}.
  \bibinfo{pages}{1492--1499}.
\newblock


\bibitem[\protect\citeauthoryear{Jin, Jin, Zhou, and Szolovits}{Jin
  et~al\mbox{.}}{2020}]%
        {jin2020bert}
\bibfield{author}{\bibinfo{person}{Di Jin}, \bibinfo{person}{Zhijing Jin},
  \bibinfo{person}{Joey~Tianyi Zhou}, {and} \bibinfo{person}{Peter Szolovits}.}
  \bibinfo{year}{2020}\natexlab{}.
\newblock \showarticletitle{{Is BERT Really Robust? A Strong Baseline for
  Natural Language Attack on Text Classification and Entailment}}. In
  \bibinfo{booktitle}{\emph{Proceedings of the AAAI conference on artificial
  intelligence}}, Vol.~\bibinfo{volume}{34}. \bibinfo{pages}{8018--8025}.
\newblock


\bibitem[\protect\citeauthoryear{Kang, Hauswald, Gao, Rovinski, Mudge, Mars,
  and Tang}{Kang et~al\mbox{.}}{2017}]%
        {kang2017neurosurgeon}
\bibfield{author}{\bibinfo{person}{Yiping Kang}, \bibinfo{person}{Johann
  Hauswald}, \bibinfo{person}{Cao Gao}, \bibinfo{person}{Austin Rovinski},
  \bibinfo{person}{Trevor Mudge}, \bibinfo{person}{Jason Mars}, {and}
  \bibinfo{person}{Lingjia Tang}.} \bibinfo{year}{2017}\natexlab{}.
\newblock \showarticletitle{{Neurosurgeon: Collaborative Intelligence Between
  the Cloud and Mobile Edge}}. In \bibinfo{booktitle}{\emph{Proceedings of the
  Twenty-Second International Conference on Architectural Support for
  Programming Languages and Operating Systems}} (Xi'an, China).
  \bibinfo{pages}{615--629}.
\newblock
\showISBNx{978-1-4503-4465-4}
\urldef\tempurl%
\url{https://doi.org/10.1145/3037697.3037698}
\showDOI{\tempurl}


\bibitem[\protect\citeauthoryear{Kim and Rush}{Kim and Rush}{2016}]%
        {kim2016sequence}
\bibfield{author}{\bibinfo{person}{Yoon Kim} {and} \bibinfo{person}{Alexander~M
  Rush}.} \bibinfo{year}{2016}\natexlab{}.
\newblock \showarticletitle{{Sequence-Level Knowledge Distillation}}. In
  \bibinfo{booktitle}{\emph{Proceedings of the 2016 Conference on Empirical
  Methods in Natural Language Processing}}. \bibinfo{pages}{1317--1327}.
\newblock


\bibitem[\protect\citeauthoryear{Kingma and Ba}{Kingma and Ba}{2015}]%
        {kingma2015adam}
\bibfield{author}{\bibinfo{person}{Diederik~P. Kingma} {and}
  \bibinfo{person}{Jimmy Ba}.} \bibinfo{year}{2015}\natexlab{}.
\newblock \showarticletitle{{Adam: A Method for Stochastic Optimization}}. In
  \bibinfo{booktitle}{\emph{Third International Conference on Learning
  Representations}}.
\newblock


\bibitem[\protect\citeauthoryear{Krizhevsky}{Krizhevsky}{2009}]%
        {krizhevsky2009learning}
\bibfield{author}{\bibinfo{person}{Alex Krizhevsky}.}
  \bibinfo{year}{2009}\natexlab{}.
\newblock \showarticletitle{{Learning Multiple Layers of Features from Tiny
  Images}}.
\newblock  (\bibinfo{year}{2009}).
\newblock


\bibitem[\protect\citeauthoryear{Krizhevsky, Sutskever, and Hinton}{Krizhevsky
  et~al\mbox{.}}{2012}]%
        {krizhevsky2012imagenet}
\bibfield{author}{\bibinfo{person}{Alex Krizhevsky}, \bibinfo{person}{Ilya
  Sutskever}, {and} \bibinfo{person}{Geoffrey~E Hinton}.}
  \bibinfo{year}{2012}\natexlab{}.
\newblock \showarticletitle{{ImageNet Classification with Deep Convolutional
  Neural Networks}}. In \bibinfo{booktitle}{\emph{Advances in Neural
  Information Processing Systems 25}},
  \bibfield{editor}{\bibinfo{person}{F.~Pereira}, \bibinfo{person}{C.~J.~C.
  Burges}, \bibinfo{person}{L.~Bottou}, {and} \bibinfo{person}{K.~Q.
  Weinberger}} (Eds.). \bibinfo{pages}{1097--1105}.
\newblock


\bibitem[\protect\citeauthoryear{Kurakin, Goodfellow, and Bengio}{Kurakin
  et~al\mbox{.}}{2016}]%
        {kurakin2016adversarial}
\bibfield{author}{\bibinfo{person}{Alexey Kurakin}, \bibinfo{person}{Ian
  Goodfellow}, {and} \bibinfo{person}{Samy Bengio}.}
  \bibinfo{year}{2016}\natexlab{}.
\newblock \showarticletitle{Adversarial examples in the physical world}.
\newblock \bibinfo{journal}{\emph{arXiv preprint arXiv:1607.02533}}
  (\bibinfo{year}{2016}).
\newblock


\bibitem[\protect\citeauthoryear{Lan, Chen, Goodman, Gimpel, Sharma, and
  Soricut}{Lan et~al\mbox{.}}{2019}]%
        {lan2019albert}
\bibfield{author}{\bibinfo{person}{Zhenzhong Lan}, \bibinfo{person}{Mingda
  Chen}, \bibinfo{person}{Sebastian Goodman}, \bibinfo{person}{Kevin Gimpel},
  \bibinfo{person}{Piyush Sharma}, {and} \bibinfo{person}{Radu Soricut}.}
  \bibinfo{year}{2019}\natexlab{}.
\newblock \showarticletitle{ALBERT: A Lite BERT for Self-supervised Learning of
  Language Representations}. In \bibinfo{booktitle}{\emph{International
  Conference on Learning Representations}}.
\newblock


\bibitem[\protect\citeauthoryear{Laskaridis, Venieris, Almeida, Leontiadis, and
  Lane}{Laskaridis et~al\mbox{.}}{2020}]%
        {laskaridis2020spinn}
\bibfield{author}{\bibinfo{person}{Stefanos Laskaridis},
  \bibinfo{person}{Stylianos~I Venieris}, \bibinfo{person}{Mario Almeida},
  \bibinfo{person}{Ilias Leontiadis}, {and} \bibinfo{person}{Nicholas~D Lane}.}
  \bibinfo{year}{2020}\natexlab{}.
\newblock \showarticletitle{{SPINN: Synergistic Progressive Inference of Neural
  Networks over Device and Cloud}}. In \bibinfo{booktitle}{\emph{Proceedings of
  the 26th Annual International Conference on Mobile Computing and
  Networking}}. \bibinfo{pages}{1--15}.
\newblock


\bibitem[\protect\citeauthoryear{LeCun, Bengio, and Hinton}{LeCun
  et~al\mbox{.}}{2015}]%
        {lecun2015deep}
\bibfield{author}{\bibinfo{person}{Yann LeCun}, \bibinfo{person}{Yoshua
  Bengio}, {and} \bibinfo{person}{Geoffrey Hinton}.}
  \bibinfo{year}{2015}\natexlab{}.
\newblock \showarticletitle{Deep learning}.
\newblock \bibinfo{journal}{\emph{Nature}} \bibinfo{volume}{521},
  \bibinfo{number}{7553} (\bibinfo{year}{2015}), \bibinfo{pages}{436}.
\newblock


\bibitem[\protect\citeauthoryear{LeCun, Bottou, Bengio, and Haffner}{LeCun
  et~al\mbox{.}}{1998}]%
        {lecun1998gradient-based}
\bibfield{author}{\bibinfo{person}{Yann LeCun}, \bibinfo{person}{L{\'e}on
  Bottou}, \bibinfo{person}{Yoshua Bengio}, {and} \bibinfo{person}{Patrick
  Haffner}.} \bibinfo{year}{1998}\natexlab{}.
\newblock \showarticletitle{{Gradient-based learning applied to document
  recognition}}.
\newblock \bibinfo{journal}{\emph{Proc. IEEE}} \bibinfo{volume}{86},
  \bibinfo{number}{11} (\bibinfo{year}{1998}), \bibinfo{pages}{2278--2324}.
\newblock


\bibitem[\protect\citeauthoryear{Lee, Kim, Moon, and Ko}{Lee
  et~al\mbox{.}}{2021}]%
        {lee2021splittable}
\bibfield{author}{\bibinfo{person}{Joo~Chan Lee}, \bibinfo{person}{Yongwoo
  Kim}, \bibinfo{person}{SungTae Moon}, {and} \bibinfo{person}{Jong~Hwan Ko}.}
  \bibinfo{year}{2021}\natexlab{}.
\newblock \showarticletitle{{A Splittable DNN-Based Object Detector for
  Edge-Cloud Collaborative Real-Time Video Inference}}. In
  \bibinfo{booktitle}{\emph{2021 17th IEEE International Conference on Advanced
  Video and Signal Based Surveillance (AVSS)}}. IEEE, \bibinfo{pages}{1--8}.
\newblock


\bibitem[\protect\citeauthoryear{Levesque, Davis, and Morgenstern}{Levesque
  et~al\mbox{.}}{2012}]%
        {levesque2012winograd}
\bibfield{author}{\bibinfo{person}{Hector~J Levesque}, \bibinfo{person}{Ernest
  Davis}, {and} \bibinfo{person}{Leora Morgenstern}.}
  \bibinfo{year}{2012}\natexlab{}.
\newblock \showarticletitle{The Winograd schema challenge}. In
  \bibinfo{booktitle}{\emph{Proceedings of the Thirteenth International
  Conference on Principles of Knowledge Representation and Reasoning}}.
  \bibinfo{pages}{552--561}.
\newblock


\bibitem[\protect\citeauthoryear{Levi and Hassner}{Levi and Hassner}{2015}]%
        {levi2015age}
\bibfield{author}{\bibinfo{person}{Gil Levi} {and} \bibinfo{person}{Tal
  Hassner}.} \bibinfo{year}{2015}\natexlab{}.
\newblock \showarticletitle{Age and gender classification using convolutional
  neural networks}. In \bibinfo{booktitle}{\emph{Proceedings of the IEEE
  conference on computer vision and pattern recognition workshops}}.
  \bibinfo{pages}{34--42}.
\newblock


\bibitem[\protect\citeauthoryear{Li, Liu, Wang, Dong, Zhao, and Feng}{Li
  et~al\mbox{.}}{2018a}]%
        {li2018auto-tuning}
\bibfield{author}{\bibinfo{person}{Guangli Li}, \bibinfo{person}{Lei Liu},
  \bibinfo{person}{Xueying Wang}, \bibinfo{person}{Xiao Dong},
  \bibinfo{person}{Peng Zhao}, {and} \bibinfo{person}{Xiaobing Feng}.}
  \bibinfo{year}{2018}\natexlab{a}.
\newblock \showarticletitle{{Auto-tuning Neural Network Quantization Framework
  for Collaborative Inference Between the Cloud and Edge}}. In
  \bibinfo{booktitle}{\emph{Int. Conference on Artificial Neural Networks}}.
  \bibinfo{pages}{402--411}.
\newblock


\bibitem[\protect\citeauthoryear{Li, Kadav, Durdanovic, Samet, and Graf}{Li
  et~al\mbox{.}}{2016}]%
        {li2016pruning}
\bibfield{author}{\bibinfo{person}{Hao Li}, \bibinfo{person}{Asim Kadav},
  \bibinfo{person}{Igor Durdanovic}, \bibinfo{person}{Hanan Samet}, {and}
  \bibinfo{person}{Hans~Peter Graf}.} \bibinfo{year}{2016}\natexlab{}.
\newblock \showarticletitle{{Pruning Filters for Efficient ConvNets}}. In
  \bibinfo{booktitle}{\emph{Fourth International Conference on Learning
  Representations}}.
\newblock


\bibitem[\protect\citeauthoryear{Li, Kadav, Durdanovic, Samet, and Graf}{Li
  et~al\mbox{.}}{2017}]%
        {li2017pruning}
\bibfield{author}{\bibinfo{person}{Hao Li}, \bibinfo{person}{Asim Kadav},
  \bibinfo{person}{Igor Durdanovic}, \bibinfo{person}{Hanan Samet}, {and}
  \bibinfo{person}{Hans~Peter Graf}.} \bibinfo{year}{2017}\natexlab{}.
\newblock \showarticletitle{{Pruning Filters for Efficient ConvNets}}. In
  \bibinfo{booktitle}{\emph{Fifth International Conference on Learning
  Representations}}.
\newblock


\bibitem[\protect\citeauthoryear{Li, Ota, and Dong}{Li et~al\mbox{.}}{2018b}]%
        {li2018learning}
\bibfield{author}{\bibinfo{person}{He Li}, \bibinfo{person}{Kaoru Ota}, {and}
  \bibinfo{person}{Mianxiong Dong}.} \bibinfo{year}{2018}\natexlab{b}.
\newblock \showarticletitle{{Learning IoT in edge: Deep learning for the
  Internet of Things with edge computing}}.
\newblock \bibinfo{journal}{\emph{IEEE network}} \bibinfo{volume}{32},
  \bibinfo{number}{1} (\bibinfo{year}{2018}), \bibinfo{pages}{96--101}.
\newblock


\bibitem[\protect\citeauthoryear{Li, Zhang, Qi, Yang, and Huang}{Li
  et~al\mbox{.}}{2019}]%
        {li2019improved}
\bibfield{author}{\bibinfo{person}{Hao Li}, \bibinfo{person}{Hong Zhang},
  \bibinfo{person}{Xiaojuan Qi}, \bibinfo{person}{Ruigang Yang}, {and}
  \bibinfo{person}{Gao Huang}.} \bibinfo{year}{2019}\natexlab{}.
\newblock \showarticletitle{{Improved Techniques for Training Adaptive Deep
  Networks}}. In \bibinfo{booktitle}{\emph{2019 IEEE/CVF International
  Conference on Computer Vision (ICCV)}}. \bibinfo{pages}{1891--1900}.
\newblock


\bibitem[\protect\citeauthoryear{Li, Zhao, Huang, and Gong}{Li
  et~al\mbox{.}}{2014}]%
        {li2014learning}
\bibfield{author}{\bibinfo{person}{Jinyu Li}, \bibinfo{person}{Rui Zhao},
  \bibinfo{person}{Jui-Ting Huang}, {and} \bibinfo{person}{Yifan Gong}.}
  \bibinfo{year}{2014}\natexlab{}.
\newblock \showarticletitle{{Learning Small-Size DNN with
  Output-Distribution-Based Criteria}}. In \bibinfo{booktitle}{\emph{Fifteenth
  annual conference of the international speech communication association}}.
\newblock


\bibitem[\protect\citeauthoryear{Li, Wallace, Shen, Lin, Keutzer, Klein, and
  Gonzalez}{Li et~al\mbox{.}}{2020}]%
        {li2020train}
\bibfield{author}{\bibinfo{person}{Zhuohan Li}, \bibinfo{person}{Eric Wallace},
  \bibinfo{person}{Sheng Shen}, \bibinfo{person}{Kevin Lin},
  \bibinfo{person}{Kurt Keutzer}, \bibinfo{person}{Dan Klein}, {and}
  \bibinfo{person}{Joey Gonzalez}.} \bibinfo{year}{2020}\natexlab{}.
\newblock \showarticletitle{Train big, then compress: Rethinking model size for
  efficient training and inference of transformers}. In
  \bibinfo{booktitle}{\emph{International Conference on Machine Learning}}.
  PMLR, \bibinfo{pages}{5958--5968}.
\newblock


\bibitem[\protect\citeauthoryear{Lin, Chen, and Yan}{Lin
  et~al\mbox{.}}{2014a}]%
        {lin2014network}
\bibfield{author}{\bibinfo{person}{Min Lin}, \bibinfo{person}{Qiang Chen},
  {and} \bibinfo{person}{Shuicheng Yan}.} \bibinfo{year}{2014}\natexlab{a}.
\newblock \showarticletitle{Network in network}.
\newblock  (\bibinfo{year}{2014}).
\newblock


\bibitem[\protect\citeauthoryear{Lin, Doll{\'a}r, Girshick, He, Hariharan, and
  Belongie}{Lin et~al\mbox{.}}{2017a}]%
        {lin2017feature}
\bibfield{author}{\bibinfo{person}{Tsung-Yi Lin}, \bibinfo{person}{Piotr
  Doll{\'a}r}, \bibinfo{person}{Ross Girshick}, \bibinfo{person}{Kaiming He},
  \bibinfo{person}{Bharath Hariharan}, {and} \bibinfo{person}{Serge Belongie}.}
  \bibinfo{year}{2017}\natexlab{a}.
\newblock \showarticletitle{{Feature pyramid networks for object detection}}.
  In \bibinfo{booktitle}{\emph{Proceedings of the IEEE Conference on Computer
  Vision and Pattern Recognition}}. \bibinfo{pages}{2117--2125}.
\newblock


\bibitem[\protect\citeauthoryear{Lin, Goyal, Girshick, He, and Doll{\'a}r}{Lin
  et~al\mbox{.}}{2017b}]%
        {lin2017focal}
\bibfield{author}{\bibinfo{person}{Tsung-Yi Lin}, \bibinfo{person}{Priya
  Goyal}, \bibinfo{person}{Ross Girshick}, \bibinfo{person}{Kaiming He}, {and}
  \bibinfo{person}{Piotr Doll{\'a}r}.} \bibinfo{year}{2017}\natexlab{b}.
\newblock \showarticletitle{{Focal Loss for Dense Object Detection}}. In
  \bibinfo{booktitle}{\emph{Proceedings of the IEEE international conference on
  computer vision}}. \bibinfo{pages}{2980--2988}.
\newblock


\bibitem[\protect\citeauthoryear{Lin, Maire, Belongie, Hays, Perona, Ramanan,
  Doll{\'a}r, and Zitnick}{Lin et~al\mbox{.}}{2014b}]%
        {lin2014microsoft}
\bibfield{author}{\bibinfo{person}{Tsung-Yi Lin}, \bibinfo{person}{Michael
  Maire}, \bibinfo{person}{Serge Belongie}, \bibinfo{person}{James Hays},
  \bibinfo{person}{Pietro Perona}, \bibinfo{person}{Deva Ramanan},
  \bibinfo{person}{Piotr Doll{\'a}r}, {and} \bibinfo{person}{C~Lawrence
  Zitnick}.} \bibinfo{year}{2014}\natexlab{b}.
\newblock \showarticletitle{Microsoft coco: Common objects in context}. In
  \bibinfo{booktitle}{\emph{European conference on computer vision}}. Springer,
  \bibinfo{pages}{740--755}.
\newblock


\bibitem[\protect\citeauthoryear{Liu, Zhou, Wang, Zhao, Deng, and JU}{Liu
  et~al\mbox{.}}{2020}]%
        {liu2020fastbert}
\bibfield{author}{\bibinfo{person}{Weijie Liu}, \bibinfo{person}{Peng Zhou},
  \bibinfo{person}{Zhiruo Wang}, \bibinfo{person}{Zhe Zhao},
  \bibinfo{person}{Haotang Deng}, {and} \bibinfo{person}{QI JU}.}
  \bibinfo{year}{2020}\natexlab{}.
\newblock \showarticletitle{FastBERT: a Self-distilling BERT with Adaptive
  Inference Time}. In \bibinfo{booktitle}{\emph{Proceedings of the 58th Annual
  Meeting of the Association for Computational Linguistics}}.
  \bibinfo{pages}{6035--6044}.
\newblock


\bibitem[\protect\citeauthoryear{Liu, Ott, Goyal, Du, Joshi, Chen, Levy, Lewis,
  Zettlemoyer, and Stoyanov}{Liu et~al\mbox{.}}{2019}]%
        {liu2019roberta}
\bibfield{author}{\bibinfo{person}{Yinhan Liu}, \bibinfo{person}{Myle Ott},
  \bibinfo{person}{Naman Goyal}, \bibinfo{person}{Jingfei Du},
  \bibinfo{person}{Mandar Joshi}, \bibinfo{person}{Danqi Chen},
  \bibinfo{person}{Omer Levy}, \bibinfo{person}{Mike Lewis},
  \bibinfo{person}{Luke Zettlemoyer}, {and} \bibinfo{person}{Veselin
  Stoyanov}.} \bibinfo{year}{2019}\natexlab{}.
\newblock \showarticletitle{{RoBERTa}: A robustly optimized bert pretraining
  approach}.
\newblock \bibinfo{journal}{\emph{arXiv preprint arXiv:1907.11692}}
  (\bibinfo{year}{2019}).
\newblock


\bibitem[\protect\citeauthoryear{Liu, Li, Li, and Cheng}{Liu
  et~al\mbox{.}}{2021}]%
        {liu2021ebert}
\bibfield{author}{\bibinfo{person}{Zejian Liu}, \bibinfo{person}{Fanrong Li},
  \bibinfo{person}{Gang Li}, {and} \bibinfo{person}{Jian Cheng}.}
  \bibinfo{year}{2021}\natexlab{}.
\newblock \showarticletitle{{EBERT: Efficient BERT Inference with Dynamic
  Structured Pruning}}. In \bibinfo{booktitle}{\emph{Findings of the
  Association for Computational Linguistics: ACL-IJCNLP 2021}}.
  \bibinfo{pages}{4814--4823}.
\newblock


\bibitem[\protect\citeauthoryear{Lo, Su, Lee, and Chang}{Lo
  et~al\mbox{.}}{2017}]%
        {lo2017adynamic}
\bibfield{author}{\bibinfo{person}{Chi Lo}, \bibinfo{person}{Yu-Yi Su},
  \bibinfo{person}{Chun-Yi Lee}, {and} \bibinfo{person}{Shih-Chieh Chang}.}
  \bibinfo{year}{2017}\natexlab{}.
\newblock \showarticletitle{{A Dynamic Deep Neural Network Design for Efficient
  Workload Allocation in Edge Computing}}. In \bibinfo{booktitle}{\emph{2017
  IEEE International Conference on Computer Design (ICCD)}}.
  \bibinfo{pages}{273--280}.
\newblock


\bibitem[\protect\citeauthoryear{Mao, You, Zhang, Huang, and Letaief}{Mao
  et~al\mbox{.}}{2017}]%
        {mao2017survey}
\bibfield{author}{\bibinfo{person}{Yuyi Mao}, \bibinfo{person}{Changsheng You},
  \bibinfo{person}{Jun Zhang}, \bibinfo{person}{Kaibin Huang}, {and}
  \bibinfo{person}{Khaled~B Letaief}.} \bibinfo{year}{2017}\natexlab{}.
\newblock \showarticletitle{A Survey on Mobile Edge Computing: The
  Communication Perspective}.
\newblock \bibinfo{journal}{\emph{IEEE Communications Surveys \& Tutorials}}
  \bibinfo{volume}{19}, \bibinfo{number}{4} (\bibinfo{year}{2017}),
  \bibinfo{pages}{2322--2358}.
\newblock


\bibitem[\protect\citeauthoryear{Mateo, Fiandrino, and Widmer}{Mateo
  et~al\mbox{.}}{2019}]%
        {mateo2019analysis}
\bibfield{author}{\bibinfo{person}{Pablo~Jim{\'e}nez Mateo},
  \bibinfo{person}{Claudio Fiandrino}, {and} \bibinfo{person}{Joerg Widmer}.}
  \bibinfo{year}{2019}\natexlab{}.
\newblock \showarticletitle{{Analysis of TCP performance in 5G mm-wave mobile
  networks}}. In \bibinfo{booktitle}{\emph{2019 IEEE International Conference
  on Communications (IEEE ICC)}}. IEEE, \bibinfo{pages}{1--7}.
\newblock


\bibitem[\protect\citeauthoryear{Matsubara, Baidya, Callegaro, Levorato, and
  Singh}{Matsubara et~al\mbox{.}}{2019}]%
        {matsubara2019distilled}
\bibfield{author}{\bibinfo{person}{Yoshitomo Matsubara}, \bibinfo{person}{Sabur
  Baidya}, \bibinfo{person}{Davide Callegaro}, \bibinfo{person}{Marco
  Levorato}, {and} \bibinfo{person}{Sameer Singh}.}
  \bibinfo{year}{2019}\natexlab{}.
\newblock \showarticletitle{{Distilled Split Deep Neural Networks for
  Edge-Assisted Real-Time Systems}}. In \bibinfo{booktitle}{\emph{Proc. of the
  2019 MobiCom Workshop on Hot Topics in Video Analytics and Intelligent
  Edges}}. \bibinfo{pages}{21--26}.
\newblock


\bibitem[\protect\citeauthoryear{Matsubara, Callegaro, Baidya, Levorato, and
  Singh}{Matsubara et~al\mbox{.}}{2020a}]%
        {matsubara2020head}
\bibfield{author}{\bibinfo{person}{Yoshitomo Matsubara},
  \bibinfo{person}{Davide Callegaro}, \bibinfo{person}{Sabur Baidya},
  \bibinfo{person}{Marco Levorato}, {and} \bibinfo{person}{Sameer Singh}.}
  \bibinfo{year}{2020}\natexlab{a}.
\newblock \showarticletitle{Head Network Distillation: Splitting Distilled Deep
  Neural Networks for Resource-Constrained Edge Computing Systems}.
\newblock \bibinfo{journal}{\emph{IEEE Access}}  \bibinfo{volume}{8}
  (\bibinfo{year}{2020}), \bibinfo{pages}{212177--212193}.
\newblock
\urldef\tempurl%
\url{https://doi.org/10.1109/ACCESS.2020.3039714}
\showDOI{\tempurl}


\bibitem[\protect\citeauthoryear{Matsubara, Callegaro, Singh, Levorato, and
  Restuccia}{Matsubara et~al\mbox{.}}{2022a}]%
        {matsubara2022bottlefit}
\bibfield{author}{\bibinfo{person}{Yoshitomo Matsubara},
  \bibinfo{person}{Davide Callegaro}, \bibinfo{person}{Sameer Singh},
  \bibinfo{person}{Marco Levorato}, {and} \bibinfo{person}{Francesco
  Restuccia}.} \bibinfo{year}{2022}\natexlab{a}.
\newblock \showarticletitle{{BottleFit: Learning Compressed Representations in
  Deep Neural Networks for Effective and Efficient Split Computing}}.
\newblock \bibinfo{journal}{\emph{arXiv preprint arXiv:2201.02693}}
  (\bibinfo{year}{2022}).
\newblock


\bibitem[\protect\citeauthoryear{Matsubara and Levorato}{Matsubara and
  Levorato}{2020}]%
        {matsubara2020split}
\bibfield{author}{\bibinfo{person}{Yoshitomo Matsubara} {and}
  \bibinfo{person}{Marco Levorato}.} \bibinfo{year}{2020}\natexlab{}.
\newblock \showarticletitle{{Split Computing for Complex Object Detectors:
  Challenges and Preliminary Results}}. In
  \bibinfo{booktitle}{\emph{Proceedings of the 4th International Workshop on
  Embedded and Mobile Deep Learning}}. \bibinfo{pages}{7--12}.
\newblock


\bibitem[\protect\citeauthoryear{Matsubara and Levorato}{Matsubara and
  Levorato}{2021}]%
        {matsubara2020neural}
\bibfield{author}{\bibinfo{person}{Yoshitomo Matsubara} {and}
  \bibinfo{person}{Marco Levorato}.} \bibinfo{year}{2021}\natexlab{}.
\newblock \showarticletitle{{Neural Compression and Filtering for Edge-assisted
  Real-time Object Detection in Challenged Networks}}. In
  \bibinfo{booktitle}{\emph{2020 25th International Conference on Pattern
  Recognition (ICPR)}}. \bibinfo{pages}{2272--2279}.
\newblock


\bibitem[\protect\citeauthoryear{Matsubara, Vu, and Moschitti}{Matsubara
  et~al\mbox{.}}{2020b}]%
        {matsubara2020reranking}
\bibfield{author}{\bibinfo{person}{Yoshitomo Matsubara}, \bibinfo{person}{Thuy
  Vu}, {and} \bibinfo{person}{Alessandro Moschitti}.}
  \bibinfo{year}{2020}\natexlab{b}.
\newblock \showarticletitle{Reranking for efficient transformer-based answer
  selection}. In \bibinfo{booktitle}{\emph{Proceedings of the 43rd
  International ACM SIGIR Conference on Research and Development in Information
  Retrieval}}. \bibinfo{pages}{1577--1580}.
\newblock


\bibitem[\protect\citeauthoryear{Matsubara, Yang, Levorato, and
  Mandt}{Matsubara et~al\mbox{.}}{2022b}]%
        {matsubara2022sc2}
\bibfield{author}{\bibinfo{person}{Yoshitomo Matsubara},
  \bibinfo{person}{Ruihan Yang}, \bibinfo{person}{Marco Levorato}, {and}
  \bibinfo{person}{Stephan Mandt}.} \bibinfo{year}{2022}\natexlab{b}.
\newblock \showarticletitle{{SC2: Supervised Compression for Split Computing}}.
\newblock \bibinfo{journal}{\emph{arXiv preprint arXiv:2203.08875}}
  (\bibinfo{year}{2022}).
\newblock


\bibitem[\protect\citeauthoryear{Matsubara, Yang, Levorato, and
  Mandt}{Matsubara et~al\mbox{.}}{2022c}]%
        {matsubara2022supervised}
\bibfield{author}{\bibinfo{person}{Yoshitomo Matsubara},
  \bibinfo{person}{Ruihan Yang}, \bibinfo{person}{Marco Levorato}, {and}
  \bibinfo{person}{Stephan Mandt}.} \bibinfo{year}{2022}\natexlab{c}.
\newblock \showarticletitle{{Supervised Compression for Resource-Constrained
  Edge Computing Systems}}. In \bibinfo{booktitle}{\emph{Proceedings of the
  IEEE/CVF Winter Conference on Applications of Computer Vision}}.
  \bibinfo{pages}{2685--2695}.
\newblock


\bibitem[\protect\citeauthoryear{Mirzadeh, Farajtabar, Li, Levine, Matsukawa,
  and Ghasemzadeh}{Mirzadeh et~al\mbox{.}}{2020}]%
        {mirzadeh2020improved}
\bibfield{author}{\bibinfo{person}{Seyed~Iman Mirzadeh},
  \bibinfo{person}{Mehrdad Farajtabar}, \bibinfo{person}{Ang Li},
  \bibinfo{person}{Nir Levine}, \bibinfo{person}{Akihiro Matsukawa}, {and}
  \bibinfo{person}{Hassan Ghasemzadeh}.} \bibinfo{year}{2020}\natexlab{}.
\newblock \showarticletitle{{Improved Knowledge Distillation via Teacher
  Assistant}}. In \bibinfo{booktitle}{\emph{Proceedings of the AAAI Conference
  on Artificial Intelligence}}, Vol.~\bibinfo{volume}{34}.
  \bibinfo{pages}{5191--5198}.
\newblock


\bibitem[\protect\citeauthoryear{Mnih, Kavukcuoglu, Silver, Graves, Antonoglou,
  Wierstra, and Riedmiller}{Mnih et~al\mbox{.}}{2013}]%
        {mnih2013playing}
\bibfield{author}{\bibinfo{person}{Volodymyr Mnih}, \bibinfo{person}{Koray
  Kavukcuoglu}, \bibinfo{person}{David Silver}, \bibinfo{person}{Alex Graves},
  \bibinfo{person}{Ioannis Antonoglou}, \bibinfo{person}{Daan Wierstra}, {and}
  \bibinfo{person}{Martin Riedmiller}.} \bibinfo{year}{2013}\natexlab{}.
\newblock \showarticletitle{{Playing Atari with Deep Reinforcement Learning}}.
\newblock \bibinfo{journal}{\emph{arXiv preprint arXiv:1312.5602}}
  (\bibinfo{year}{2013}).
\newblock


\bibitem[\protect\citeauthoryear{Nair and Hinton}{Nair and Hinton}{2010}]%
        {nair2010rectified}
\bibfield{author}{\bibinfo{person}{Vinod Nair} {and}
  \bibinfo{person}{Geoffrey~E Hinton}.} \bibinfo{year}{2010}\natexlab{}.
\newblock \showarticletitle{Rectified linear units improve restricted boltzmann
  machines}. In \bibinfo{booktitle}{\emph{Proceedings of the 27th International
  Conference on International Conference on Machine Learning}}.
  \bibinfo{pages}{807--814}.
\newblock


\bibitem[\protect\citeauthoryear{Nakahara, Hisano, Nishimura, Ushiku, Maruta,
  and Nakayama}{Nakahara et~al\mbox{.}}{2021}]%
        {nakahara2021retransmission}
\bibfield{author}{\bibinfo{person}{Mutsuki Nakahara}, \bibinfo{person}{Daisuke
  Hisano}, \bibinfo{person}{Mai Nishimura}, \bibinfo{person}{Yoshitaka Ushiku},
  \bibinfo{person}{Kazuki Maruta}, {and} \bibinfo{person}{Yu Nakayama}.}
  \bibinfo{year}{2021}\natexlab{}.
\newblock \showarticletitle{Retransmission Edge Computing System Conducting
  Adaptive Image Compression Based on Image Recognition Accuracy}. In
  \bibinfo{booktitle}{\emph{2021 IEEE 94rd Vehicular Technology Conference
  (VTC2021-Fall)}}. IEEE, \bibinfo{pages}{1--5}.
\newblock


\bibitem[\protect\citeauthoryear{Neshatpour, Behnia, Homayoun, and
  Sasan}{Neshatpour et~al\mbox{.}}{2019}]%
        {neshatpour2019exploiting}
\bibfield{author}{\bibinfo{person}{Katayoun Neshatpour},
  \bibinfo{person}{Farnaz Behnia}, \bibinfo{person}{Houman Homayoun}, {and}
  \bibinfo{person}{Avesta Sasan}.} \bibinfo{year}{2019}\natexlab{}.
\newblock \showarticletitle{{Exploiting Energy-Accuracy Trade-off through
  Contextual Awareness in Multi-Stage Convolutional Neural Networks}}. In
  \bibinfo{booktitle}{\emph{20th International Symposium on Quality Electronic
  Design (ISQED)}}. \bibinfo{pages}{265--270}.
\newblock


\bibitem[\protect\citeauthoryear{Netzer, Wang, Coates, Bissacco, Wu, and
  Ng}{Netzer et~al\mbox{.}}{[n.d.]}]%
        {netzer2011reading}
\bibfield{author}{\bibinfo{person}{Yuval Netzer}, \bibinfo{person}{Tao Wang},
  \bibinfo{person}{Adam Coates}, \bibinfo{person}{Alessandro Bissacco},
  \bibinfo{person}{Bo Wu}, {and} \bibinfo{person}{Andrew~Y Ng}.}
  \bibinfo{year}{[n.d.]}\natexlab{}.
\newblock \showarticletitle{Reading digits in natural images with unsupervised
  feature learning}. In \bibinfo{booktitle}{\emph{NIPS Workshop on Deep
  Learning and Unsupervised Feature Learning 2011}}.
\newblock


\bibitem[\protect\citeauthoryear{Nguyen, Rosenberg, Song, Gao, Tiwary,
  Majumder, and Deng}{Nguyen et~al\mbox{.}}{2016}]%
        {nguyen2016ms}
\bibfield{author}{\bibinfo{person}{Tri Nguyen}, \bibinfo{person}{Mir
  Rosenberg}, \bibinfo{person}{Xia Song}, \bibinfo{person}{Jianfeng Gao},
  \bibinfo{person}{Saurabh Tiwary}, \bibinfo{person}{Rangan Majumder}, {and}
  \bibinfo{person}{Li Deng}.} \bibinfo{year}{2016}\natexlab{}.
\newblock \showarticletitle{{MS MARCO}: A human generated machine reading
  comprehension dataset}. In \bibinfo{booktitle}{\emph{CoCo@ NIPS}}.
\newblock


\bibitem[\protect\citeauthoryear{Padhy, Verma, Ahmad, Choudhury, and Sa}{Padhy
  et~al\mbox{.}}{2018}]%
        {padhy2018deep}
\bibfield{author}{\bibinfo{person}{Ram~Prasad Padhy}, \bibinfo{person}{Sachin
  Verma}, \bibinfo{person}{Shahzad Ahmad}, \bibinfo{person}{Suman~Kumar
  Choudhury}, {and} \bibinfo{person}{Pankaj~Kumar Sa}.}
  \bibinfo{year}{2018}\natexlab{}.
\newblock \showarticletitle{{Deep Neural Network for Autonomous UAV Navigation
  in Indoor Corridor Environments}}.
\newblock \bibinfo{journal}{\emph{Procedia computer science}}
  \bibinfo{volume}{133} (\bibinfo{year}{2018}), \bibinfo{pages}{643--650}.
\newblock


\bibitem[\protect\citeauthoryear{Pagliari, Chiaro, Macii, and Poncino}{Pagliari
  et~al\mbox{.}}{2020}]%
        {pagliari2020crime}
\bibfield{author}{\bibinfo{person}{Daniele~Jahier Pagliari},
  \bibinfo{person}{Roberta Chiaro}, \bibinfo{person}{Enrico Macii}, {and}
  \bibinfo{person}{Massimo Poncino}.} \bibinfo{year}{2020}\natexlab{}.
\newblock \showarticletitle{{CRIME: Input-Dependent Collaborative Inference for
  Recurrent Neural Networks}}.
\newblock \bibinfo{journal}{\emph{IEEE Trans. Comput.}} (\bibinfo{year}{2020}).
\newblock


\bibitem[\protect\citeauthoryear{Panayotov, Chen, Povey, and
  Khudanpur}{Panayotov et~al\mbox{.}}{2015}]%
        {panayotov2015librispeech}
\bibfield{author}{\bibinfo{person}{Vassil Panayotov}, \bibinfo{person}{Guoguo
  Chen}, \bibinfo{person}{Daniel Povey}, {and} \bibinfo{person}{Sanjeev
  Khudanpur}.} \bibinfo{year}{2015}\natexlab{}.
\newblock \showarticletitle{{LibriSpeech}: An {ASR} corpus based on public
  domain audio books}. In \bibinfo{booktitle}{\emph{2015 IEEE International
  Conference on Acoustics, Speech and Signal Processing (ICASSP)}}. IEEE,
  \bibinfo{pages}{5206--5210}.
\newblock


\bibitem[\protect\citeauthoryear{Phuong and Lampert}{Phuong and
  Lampert}{2019}]%
        {phuong2019distillationbased}
\bibfield{author}{\bibinfo{person}{Mary Phuong} {and}
  \bibinfo{person}{Christoph~H. Lampert}.} \bibinfo{year}{2019}\natexlab{}.
\newblock \showarticletitle{{Distillation-Based Training for Multi-Exit
  Architectures}}. In \bibinfo{booktitle}{\emph{2019 IEEE/CVF International
  Conference on Computer Vision (ICCV)}}. \bibinfo{pages}{1355--1364}.
\newblock


\bibitem[\protect\citeauthoryear{Pomponi, Scardapane, and Uncini}{Pomponi
  et~al\mbox{.}}{2021}]%
        {pomponi2021probabilistic}
\bibfield{author}{\bibinfo{person}{Jary Pomponi}, \bibinfo{person}{Simone
  Scardapane}, {and} \bibinfo{person}{Aurelio Uncini}.}
  \bibinfo{year}{2021}\natexlab{}.
\newblock \showarticletitle{{A Probabilistic Re-Intepretation of Confidence
  Scores in Multi-Exit Models}}.
\newblock \bibinfo{journal}{\emph{Entropy}} \bibinfo{volume}{24},
  \bibinfo{number}{1} (\bibinfo{year}{2021}), \bibinfo{pages}{1}.
\newblock


\bibitem[\protect\citeauthoryear{Pouyanfar, Sadiq, Yan, Tian, Tao, Reyes, Shyu,
  Chen, and Iyengar}{Pouyanfar et~al\mbox{.}}{2018}]%
        {pouyanfar2018survey}
\bibfield{author}{\bibinfo{person}{Samira Pouyanfar}, \bibinfo{person}{Saad
  Sadiq}, \bibinfo{person}{Yilin Yan}, \bibinfo{person}{Haiman Tian},
  \bibinfo{person}{Yudong Tao}, \bibinfo{person}{Maria~Presa Reyes},
  \bibinfo{person}{Mei-Ling Shyu}, \bibinfo{person}{Shu-Ching Chen}, {and}
  \bibinfo{person}{SS Iyengar}.} \bibinfo{year}{2018}\natexlab{}.
\newblock \showarticletitle{{A Survey on Deep Learning: Algorithms, Techniques,
  and Applications}}.
\newblock \bibinfo{journal}{\emph{ACM Computing Surveys (CSUR)}}
  \bibinfo{volume}{51}, \bibinfo{number}{5} (\bibinfo{year}{2018}),
  \bibinfo{pages}{1--36}.
\newblock


\bibitem[\protect\citeauthoryear{Povey, Ghoshal, Boulianne, Burget, Glembek,
  Goel, Hannemann, Motlicek, Qian, Schwarz, et~al\mbox{.}}{Povey
  et~al\mbox{.}}{2011}]%
        {povey2011kaldi}
\bibfield{author}{\bibinfo{person}{Daniel Povey}, \bibinfo{person}{Arnab
  Ghoshal}, \bibinfo{person}{Gilles Boulianne}, \bibinfo{person}{Lukas Burget},
  \bibinfo{person}{Ondrej Glembek}, \bibinfo{person}{Nagendra Goel},
  \bibinfo{person}{Mirko Hannemann}, \bibinfo{person}{Petr Motlicek},
  \bibinfo{person}{Yanmin Qian}, \bibinfo{person}{Petr Schwarz},
  {et~al\mbox{.}}} \bibinfo{year}{2011}\natexlab{}.
\newblock \showarticletitle{The Kaldi speech recognition toolkit}. In
  \bibinfo{booktitle}{\emph{IEEE 2011 workshop on automatic speech recognition
  and understanding}}. IEEE Signal Processing Society.
\newblock


\bibitem[\protect\citeauthoryear{Qiu, Li, Li, Jiang, Hu, and Yang}{Qiu
  et~al\mbox{.}}{2018}]%
        {qiu2018reevisiting}
\bibfield{author}{\bibinfo{person}{Y. Qiu}, \bibinfo{person}{Hongzheng Li},
  \bibinfo{person}{Shen Li}, \bibinfo{person}{Yingdi Jiang},
  \bibinfo{person}{Renfen Hu}, {and} \bibinfo{person}{L. Yang}.}
  \bibinfo{year}{2018}\natexlab{}.
\newblock \showarticletitle{Revisiting Correlations between Intrinsic and
  Extrinsic Evaluations of Word Embeddings}. In
  \bibinfo{booktitle}{\emph{CCL}}.
\newblock


\bibitem[\protect\citeauthoryear{Radosavovic, Kosaraju, Girshick, He, and
  Doll{\'a}r}{Radosavovic et~al\mbox{.}}{2020}]%
        {radosavovic2020designing}
\bibfield{author}{\bibinfo{person}{Ilija Radosavovic},
  \bibinfo{person}{Raj~Prateek Kosaraju}, \bibinfo{person}{Ross Girshick},
  \bibinfo{person}{Kaiming He}, {and} \bibinfo{person}{Piotr Doll{\'a}r}.}
  \bibinfo{year}{2020}\natexlab{}.
\newblock \showarticletitle{{Designing Network Design Spaces}}. In
  \bibinfo{booktitle}{\emph{Proceedings of the IEEE/CVF Conference on Computer
  Vision and Pattern Recognition}}. \bibinfo{pages}{10428--10436}.
\newblock


\bibitem[\protect\citeauthoryear{Rajpurkar, Zhang, Lopyrev, and
  Liang}{Rajpurkar et~al\mbox{.}}{2016}]%
        {rajpurkar2016squad}
\bibfield{author}{\bibinfo{person}{Pranav Rajpurkar}, \bibinfo{person}{Jian
  Zhang}, \bibinfo{person}{Konstantin Lopyrev}, {and} \bibinfo{person}{Percy
  Liang}.} \bibinfo{year}{2016}\natexlab{}.
\newblock \showarticletitle{{SQuAD: 100,000+ Questions for Machine
  Comprehension of Text}}. In \bibinfo{booktitle}{\emph{Proceedings of the 2016
  Conference on Empirical Methods in Natural Language Processing}}.
  \bibinfo{pages}{2383--2392}.
\newblock


\bibitem[\protect\citeauthoryear{Redmon and Farhadi}{Redmon and
  Farhadi}{2017}]%
        {redmon2017yolo9000}
\bibfield{author}{\bibinfo{person}{Joseph Redmon} {and} \bibinfo{person}{Ali
  Farhadi}.} \bibinfo{year}{2017}\natexlab{}.
\newblock \showarticletitle{{YOLO9000}: better, faster, stronger}. In
  \bibinfo{booktitle}{\emph{Proceedings of the IEEE conference on computer
  vision and pattern recognition}}. \bibinfo{pages}{7263--7271}.
\newblock


\bibitem[\protect\citeauthoryear{Redmon and Farhadi}{Redmon and
  Farhadi}{2018}]%
        {redmon2018yolov3}
\bibfield{author}{\bibinfo{person}{Joseph Redmon} {and} \bibinfo{person}{Ali
  Farhadi}.} \bibinfo{year}{2018}\natexlab{}.
\newblock \showarticletitle{{YOLOv3: An incremental improvement}}.
\newblock \bibinfo{journal}{\emph{arXiv preprint arXiv:1804.02767}}
  (\bibinfo{year}{2018}).
\newblock


\bibitem[\protect\citeauthoryear{Ren, He, Girshick, and Sun}{Ren
  et~al\mbox{.}}{2015}]%
        {ren2015faster}
\bibfield{author}{\bibinfo{person}{Shaoqing Ren}, \bibinfo{person}{Kaiming He},
  \bibinfo{person}{Ross Girshick}, {and} \bibinfo{person}{Jian Sun}.}
  \bibinfo{year}{2015}\natexlab{}.
\newblock \showarticletitle{{Faster R-CNN: Towards Real-Time Object Detection
  with Region Proposal Networks}}. In \bibinfo{booktitle}{\emph{Advances in
  neural information processing systems}}. \bibinfo{pages}{91--99}.
\newblock


\bibitem[\protect\citeauthoryear{Restuccia and Melodia}{Restuccia and
  Melodia}{2020}]%
        {restuccia2020deep}
\bibfield{author}{\bibinfo{person}{Francesco Restuccia} {and}
  \bibinfo{person}{Tommaso Melodia}.} \bibinfo{year}{2020}\natexlab{}.
\newblock \showarticletitle{{Deep Learning at the Physical Layer: System
  Challenges and Applications to 5G and Beyond}}.
\newblock \bibinfo{journal}{\emph{IEEE Communications Magazine}}
  \bibinfo{volume}{58}, \bibinfo{number}{10} (\bibinfo{year}{2020}),
  \bibinfo{pages}{58--64}.
\newblock
\urldef\tempurl%
\url{https://doi.org/10.1109/MCOM.001.2000243}
\showDOI{\tempurl}


\bibitem[\protect\citeauthoryear{Roig, Boix, Shitrit, and Fua}{Roig
  et~al\mbox{.}}{2011}]%
        {roig2011conditional}
\bibfield{author}{\bibinfo{person}{Gemma Roig}, \bibinfo{person}{Xavier Boix},
  \bibinfo{person}{Horesh~Ben Shitrit}, {and} \bibinfo{person}{Pascal Fua}.}
  \bibinfo{year}{2011}\natexlab{}.
\newblock \showarticletitle{Conditional random fields for multi-camera object
  detection}. In \bibinfo{booktitle}{\emph{2011 International Conference on
  Computer Vision}}. IEEE, \bibinfo{pages}{563--570}.
\newblock


\bibitem[\protect\citeauthoryear{Russakovsky, Deng, Su, Krause, Satheesh, Ma,
  Huang, Karpathy, Khosla, Bernstein, Berg, and Fei-Fei}{Russakovsky
  et~al\mbox{.}}{2015}]%
        {russakovsky2015imagenet}
\bibfield{author}{\bibinfo{person}{Olga Russakovsky}, \bibinfo{person}{Jia
  Deng}, \bibinfo{person}{Hao Su}, \bibinfo{person}{Jonathan Krause},
  \bibinfo{person}{Sanjeev Satheesh}, \bibinfo{person}{Sean Ma},
  \bibinfo{person}{Zhiheng Huang}, \bibinfo{person}{Andrej Karpathy},
  \bibinfo{person}{Aditya Khosla}, \bibinfo{person}{Michael Bernstein},
  \bibinfo{person}{Alexander~C. Berg}, {and} \bibinfo{person}{Li Fei-Fei}.}
  \bibinfo{year}{2015}\natexlab{}.
\newblock \showarticletitle{{ImageNet Large Scale Visual Recognition
  Challenge}}.
\newblock \bibinfo{journal}{\emph{International Journal of Computer Vision}}
  \bibinfo{volume}{115}, \bibinfo{number}{3} (\bibinfo{year}{2015}),
  \bibinfo{pages}{211--252}.
\newblock


\bibitem[\protect\citeauthoryear{Samie, Bauer, and Henkel}{Samie
  et~al\mbox{.}}{2016}]%
        {samie2016iot}
\bibfield{author}{\bibinfo{person}{Farzad Samie}, \bibinfo{person}{Lars Bauer},
  {and} \bibinfo{person}{J{\"o}rg Henkel}.} \bibinfo{year}{2016}\natexlab{}.
\newblock \showarticletitle{{IoT Technologies for Embedded Computing: A
  Survey}}. In \bibinfo{booktitle}{\emph{2016 International Conference on
  Hardware/Software Codesign and System Synthesis (CODES+ ISSS)}}. IEEE,
  \bibinfo{pages}{1--10}.
\newblock


\bibitem[\protect\citeauthoryear{Sandler, Howard, Zhu, Zhmoginov, and
  Chen}{Sandler et~al\mbox{.}}{2018}]%
        {sandler2018mobilenetv2}
\bibfield{author}{\bibinfo{person}{Mark Sandler}, \bibinfo{person}{Andrew
  Howard}, \bibinfo{person}{Menglong Zhu}, \bibinfo{person}{Andrey Zhmoginov},
  {and} \bibinfo{person}{Liang-Chieh Chen}.} \bibinfo{year}{2018}\natexlab{}.
\newblock \showarticletitle{{MobileNetV2: Inverted Residuals and Linear
  Bottlenecks}}. In \bibinfo{booktitle}{\emph{Proceedings of the IEEE
  Conference on Computer Vision and Pattern Recognition}}.
  \bibinfo{pages}{4510--4520}.
\newblock


\bibitem[\protect\citeauthoryear{Saxe, Bansal, Dapello, Advani, Kolchinsky,
  Tracey, and Cox}{Saxe et~al\mbox{.}}{2019}]%
        {saxe2019information}
\bibfield{author}{\bibinfo{person}{Andrew~M Saxe}, \bibinfo{person}{Yamini
  Bansal}, \bibinfo{person}{Joel Dapello}, \bibinfo{person}{Madhu Advani},
  \bibinfo{person}{Artemy Kolchinsky}, \bibinfo{person}{Brendan~D Tracey},
  {and} \bibinfo{person}{David~D Cox}.} \bibinfo{year}{2019}\natexlab{}.
\newblock \showarticletitle{On the information bottleneck theory of deep
  learning}.
\newblock \bibinfo{journal}{\emph{Journal of Statistical Mechanics: Theory and
  Experiment}} \bibinfo{volume}{2019}, \bibinfo{number}{12}
  (\bibinfo{year}{2019}), \bibinfo{pages}{124020}.
\newblock


\bibitem[\protect\citeauthoryear{Sbai, Saputra, Trigoni, and Markham}{Sbai
  et~al\mbox{.}}{2021}]%
        {sbai2021cut}
\bibfield{author}{\bibinfo{person}{Marion Sbai}, \bibinfo{person}{Muhamad
  Risqi~U Saputra}, \bibinfo{person}{Niki Trigoni}, {and}
  \bibinfo{person}{Andrew Markham}.} \bibinfo{year}{2021}\natexlab{}.
\newblock \showarticletitle{{Cut, Distil and Encode (CDE): Split Cloud-Edge
  Deep Inference}}. In \bibinfo{booktitle}{\emph{2021 18th Annual IEEE
  International Conference on Sensing, Communication, and Networking (SECON)}}.
  IEEE, \bibinfo{pages}{1--9}.
\newblock


\bibitem[\protect\citeauthoryear{Sermanet, Eigen, Zhang, Mathieu, Fergus, and
  LeCun}{Sermanet et~al\mbox{.}}{2014}]%
        {sermanet2014overfeat}
\bibfield{author}{\bibinfo{person}{Pierre Sermanet}, \bibinfo{person}{David
  Eigen}, \bibinfo{person}{Xiang Zhang}, \bibinfo{person}{Micha{\"e}l Mathieu},
  \bibinfo{person}{Rob Fergus}, {and} \bibinfo{person}{Yann LeCun}.}
  \bibinfo{year}{2014}\natexlab{}.
\newblock \showarticletitle{{OverFeat}: Integrated recognition, localization
  and detection using convolutional networks}. In
  \bibinfo{booktitle}{\emph{Second International Conference on Learning
  Representations}}.
\newblock


\bibitem[\protect\citeauthoryear{Shao and Zhang}{Shao and Zhang}{2020}]%
        {shao2020bottlenet++}
\bibfield{author}{\bibinfo{person}{Jiawei Shao} {and} \bibinfo{person}{Jun
  Zhang}.} \bibinfo{year}{2020}\natexlab{}.
\newblock \showarticletitle{{BottleNet++: An End-to-End Approach for Feature
  Compression in Device-Edge Co-Inference Systems}}. In
  \bibinfo{booktitle}{\emph{2020 IEEE International Conference on
  Communications Workshops (ICC Workshops)}}. IEEE, \bibinfo{pages}{1--6}.
\newblock


\bibitem[\protect\citeauthoryear{Silver, Schrittwieser, Simonyan, Antonoglou,
  Huang, Guez, Hubert, Baker, Lai, Bolton, et~al\mbox{.}}{Silver
  et~al\mbox{.}}{2017}]%
        {silver2017mastering}
\bibfield{author}{\bibinfo{person}{David Silver}, \bibinfo{person}{Julian
  Schrittwieser}, \bibinfo{person}{Karen Simonyan}, \bibinfo{person}{Ioannis
  Antonoglou}, \bibinfo{person}{Aja Huang}, \bibinfo{person}{Arthur Guez},
  \bibinfo{person}{Thomas Hubert}, \bibinfo{person}{Lucas Baker},
  \bibinfo{person}{Matthew Lai}, \bibinfo{person}{Adrian Bolton},
  {et~al\mbox{.}}} \bibinfo{year}{2017}\natexlab{}.
\newblock \showarticletitle{{Mastering the Game of Go Without Human
  Knowledge}}.
\newblock \bibinfo{journal}{\emph{Nature}} \bibinfo{volume}{550},
  \bibinfo{number}{7676} (\bibinfo{year}{2017}), \bibinfo{pages}{354}.
\newblock


\bibitem[\protect\citeauthoryear{Simonyan and Zisserman}{Simonyan and
  Zisserman}{2015}]%
        {simonyan2014very}
\bibfield{author}{\bibinfo{person}{Karen Simonyan} {and}
  \bibinfo{person}{Andrew Zisserman}.} \bibinfo{year}{2015}\natexlab{}.
\newblock \showarticletitle{{Very deep convolutional networks for large-scale
  image recognition}}. In \bibinfo{booktitle}{\emph{Third International
  Conference on Learning Representations}}.
\newblock


\bibitem[\protect\citeauthoryear{Singh, Patil, and Omkar}{Singh
  et~al\mbox{.}}{2018}]%
        {singh2018eye}
\bibfield{author}{\bibinfo{person}{Amarjot Singh}, \bibinfo{person}{Devendra
  Patil}, {and} \bibinfo{person}{SN Omkar}.} \bibinfo{year}{2018}\natexlab{}.
\newblock \showarticletitle{Eye in the sky: Real-time Drone Surveillance System
  (DSS) for violent individuals identification using ScatterNet Hybrid Deep
  Learning network}. In \bibinfo{booktitle}{\emph{Proceedings of the IEEE
  Conference on Computer Vision and Pattern Recognition Workshops}}.
  \bibinfo{pages}{1629--1637}.
\newblock


\bibitem[\protect\citeauthoryear{Snell, Swersky, and Zemel}{Snell
  et~al\mbox{.}}{2017}]%
        {snell2017prototypical}
\bibfield{author}{\bibinfo{person}{Jake Snell}, \bibinfo{person}{Kevin
  Swersky}, {and} \bibinfo{person}{Richard Zemel}.}
  \bibinfo{year}{2017}\natexlab{}.
\newblock \showarticletitle{Prototypical networks for few-shot learning}. In
  \bibinfo{booktitle}{\emph{Advances in neural information processing
  systems}}. \bibinfo{pages}{4077--4087}.
\newblock


\bibitem[\protect\citeauthoryear{Socher, Perelygin, Wu, Chuang, Manning, Ng,
  and Potts}{Socher et~al\mbox{.}}{2013}]%
        {socher2013recursive}
\bibfield{author}{\bibinfo{person}{Richard Socher}, \bibinfo{person}{Alex
  Perelygin}, \bibinfo{person}{Jean Wu}, \bibinfo{person}{Jason Chuang},
  \bibinfo{person}{Christopher~D Manning}, \bibinfo{person}{Andrew~Y Ng}, {and}
  \bibinfo{person}{Christopher Potts}.} \bibinfo{year}{2013}\natexlab{}.
\newblock \showarticletitle{Recursive deep models for semantic compositionality
  over a sentiment treebank}. In \bibinfo{booktitle}{\emph{Proceedings of the
  2013 conference on empirical methods in natural language processing}}.
  \bibinfo{pages}{1631--1642}.
\newblock


\bibitem[\protect\citeauthoryear{Soldaini and Moschitti}{Soldaini and
  Moschitti}{2020}]%
        {soldaini2020cascade}
\bibfield{author}{\bibinfo{person}{Luca Soldaini} {and}
  \bibinfo{person}{Alessandro Moschitti}.} \bibinfo{year}{2020}\natexlab{}.
\newblock \showarticletitle{{The Cascade Transformer: an Application for
  Efficient Answer Sentence Selection}}. In
  \bibinfo{booktitle}{\emph{Proceedings of the 58th Annual Meeting of the
  Association for Computational Linguistics}}. \bibinfo{pages}{5697--5708}.
\newblock


\bibitem[\protect\citeauthoryear{Srivastava, Hinton, Krizhevsky, Sutskever, and
  Salakhutdinov}{Srivastava et~al\mbox{.}}{2014}]%
        {srivastava2014dropout}
\bibfield{author}{\bibinfo{person}{Nitish Srivastava},
  \bibinfo{person}{Geoffrey Hinton}, \bibinfo{person}{Alex Krizhevsky},
  \bibinfo{person}{Ilya Sutskever}, {and} \bibinfo{person}{Ruslan
  Salakhutdinov}.} \bibinfo{year}{2014}\natexlab{}.
\newblock \showarticletitle{{Dropout: A Simple Way to Prevent Neural Networks
  from Overfitting}}.
\newblock \bibinfo{journal}{\emph{The journal of machine learning research}}
  \bibinfo{volume}{15}, \bibinfo{number}{1} (\bibinfo{year}{2014}),
  \bibinfo{pages}{1929--1958}.
\newblock


\bibitem[\protect\citeauthoryear{Steiner, Kolesnikov, Zhai, Wightman,
  Uszkoreit, and Beyer}{Steiner et~al\mbox{.}}{2021}]%
        {steiner2021train}
\bibfield{author}{\bibinfo{person}{Andreas Steiner}, \bibinfo{person}{Alexander
  Kolesnikov}, \bibinfo{person}{Xiaohua Zhai}, \bibinfo{person}{Ross Wightman},
  \bibinfo{person}{Jakob Uszkoreit}, {and} \bibinfo{person}{Lucas Beyer}.}
  \bibinfo{year}{2021}\natexlab{}.
\newblock \showarticletitle{{How to train your ViT? Data, Augmentation, and
  Regularization in Vision Transformers}}.
\newblock \bibinfo{journal}{\emph{arXiv preprint arXiv:2106.10270}}
  (\bibinfo{year}{2021}).
\newblock


\bibitem[\protect\citeauthoryear{Szegedy, Liu, Jia, Sermanet, Reed, Anguelov,
  Erhan, Vanhoucke, and Rabinovich}{Szegedy et~al\mbox{.}}{2015}]%
        {szegedy2015going}
\bibfield{author}{\bibinfo{person}{Christian Szegedy}, \bibinfo{person}{Wei
  Liu}, \bibinfo{person}{Yangqing Jia}, \bibinfo{person}{Pierre Sermanet},
  \bibinfo{person}{Scott Reed}, \bibinfo{person}{Dragomir Anguelov},
  \bibinfo{person}{Dumitru Erhan}, \bibinfo{person}{Vincent Vanhoucke}, {and}
  \bibinfo{person}{Andrew Rabinovich}.} \bibinfo{year}{2015}\natexlab{}.
\newblock \showarticletitle{Going deeper with convolutions}. In
  \bibinfo{booktitle}{\emph{Proceedings of the IEEE conference on computer
  vision and pattern recognition}}. \bibinfo{pages}{1--9}.
\newblock


\bibitem[\protect\citeauthoryear{Szegedy, Vanhoucke, Ioffe, Shlens, and
  Wojna}{Szegedy et~al\mbox{.}}{2016}]%
        {szegedy2016rethinking}
\bibfield{author}{\bibinfo{person}{Christian Szegedy}, \bibinfo{person}{Vincent
  Vanhoucke}, \bibinfo{person}{Sergey Ioffe}, \bibinfo{person}{Jon Shlens},
  {and} \bibinfo{person}{Zbigniew Wojna}.} \bibinfo{year}{2016}\natexlab{}.
\newblock \showarticletitle{{Rethinking the inception architecture for computer
  vision}}. In \bibinfo{booktitle}{\emph{Proceedings of the IEEE conference on
  computer vision and pattern recognition}}. \bibinfo{pages}{2818--2826}.
\newblock


\bibitem[\protect\citeauthoryear{Taigman, Yang, Ranzato, and Wolf}{Taigman
  et~al\mbox{.}}{2014}]%
        {taigman2014deepface}
\bibfield{author}{\bibinfo{person}{Yaniv Taigman}, \bibinfo{person}{Ming Yang},
  \bibinfo{person}{Marc'Aurelio Ranzato}, {and} \bibinfo{person}{Lior Wolf}.}
  \bibinfo{year}{2014}\natexlab{}.
\newblock \showarticletitle{Deepface: Closing the gap to human-level
  performance in face verification}. In \bibinfo{booktitle}{\emph{Proceedings
  of the IEEE conference on computer vision and pattern recognition}}.
  \bibinfo{pages}{1701--1708}.
\newblock


\bibitem[\protect\citeauthoryear{Tan, Chen, Pang, Vasudevan, Sandler, Howard,
  and Le}{Tan et~al\mbox{.}}{2019}]%
        {tan2019mnasnet}
\bibfield{author}{\bibinfo{person}{Mingxing Tan}, \bibinfo{person}{Bo Chen},
  \bibinfo{person}{Ruoming Pang}, \bibinfo{person}{Vijay Vasudevan},
  \bibinfo{person}{Mark Sandler}, \bibinfo{person}{Andrew Howard}, {and}
  \bibinfo{person}{Quoc~V Le}.} \bibinfo{year}{2019}\natexlab{}.
\newblock \showarticletitle{{MnasNet: Platform-Aware Neural Architecture Search
  for Mobile}}. In \bibinfo{booktitle}{\emph{Proceedings of the IEEE Conf. on
  Computer Vision and Pattern Recognition}}. \bibinfo{pages}{2820--2828}.
\newblock


\bibitem[\protect\citeauthoryear{Tan, Pang, and Le}{Tan et~al\mbox{.}}{2020}]%
        {tan2020efficientdet}
\bibfield{author}{\bibinfo{person}{Mingxing Tan}, \bibinfo{person}{Ruoming
  Pang}, {and} \bibinfo{person}{Quoc~V Le}.} \bibinfo{year}{2020}\natexlab{}.
\newblock \showarticletitle{{EfficientDet: Scalable and Efficient Object
  Detection}}. In \bibinfo{booktitle}{\emph{Proceedings of the IEEE/CVF
  conference on computer vision and pattern recognition}}.
  \bibinfo{pages}{10781--10790}.
\newblock


\bibitem[\protect\citeauthoryear{Teerapittayanon, McDanel, and
  Kung}{Teerapittayanon et~al\mbox{.}}{2016}]%
        {teerapittayanon2016branchynet}
\bibfield{author}{\bibinfo{person}{Surat Teerapittayanon},
  \bibinfo{person}{Bradley McDanel}, {and} \bibinfo{person}{Hsiang-Tsung
  Kung}.} \bibinfo{year}{2016}\natexlab{}.
\newblock \showarticletitle{{BranchyNet: Fast Inference via Early Exiting from
  Deep Neural Networks}}. In \bibinfo{booktitle}{\emph{2016 23rd International
  Conference on Pattern Recognition (ICPR)}}. IEEE,
  \bibinfo{pages}{2464--2469}.
\newblock


\bibitem[\protect\citeauthoryear{Teerapittayanon, McDanel, and
  Kung}{Teerapittayanon et~al\mbox{.}}{2017}]%
        {teerapittayanon2017distributed}
\bibfield{author}{\bibinfo{person}{Surat Teerapittayanon},
  \bibinfo{person}{Bradley McDanel}, {and} \bibinfo{person}{H.~T. Kung}.}
  \bibinfo{year}{2017}\natexlab{}.
\newblock \showarticletitle{{Distributed Deep Neural Networks Over the Cloud,
  the Edge and End Devices}}. In \bibinfo{booktitle}{\emph{2017 IEEE 37th
  International Conference on Distributed Computing Systems (ICDCS)}}.
  \bibinfo{pages}{328--339}.
\newblock


\bibitem[\protect\citeauthoryear{Tishby, Pereira, and Bialek}{Tishby
  et~al\mbox{.}}{2000}]%
        {tishby2000information}
\bibfield{author}{\bibinfo{person}{Naftali Tishby}, \bibinfo{person}{Fernando~C
  Pereira}, {and} \bibinfo{person}{William Bialek}.}
  \bibinfo{year}{2000}\natexlab{}.
\newblock \showarticletitle{The information bottleneck method}.
\newblock \bibinfo{journal}{\emph{arXiv preprint physics/0004057}}
  (\bibinfo{year}{2000}).
\newblock


\bibitem[\protect\citeauthoryear{Tishby and Zaslavsky}{Tishby and
  Zaslavsky}{2015}]%
        {tishby2015deep}
\bibfield{author}{\bibinfo{person}{Naftali Tishby} {and} \bibinfo{person}{Noga
  Zaslavsky}.} \bibinfo{year}{2015}\natexlab{}.
\newblock \showarticletitle{Deep learning and the information bottleneck
  principle}. In \bibinfo{booktitle}{\emph{2015 IEEE Information Theory
  Workshop (ITW)}}. IEEE, \bibinfo{pages}{1--5}.
\newblock


\bibitem[\protect\citeauthoryear{Ultralytics}{Ultralytics}{[n.d.]}]%
        {ultralytics2021yolov5}
\bibfield{author}{\bibinfo{person}{Ultralytics}.}
  \bibinfo{year}{[n.d.]}\natexlab{}.
\newblock \bibinfo{title}{{YOLOv5}}.
\newblock \bibinfo{howpublished}{https://github.com/ultralytics/yolov5}.
\newblock


\bibitem[\protect\citeauthoryear{Vaswani, Shazeer, Parmar, Uszkoreit, Jones,
  Gomez, Kaiser, and Polosukhin}{Vaswani et~al\mbox{.}}{2017}]%
        {vaswani2017attention}
\bibfield{author}{\bibinfo{person}{Ashish Vaswani}, \bibinfo{person}{Noam
  Shazeer}, \bibinfo{person}{Niki Parmar}, \bibinfo{person}{Jakob Uszkoreit},
  \bibinfo{person}{Llion Jones}, \bibinfo{person}{Aidan~N Gomez},
  \bibinfo{person}{\L~ukasz Kaiser}, {and} \bibinfo{person}{Illia Polosukhin}.}
  \bibinfo{year}{2017}\natexlab{}.
\newblock \showarticletitle{Attention is All you Need}. In
  \bibinfo{booktitle}{\emph{Advances in Neural Information Processing
  Systems}}, \bibfield{editor}{\bibinfo{person}{I.~Guyon},
  \bibinfo{person}{U.~V. Luxburg}, \bibinfo{person}{S.~Bengio},
  \bibinfo{person}{H.~Wallach}, \bibinfo{person}{R.~Fergus},
  \bibinfo{person}{S.~Vishwanathan}, {and} \bibinfo{person}{R.~Garnett}}
  (Eds.), Vol.~\bibinfo{volume}{30}. \bibinfo{publisher}{Curran Associates,
  Inc.}, \bibinfo{pages}{5998--6008}.
\newblock


\bibitem[\protect\citeauthoryear{Wang, Singh, Michael, Hill, Levy, and
  Bowman}{Wang et~al\mbox{.}}{2019b}]%
        {wang2019glue}
\bibfield{author}{\bibinfo{person}{Alex Wang}, \bibinfo{person}{Amanpreet
  Singh}, \bibinfo{person}{Julian Michael}, \bibinfo{person}{Felix Hill},
  \bibinfo{person}{Omer Levy}, {and} \bibinfo{person}{Samuel~R Bowman}.}
  \bibinfo{year}{2019}\natexlab{b}.
\newblock \showarticletitle{{GLUE: A Multi-Task Benchmark and Analysis Platform
  for Natural Language Understanding}}. In
  \bibinfo{booktitle}{\emph{International Conference on Learning
  Representations}}.
\newblock


\bibitem[\protect\citeauthoryear{Wang, Diao, Sun, and Xu}{Wang
  et~al\mbox{.}}{2020a}]%
        {wang2020data}
\bibfield{author}{\bibinfo{person}{Fei Wang}, \bibinfo{person}{Boyu Diao},
  \bibinfo{person}{Tao Sun}, {and} \bibinfo{person}{Yongjun Xu}.}
  \bibinfo{year}{2020}\natexlab{a}.
\newblock \showarticletitle{Data security and privacy challenges of computing
  offloading in fins}.
\newblock \bibinfo{journal}{\emph{IEEE Network}} \bibinfo{volume}{34},
  \bibinfo{number}{2} (\bibinfo{year}{2020}), \bibinfo{pages}{14--20}.
\newblock


\bibitem[\protect\citeauthoryear{Wang, Mo, Lin, Wang, and Du}{Wang
  et~al\mbox{.}}{2019a}]%
        {wang2019dynexit}
\bibfield{author}{\bibinfo{person}{Meiqi Wang}, \bibinfo{person}{Jianqiao Mo},
  \bibinfo{person}{Jun Lin}, \bibinfo{person}{Zhongfeng Wang}, {and}
  \bibinfo{person}{Li Du}.} \bibinfo{year}{2019}\natexlab{a}.
\newblock \showarticletitle{{DynExit: A Dynamic Early-Exit Strategy for Deep
  Residual Networks}}. In \bibinfo{booktitle}{\emph{2019 IEEE International
  Workshop on Signal Processing Systems (SiPS)}}. IEEE,
  \bibinfo{pages}{178--183}.
\newblock


\bibitem[\protect\citeauthoryear{Wang, Smith, and Mitamura}{Wang
  et~al\mbox{.}}{2007}]%
        {wang2007jeopardy}
\bibfield{author}{\bibinfo{person}{Mengqiu Wang}, \bibinfo{person}{Noah~A
  Smith}, {and} \bibinfo{person}{Teruko Mitamura}.}
  \bibinfo{year}{2007}\natexlab{}.
\newblock \showarticletitle{What is the Jeopardy model? A quasi-synchronous
  grammar for QA}. In \bibinfo{booktitle}{\emph{Proceedings of the 2007 Joint
  Conference on Empirical Methods in Natural Language Processing and
  Computational Natural Language Learning (EMNLP-CoNLL)}}.
  \bibinfo{pages}{22--32}.
\newblock


\bibitem[\protect\citeauthoryear{Wang, Shen, Hu, Xu, Nguyen, Baraniuk, Wang,
  and Lin}{Wang et~al\mbox{.}}{2020b}]%
        {wang2020dual}
\bibfield{author}{\bibinfo{person}{Yue Wang}, \bibinfo{person}{Jianghao Shen},
  \bibinfo{person}{Ting-Kuei Hu}, \bibinfo{person}{P. Xu}, \bibinfo{person}{Tan
  Nguyen}, \bibinfo{person}{Richard Baraniuk}, \bibinfo{person}{Zhangyang
  Wang}, {and} \bibinfo{person}{Yingyan Lin}.}
  \bibinfo{year}{2020}\natexlab{b}.
\newblock \showarticletitle{{Dual Dynamic Inference: Enabling More Efficient,
  Adaptive, and Controllable Deep Inference}}.
\newblock \bibinfo{journal}{\emph{IEEE Journal of Selected Topics in Signal
  Processing}}  \bibinfo{volume}{14} (\bibinfo{year}{2020}),
  \bibinfo{pages}{623--633}.
\newblock


\bibitem[\protect\citeauthoryear{Warstadt, Singh, and Bowman}{Warstadt
  et~al\mbox{.}}{2019}]%
        {warstadt2019neural}
\bibfield{author}{\bibinfo{person}{Alex Warstadt}, \bibinfo{person}{Amanpreet
  Singh}, {and} \bibinfo{person}{Samuel Bowman}.}
  \bibinfo{year}{2019}\natexlab{}.
\newblock \showarticletitle{Neural Network Acceptability Judgments}.
\newblock \bibinfo{journal}{\emph{Transactions of the Association for
  Computational Linguistics}}  \bibinfo{volume}{7} (\bibinfo{year}{2019}),
  \bibinfo{pages}{625--641}.
\newblock


\bibitem[\protect\citeauthoryear{Williams, Nangia, and Bowman}{Williams
  et~al\mbox{.}}{2018}]%
        {williams2018broad}
\bibfield{author}{\bibinfo{person}{Adina Williams}, \bibinfo{person}{Nikita
  Nangia}, {and} \bibinfo{person}{Samuel Bowman}.}
  \bibinfo{year}{2018}\natexlab{}.
\newblock \showarticletitle{A Broad-Coverage Challenge Corpus for Sentence
  Understanding through Inference}. In \bibinfo{booktitle}{\emph{Proceedings of
  the 2018 Conference of the North American Chapter of the Association for
  Computational Linguistics: Human Language Technologies, Volume 1 (Long
  Papers)}}. \bibinfo{pages}{1112--1122}.
\newblock


\bibitem[\protect\citeauthoryear{Wo{\l}czyk, W{\'o}jcik, Ba{\l}azy, Podolak,
  Tabor, {\'S}mieja, and Trzcinski}{Wo{\l}czyk et~al\mbox{.}}{2021}]%
        {wolczyk2021zero}
\bibfield{author}{\bibinfo{person}{Maciej Wo{\l}czyk}, \bibinfo{person}{Bartosz
  W{\'o}jcik}, \bibinfo{person}{Klaudia Ba{\l}azy}, \bibinfo{person}{Igor
  Podolak}, \bibinfo{person}{Jacek Tabor}, \bibinfo{person}{Marek {\'S}mieja},
  {and} \bibinfo{person}{Tomasz Trzcinski}.} \bibinfo{year}{2021}\natexlab{}.
\newblock \showarticletitle{{Zero Time Waste: Recycling Predictions in Early
  Exit Neural Networks}}.
\newblock \bibinfo{journal}{\emph{Advances in Neural Information Processing
  Systems}}  \bibinfo{volume}{34} (\bibinfo{year}{2021}).
\newblock


\bibitem[\protect\citeauthoryear{Wolf, Chaumond, Debut, Sanh, Delangue, Moi,
  Cistac, Funtowicz, Davison, Shleifer, et~al\mbox{.}}{Wolf
  et~al\mbox{.}}{2020}]%
        {wolf2020transformers}
\bibfield{author}{\bibinfo{person}{Thomas Wolf}, \bibinfo{person}{Julien
  Chaumond}, \bibinfo{person}{Lysandre Debut}, \bibinfo{person}{Victor Sanh},
  \bibinfo{person}{Clement Delangue}, \bibinfo{person}{Anthony Moi},
  \bibinfo{person}{Pierric Cistac}, \bibinfo{person}{Morgan Funtowicz},
  \bibinfo{person}{Joe Davison}, \bibinfo{person}{Sam Shleifer},
  {et~al\mbox{.}}} \bibinfo{year}{2020}\natexlab{}.
\newblock \showarticletitle{Transformers: State-of-the-art natural language
  processing}. In \bibinfo{booktitle}{\emph{Proceedings of the 2020 Conference
  on Empirical Methods in Natural Language Processing: System Demonstrations}}.
  \bibinfo{pages}{38--45}.
\newblock


\bibitem[\protect\citeauthoryear{Xin, Nogueira, Yu, and Lin}{Xin
  et~al\mbox{.}}{2020a}]%
        {xin2020early}
\bibfield{author}{\bibinfo{person}{Ji Xin}, \bibinfo{person}{Rodrigo Nogueira},
  \bibinfo{person}{Yaoliang Yu}, {and} \bibinfo{person}{Jimmy Lin}.}
  \bibinfo{year}{2020}\natexlab{a}.
\newblock \showarticletitle{{Early Exiting BERT for Efficient Document
  Ranking}}. In \bibinfo{booktitle}{\emph{Proceedings of SustaiNLP: Workshop on
  Simple and Efficient Natural Language Processing}}. \bibinfo{pages}{83--88}.
\newblock


\bibitem[\protect\citeauthoryear{Xin, Tang, Lee, Yu, and Lin}{Xin
  et~al\mbox{.}}{2020b}]%
        {xin2020deebert}
\bibfield{author}{\bibinfo{person}{Ji Xin}, \bibinfo{person}{Raphael Tang},
  \bibinfo{person}{Jaejun Lee}, \bibinfo{person}{Yaoliang Yu}, {and}
  \bibinfo{person}{Jimmy Lin}.} \bibinfo{year}{2020}\natexlab{b}.
\newblock \showarticletitle{{DeeBERT: Dynamic Early Exiting for Accelerating
  BERT Inference}}. In \bibinfo{booktitle}{\emph{Proceedings of the 58th Annual
  Meeting of the Association for Computational Linguistics}}.
  \bibinfo{pages}{2246--2251}.
\newblock


\bibitem[\protect\citeauthoryear{Xing, Xu, Li, and Guan}{Xing
  et~al\mbox{.}}{2020}]%
        {xing2020early}
\bibfield{author}{\bibinfo{person}{Qunliang Xing}, \bibinfo{person}{Mai Xu},
  \bibinfo{person}{Tianyi Li}, {and} \bibinfo{person}{Zhenyu Guan}.}
  \bibinfo{year}{2020}\natexlab{}.
\newblock \showarticletitle{{Early Exit Or Not: Resource-Efficient Blind
  Quality Enhancement for Compressed Images}}. In
  \bibinfo{booktitle}{\emph{Computer Vision -- ECCV 2020}}.
  \bibinfo{publisher}{Springer International Publishing}.
\newblock


\bibitem[\protect\citeauthoryear{Yang, Han, Chen, Song, Dai, and Huang}{Yang
  et~al\mbox{.}}{2020b}]%
        {yang2020resolution}
\bibfield{author}{\bibinfo{person}{L. Yang}, \bibinfo{person}{Yizeng Han},
  \bibinfo{person}{X. Chen}, \bibinfo{person}{Shiji Song},
  \bibinfo{person}{Jifeng Dai}, {and} \bibinfo{person}{Gao Huang}.}
  \bibinfo{year}{2020}\natexlab{b}.
\newblock \showarticletitle{{Resolution Adaptive Networks for Efficient
  Inference}}. In \bibinfo{booktitle}{\emph{2020 IEEE/CVF Conference on
  Computer Vision and Pattern Recognition (CVPR)}}.
  \bibinfo{pages}{2366--2375}.
\newblock


\bibitem[\protect\citeauthoryear{Yang, Zhu, Chen, Yan, Zhang, and Willis}{Yang
  et~al\mbox{.}}{2020c}]%
        {yang2020mutualnet}
\bibfield{author}{\bibinfo{person}{Taojiannan Yang}, \bibinfo{person}{Sijie
  Zhu}, \bibinfo{person}{Chen Chen}, \bibinfo{person}{Shen Yan},
  \bibinfo{person}{Mi Zhang}, {and} \bibinfo{person}{Andrew Willis}.}
  \bibinfo{year}{2020}\natexlab{c}.
\newblock \showarticletitle{{MutualNet}: Adaptive convnet via mutual learning
  from network width and resolution}. In \bibinfo{booktitle}{\emph{European
  conference on computer vision}}. Springer, \bibinfo{pages}{299--315}.
\newblock


\bibitem[\protect\citeauthoryear{Yang, Chen, and Sze}{Yang
  et~al\mbox{.}}{2017}]%
        {yang2017designing}
\bibfield{author}{\bibinfo{person}{Tien-Ju Yang}, \bibinfo{person}{Yu-Hsin
  Chen}, {and} \bibinfo{person}{Vivienne Sze}.}
  \bibinfo{year}{2017}\natexlab{}.
\newblock \showarticletitle{Designing energy-efficient convolutional neural
  networks using energy-aware pruning}. In
  \bibinfo{booktitle}{\emph{Proceedings of the IEEE Conference on Computer
  Vision and Pattern Recognition}}. \bibinfo{pages}{5687--5695}.
\newblock


\bibitem[\protect\citeauthoryear{Yang, Howard, Chen, Zhang, Go, Sandler, Sze,
  and Adam}{Yang et~al\mbox{.}}{2018}]%
        {yang2018netadapt}
\bibfield{author}{\bibinfo{person}{Tien-Ju Yang}, \bibinfo{person}{Andrew
  Howard}, \bibinfo{person}{Bo Chen}, \bibinfo{person}{Xiao Zhang},
  \bibinfo{person}{Alec Go}, \bibinfo{person}{Mark Sandler},
  \bibinfo{person}{Vivienne Sze}, {and} \bibinfo{person}{Hartwig Adam}.}
  \bibinfo{year}{2018}\natexlab{}.
\newblock \showarticletitle{{NetAdapt: Platform-Aware Neural Network Adaptation
  for Mobile Applications}}. In \bibinfo{booktitle}{\emph{Proceedings of the
  European Conference on Computer Vision (ECCV)}}. \bibinfo{pages}{285--300}.
\newblock


\bibitem[\protect\citeauthoryear{Yang, Bamler, and Mandt}{Yang
  et~al\mbox{.}}{2020a}]%
        {yang2020variational}
\bibfield{author}{\bibinfo{person}{Yibo Yang}, \bibinfo{person}{Robert Bamler},
  {and} \bibinfo{person}{Stephan Mandt}.} \bibinfo{year}{2020}\natexlab{a}.
\newblock \showarticletitle{Variational bayesian quantization}. In
  \bibinfo{booktitle}{\emph{International Conference on Machine Learning}}.
  PMLR, \bibinfo{pages}{10670--10680}.
\newblock


\bibitem[\protect\citeauthoryear{Yang, Yih, and Meek}{Yang
  et~al\mbox{.}}{2015}]%
        {yang2015wikiqa}
\bibfield{author}{\bibinfo{person}{Yi Yang}, \bibinfo{person}{Wen-tau Yih},
  {and} \bibinfo{person}{Christopher Meek}.} \bibinfo{year}{2015}\natexlab{}.
\newblock \showarticletitle{{WikiQA}: A challenge dataset for open-domain
  question answering}. In \bibinfo{booktitle}{\emph{Proceedings of the 2015
  conference on empirical methods in natural language processing}}.
  \bibinfo{pages}{2013--2018}.
\newblock


\bibitem[\protect\citeauthoryear{Yao, Li, Liu, Wang, Liu, Shao, and
  Abdelzaher}{Yao et~al\mbox{.}}{2020}]%
        {yao2020deep}
\bibfield{author}{\bibinfo{person}{Shuochao Yao}, \bibinfo{person}{Jinyang Li},
  \bibinfo{person}{Dongxin Liu}, \bibinfo{person}{Tianshi Wang},
  \bibinfo{person}{Shengzhong Liu}, \bibinfo{person}{Huajie Shao}, {and}
  \bibinfo{person}{Tarek Abdelzaher}.} \bibinfo{year}{2020}\natexlab{}.
\newblock \showarticletitle{Deep compressive offloading: speeding up neural
  network inference by trading edge computation for network latency}. In
  \bibinfo{booktitle}{\emph{Proceedings of the 18th Conference on Embedded
  Networked Sensor Systems}}. \bibinfo{pages}{476--488}.
\newblock


\bibitem[\protect\citeauthoryear{Yu and Principe}{Yu and Principe}{2019}]%
        {yu2019understanding}
\bibfield{author}{\bibinfo{person}{Shujian Yu} {and} \bibinfo{person}{Jose~C
  Principe}.} \bibinfo{year}{2019}\natexlab{}.
\newblock \showarticletitle{{Understanding autoencoders with information
  theoretic concepts}}.
\newblock \bibinfo{journal}{\emph{Neural Networks}}  \bibinfo{volume}{117}
  (\bibinfo{year}{2019}), \bibinfo{pages}{104--123}.
\newblock


\bibitem[\protect\citeauthoryear{Yu, Wickstr{\o}m, Jenssen, and
  Pr{\'\i}ncipe}{Yu et~al\mbox{.}}{2020}]%
        {yu2020understanding}
\bibfield{author}{\bibinfo{person}{Shujian Yu}, \bibinfo{person}{Kristoffer
  Wickstr{\o}m}, \bibinfo{person}{Robert Jenssen}, {and}
  \bibinfo{person}{Jos{\'e}~C Pr{\'\i}ncipe}.} \bibinfo{year}{2020}\natexlab{}.
\newblock \showarticletitle{{Understanding convolutional neural networks with
  information theory: An initial exploration}}.
\newblock \bibinfo{journal}{\emph{IEEE transactions on neural networks and
  learning systems}} (\bibinfo{year}{2020}).
\newblock


\bibitem[\protect\citeauthoryear{Zagoruyko and Komodakis}{Zagoruyko and
  Komodakis}{2016}]%
        {zagoruyko2016wide}
\bibfield{author}{\bibinfo{person}{Sergey Zagoruyko} {and}
  \bibinfo{person}{Nikos Komodakis}.} \bibinfo{year}{2016}\natexlab{}.
\newblock \showarticletitle{Wide Residual Networks}. In
  \bibinfo{booktitle}{\emph{Proceedings of the British Machine Vision
  Conference (BMVC)}}. \bibinfo{publisher}{BMVA Press},
  \bibinfo{pages}{87.1--87.12}.
\newblock


\bibitem[\protect\citeauthoryear{Zeng, Li, Zhou, and Chen}{Zeng
  et~al\mbox{.}}{2019}]%
        {zeng2019boomerang}
\bibfield{author}{\bibinfo{person}{Liekang Zeng}, \bibinfo{person}{En Li},
  \bibinfo{person}{Zhi Zhou}, {and} \bibinfo{person}{X. Chen}.}
  \bibinfo{year}{2019}\natexlab{}.
\newblock \showarticletitle{{Boomerang: On-Demand Cooperative Deep Neural
  Network Inference for Edge Intelligence on the Industrial Internet of
  Things}}.
\newblock \bibinfo{journal}{\emph{IEEE Network}}  \bibinfo{volume}{33}
  (\bibinfo{year}{2019}), \bibinfo{pages}{96--103}.
\newblock


\bibitem[\protect\citeauthoryear{Zhang, Polese, Mezzavilla, Zhu, Rangan,
  Panwar, and Zorzi}{Zhang et~al\mbox{.}}{2019}]%
        {zhang2019will}
\bibfield{author}{\bibinfo{person}{Menglei Zhang}, \bibinfo{person}{Michele
  Polese}, \bibinfo{person}{Marco Mezzavilla}, \bibinfo{person}{Jing Zhu},
  \bibinfo{person}{Sundeep Rangan}, \bibinfo{person}{Shivendra Panwar}, {and}
  \bibinfo{person}{Michele Zorzi}.} \bibinfo{year}{2019}\natexlab{}.
\newblock \showarticletitle{{Will TCP work in mmWave 5G cellular networks?}}
\newblock \bibinfo{journal}{\emph{IEEE Communications Magazine}}
  \bibinfo{volume}{57}, \bibinfo{number}{1} (\bibinfo{year}{2019}),
  \bibinfo{pages}{65--71}.
\newblock


\bibitem[\protect\citeauthoryear{Zhang, Zhang, Yang, Wei, Wang, Jiao, and
  Zhang}{Zhang et~al\mbox{.}}{2020}]%
        {zhang2020person}
\bibfield{author}{\bibinfo{person}{Shizhou Zhang}, \bibinfo{person}{Qi Zhang},
  \bibinfo{person}{Yifei Yang}, \bibinfo{person}{Xing Wei},
  \bibinfo{person}{Peng Wang}, \bibinfo{person}{Bingliang Jiao}, {and}
  \bibinfo{person}{Yanning Zhang}.} \bibinfo{year}{2020}\natexlab{}.
\newblock \showarticletitle{Person Re-identification in Aerial imagery}.
\newblock \bibinfo{journal}{\emph{IEEE Transactions on Multimedia}}
  \bibinfo{volume}{23} (\bibinfo{year}{2020}), \bibinfo{pages}{281--291}.
\newblock


\bibitem[\protect\citeauthoryear{Zhang, Zhao, and LeCun}{Zhang
  et~al\mbox{.}}{2015}]%
        {zhang2015characterlevel}
\bibfield{author}{\bibinfo{person}{X. Zhang}, \bibinfo{person}{J. Zhao}, {and}
  \bibinfo{person}{Y. LeCun}.} \bibinfo{year}{2015}\natexlab{}.
\newblock \showarticletitle{Character-level Convolutional Networks for Text
  Classification}. In \bibinfo{booktitle}{\emph{NIPS}}.
\newblock


\bibitem[\protect\citeauthoryear{Zhou, Xu, Ge, McAuley, Xu, and Wei}{Zhou
  et~al\mbox{.}}{2020}]%
        {zhou2020bert}
\bibfield{author}{\bibinfo{person}{Wangchunshu Zhou}, \bibinfo{person}{Canwen
  Xu}, \bibinfo{person}{Tao Ge}, \bibinfo{person}{Julian McAuley},
  \bibinfo{person}{Ke Xu}, {and} \bibinfo{person}{Furu Wei}.}
  \bibinfo{year}{2020}\natexlab{}.
\newblock \showarticletitle{{BERT Loses Patience: Fast and Robust Inference
  with Early Exit}}.
\newblock \bibinfo{journal}{\emph{Advances in Neural Information Processing
  Systems}}  \bibinfo{volume}{33} (\bibinfo{year}{2020}).
\newblock


\bibitem[\protect\citeauthoryear{Zoph and Le}{Zoph and Le}{2017}]%
        {zoph2017neural}
\bibfield{author}{\bibinfo{person}{Barret Zoph} {and} \bibinfo{person}{Quoc
  Le}.} \bibinfo{year}{2017}\natexlab{}.
\newblock \showarticletitle{Neural Architecture Search with Reinforcement
  Learning}. In \bibinfo{booktitle}{\emph{International Conference on Learning
  Representations}}.
\newblock


\bibitem[\protect\citeauthoryear{Zoph, Vasudevan, Shlens, and Le}{Zoph
  et~al\mbox{.}}{2018}]%
        {zoph2018learning}
\bibfield{author}{\bibinfo{person}{Barret Zoph}, \bibinfo{person}{Vijay
  Vasudevan}, \bibinfo{person}{Jonathon Shlens}, {and} \bibinfo{person}{Quoc~V
  Le}.} \bibinfo{year}{2018}\natexlab{}.
\newblock \showarticletitle{{Learning Transferable Architectures for Scalable
  Image Recognition}}. In \bibinfo{booktitle}{\emph{Proceedings of the IEEE
  conference on computer vision and pattern recognition}}.
  \bibinfo{pages}{8697--8710}.
\newblock


\end{thebibliography}

\end{document}